\documentclass[a4paper,british,english]{IEEEtran}
\usepackage[T1]{fontenc}
\usepackage[latin9]{inputenc}
\usepackage{bm}
\usepackage{amsmath}
\usepackage{amssymb}
\usepackage{graphicx}

\makeatletter

\pdfpageheight\paperheight
\pdfpagewidth\paperwidth

\providecommand{\tabularnewline}{\\}

\usepackage{fancyhdr}
\usepackage{amsmath,graphicx}

\usepackage{graphicx}

\usepackage{epstopdf}

\usepackage{tikz}
\usetikzlibrary{matrix,shapes,arrows,positioning,chains}

\usepackage{graphicx}
\usetikzlibrary{ arrows,  shapes.misc,  shapes.arrows,  chains,  matrix,  positioning,  scopes,  decorations.pathmorphing,  shadows, calc}
\tikzset{
nonterminal/.style={rectangle, minimum size=6mm, very thick, draw=red!50!black!20, top color=white, bottom color=red!50!black!50, font=\itshape },
terminal/.style={rounded rectangle,  minimum size=6mm, very thick,draw=black!50, top color=white, bottom color=black!20, font=\ttfamily},
junction/.style={circle, draw }}
\pgfsetlinewidth{.4mm}

\usetikzlibrary{decorations.pathmorphing} 
\usetikzlibrary{matrix} 
\usetikzlibrary{arrows} 
\usetikzlibrary{calc} 

\tikzstyle{block} = [draw,rectangle,thick,minimum height=2em,minimum width=2em]
\tikzstyle{sum} = [draw,circle,inner sep=0mm,minimum size=2mm]
\tikzstyle{connector} = [->,thick]
\tikzstyle{line} = [thick]
\tikzstyle{branch} = [circle,inner sep=0pt,minimum size=1mm,fill=black,draw=black]
\tikzstyle{branch2} = [circle,inner sep=0pt,minimum size=0mm,fill=black,draw=black]
\tikzstyle{guide} = []
\tikzstyle{snakeline} = [connector, decorate, decoration={pre length=0.2cm,
                         post length=0.2cm, snake, amplitude=.4mm,
                         segment length=2mm},thick, magenta, ->]

\tikzset{
    state/.style={
           rectangle,
           rounded corners,
           draw=black, very thick,
           minimum height=1em,
           inner sep=1pt,
           text centered,
           },
}

\usetikzlibrary{calc,backgrounds}
\usepackage{verbatim}

\usetikzlibrary{positioning}
\usetikzlibrary{decorations.text}
\usetikzlibrary{decorations.pathmorphing}

\usepackage{bm}

\usepackage{ctable}

\usetikzlibrary{fit}					
\usetikzlibrary{backgrounds}	

\usepackage{lettrine}

\usepackage{url}

\fancypagestyle{firstpage}{
  \fancyhf{}
  \fancyhead[L]{Nikolaos Dionelis, Mike Brookes, Submitted to IEEE Transactions on Audio, Speech and Language Processing}
  \fancyhead[R]{1}
}

\makeatother

\usepackage{babel}
\begin{document}

\title{Modulation-Domain Kalman Filtering for Monaural Blind Speech Denoising
and Dereverberation }

\author{N. Dionelis, \url{https://www.commsp.ee.ic.ac.uk/~sap/people-nikolaos-dionelis/},
M. Brookes, Member, IEEE}
\maketitle
\begin{abstract}
We describe a monaural speech enhancement algorithm based on modulation-domain
Kalman filtering to blindly track the time-frequency log-magnitude
spectra of speech and reverberation. We propose an adaptive algorithm
that performs blind joint denoising and dereverberation, while accounting
for the inter-frame speech dynamics, by estimating the posterior distribution
of the speech log-magnitude spectrum given the log-magnitude spectrum
of the noisy reverberant speech. The Kalman filter update step models
the non-linear relations between the speech, noise and reverberation
log-spectra. The Kalman filtering algorithm uses a signal model that
takes into account the reverberation parameters of the reverberation
time, $T_{60}$, and the direct-to-reverberant energy ratio (DRR)
and also estimates and tracks the $T_{60}$ and the DRR in every frequency
bin in order to improve the estimation of the speech log-magnitude
spectrum. The Kalman filtering algorithm is tested and graphs that
depict the estimated reverberation features over time are examined.
The proposed algorithm is evaluated in terms of speech quality, speech
intelligibility and dereverberation performance for a range of reverberation
parameters and SNRs, in different noise types, and is also compared
to competing denoising and dereverberation techniques. Experimental
results using noisy reverberant speech demonstrate the effectiveness
of the enhancement algorithm.
\end{abstract}


\IEEEpeerreviewmaketitle{
\begin{table}[b]

\IEEEpeerreviewmaketitle{\quad N. Dionelis and M. Brookes are with the Department of Electrical and Electronic Engineering, Imperial College London, London SW7 2AZ, U.K. (e-mail: nikolaos.dionelis11@imperial.ac.uk; mike.brookes@imperial.ac.uk).}
\end{table}
}

\section{Introduction }

\lettrine[lines=2]{N}{owadays}, technology is ever evolving with
tremendous haste and the demand for speech enhancement systems is
evident. Speech enhancement in noisy reverberant environments, for
human listeners, is challenging. Speech is degraded by noise and reverberation
when captured using a near-field or far-field distant microphone \cite{a257}
\cite{a333}. A room impulse response (RIR) can include components
at long delays, hence resulting in reverberation and echoes \cite{a262}
\cite{a261}. Reverberation is a convolutive distortion that can be
quite long with a reverberation time, $T_{60}$, of more than $0.8$
s. Due to convolution, reverberation induces long-term correlation
between consecutive observations. Reverberation and noise, which can
be stationary or non-stationary, have a detrimental impact on speech
quality and intelligibility. Reverberation, especially in the presence
of non-stationary noise, damages the intelligibility of speech.\thispagestyle{firstpage}

The direct to reverberant energy ratio (DRR) and the reverberation
time, $T_{60}$, are the two main parameters of a reverberation model
\cite{a352} \cite{a257}. The DRR describes reverberation in the
space domain, depending on the positions of the sound source and the
receiver. The $T_{60}$ is the time interval required for a sound
level to decay $60$ dB after ceasing its original stimulus. The reverberation
time, when measured in the diffuse sound field, is independent of
the source to microphone configuration and mainly depends on the room.
The impact of reverberation on auditory perception depends on the
$T_{60}$. If the $T_{60}$ is short, the environment reinforces the
sound which may enhance the sound perception \cite{Lebart2001} \cite{a286}.
On the contrary, if the $T_{60}$ is long, spoken syllables interfere
with future spoken syllables. Reverberation spreads energy over time
and this smearing across time has two effects: (a) the energy of individual
phonemes spreads out in time and, hence, plosives have a delayed decay
and fricatives are smoothed, and (b) preceding phonemes blur into
the current phonemes.

The aim of speech enhancement is to reduce and ideally eliminate the
effects of both noise and reverberation without distorting the speech
signal \cite{a361}. Enhancement algorithms typically aim to suppress
noise and late reverberation because early reverberation is not perceived
as separate sound sources and usually improves the quality and intelligibility
of speech. Noise is assumed to be uncorrelated with speech, early
reverberation is correlated with speech and late reverberation is
commonly assumed to be uncorrelated with speech \cite{Habets2007}
\cite{Lebart2001}.

Speech enhancement can be performed in different domains. The ideal
domain should be chosen such that (a) good statistical models of speech
and noise exist in this domain, and (b) speech and noise are separable
in this domain. Speech and noise are additive in the time domain and
the Short Time Fourier Transform (STFT) domain \cite{a278} \cite{a126}.
The relation between speech and noise becomes progressively more complicated
in the amplitude, power and log-power spectral domains. Noise suppression
algorithms usually operate in a time-frequency STFT domain and these
techniques have been extended to address dereverberation. In \cite{Habets2007},
spectral enhancement methods based on a time-frequency gain, originally
developed for noise suppression, have been modified and employed for
dereverberation. Such algorithms suppress late reverberation assuming
that the early and late reverberation components are uncorrelated.
The spectral enhancement methods in \cite{a365} \cite{Habets2007}
estimate the late reverberant spectral variance (LRSV) and use it
in the place of the noise spectral variance, reducing the problem
of late reverberation suppression to that of estimating the LRSV \cite{Lebart2001}.
In \cite{a286}, blind spectral weighting is employed to reduce the
overlap-masking effect of reverberation using an uncorrelated and
additive assumption for late reverberation.

Dereverberation algorithms that leave the phase unaltered and operate
in the amplitude, power or log-power spectral domains are relatively
insensitive to minor variations in the spatial placement of sources
\cite{a285}. Two criticisms of spectral enhancement algorithms based
on LRSV reverberation noise estimation are that they introduce musical
noise and suppress speech onsets when they over-estimate reverberation
\cite{a253}. The LRSV estimator in \cite{a277}, which is a continuation
of \cite{Lebart2001}, models the RIR in the STFT domain and not in
the time domain \cite{Lebart2001} \cite{a363}, using the same model
of the RIR that is attributed to J. Polack or J. Moorer \cite{a363}.
Reverberation is estimated in \cite{a277} considering the STFT energy
contribution of the direct path of speech and an external $T_{60}$
estimate.

Modelling the speech temporal dynamics is beneficial when the $T_{60}$
is long and the DRR is low \cite{a241} \cite{a333}. Joint denoising
and dereverberation using speech and noise tracking is performed in
\cite{a241}. The SPENDRED algorithm \cite{a179} \cite{a175}, which
is a model-based method with a convolution model for reverberation
based on the $T_{60}$ and the DRR, considers the speech temporal
dynamics. SPENDRED employs a parametric model of the RIR \cite{a237}
and performs frequency-dependent and time-varying $T_{60}$ and DRR
estimation. However, unless the source or the microphone are moving,
the $T_{60}$ and the DRR will be constant throughout the recording.
The SPENDRED algorithm assumes that $\text{DRR} \geq 1 - (10^{-6})^{\frac{L}{T_{60}}}$
where $L$ is the acoustic frame increment. For example, when $T_{60}=0.4$
s and $L=5$ ms, then $\text{DRR} \geq -8$ dB is assumed. In addition,
SPENDRED performs intra-frame correlation modelling, which can be
beneficial in adverse conditions, while typical algorithms decouple
different frequency dimensions \cite{a278}.

Statistical-based models, such as the SPENDRED algorithm, describe
reverberation by a convolution in the power spectral domain while
LRSV models describe reverberation as an additive distortion in the
power spectral domain \cite{a237} \cite{a181}. A model with an infinite
impulse response is used either with the two parameters of the $T_{60}$
and the DRR, as in \cite{a179} \cite{a175}, or with a finite number
of parameters. The infinite-order convolution model of reverberation
with the $T_{60}$ and the DRR is sparse and contrasts with the higher-order
autoregressive processes in the complex STFT domain, used in \cite{a339}
\cite{a354}.

The algorithms described in \cite{a282}, \cite{a237} and \cite{a181}
create non-linear observation models of noisy reverberant speech in
the log Mel-power spectral domain, using the reverberation-to-noise
ratio (RNR). As discussed in \cite{a238}, phase differences in Mel-frequency
bands have different properties from phase differences in STFT bins.
The phase factor between reverberant speech and noise is different
from that between speech and noise \cite{a282}. In \cite{a237},
the phase factor between reverberant speech and noise in Mel-frequency
bands is examined.

In noisy reverberant conditions, finding the onset of speech phonemes
and determining which frames are unvoiced/silence is difficult, due
to the smearing across time, often leading to noise over-estimation.
The concatenation of different techniques for denoising and dereverberation
has lower performance than unified methods due to over-estimating
noise when estimating noise and reverberation separately \cite{a241}
\cite{a358}.

Despite the claim that it is inefficient to perform a two step procedure
that is comprised of denoising followed by dereverberation \cite{a241}
\cite{a358}, long-term linear prediction with pre-denoising can be
used to suppress noise and reverberation. With the weighted prediction
error (WPE) algorithm \cite{a254} \cite{a333}, reverberation is
represented as a one-dimensional convolution in each frequency bin.
In \cite{a250}, the WPE algorithm is discussed along with inter-frame
correlation. In \cite{a248}, the WPE algorithm is used in the complex
STFT domain performing batch processing and iteratively estimating,
first, the reverberation prediction coefficients and, then, the speech
spectral variance. The WPE linear filtering approach, which can be
employed in the power spectral domain \cite{a249} \cite{a284}, takes
into account past frames, from the $3$-rd to the $40$-th past frame
\cite{a248} \cite{a249}.

This paper presents an adaptive denoising and dereverberation Kalman
filtering framework that tracks the speech and reverberation spectral
log-magnitudes. In this paper, we extend the enhancer in \cite{a225}
to include dereverberation. Enhancement is performed using a Kalman
filter (KF) to model inter-frame correlations. We use an integrated
structure of two parallel signal models to track speech, reverberation
and the $T_{60}$ and DRR reverberation parameters. The $T_{60}$
and the DRR are updated in every frame to improve the estimation of
the speech log-magnitude spectrum. We create an observation model
and a series of non-linear KF update steps performing joint noise
and reverberation suppression by estimating the first two moments
of the posterior distribution of the speech log-spectrum given the
noisy reverberant log-spectrum. The log-spectral domain is chosen,
as in \cite{a268} \cite{a225}, because good speech models exist
in this domain. Modelling spectral log-amplitudes as Gaussian distributions
leads to good speech modelling in noisy reverberant environments since
super-Gaussian distributions that resemble the log-normal, such as
the Gamma \cite{a126} \cite{a123}, are used to model the speech
amplitude spectrum. Mean squared errors (MSEs) in the log-spectral
domain are a good measure to use for perceptual quality and speech
log-spectra are well modelled by Gaussian distributions, as in \cite{a236}
and \cite{a360}.

The structure of this paper is as follows. Section II describes the
signal model and Sec. III presents the enhancement algorithm and its
non-linear KF. The implementation and the validation of the algorithm
are in Sec. IV. The algorithm's evaluation is in Sec. V. Conclusions
are drawn in Sec. VI.

\section{Signal model and notation}

In the complex STFT domain, the noisy speech, $Y_t(k)$, is given
by $Y_t(k) = S_t(k) + R_t(k) + N_t(k)$ where $S_t(k)$ is the direct
speech component, $R_t(k)$ is the reverberant speech component and
$N_t(k)$ is the noise, as for example in \cite{a339} \cite{a354}.
The time-frame index is $t$ and the frequency bin index is $k$.
For clarity, we also define $Z_t(k)=R_t(k)+N_t(k)$. We drop the time
and frequency indexes and we obtain $Y = S + Z = S + R + N$. We define
the log-magnitude spectrum of $S$ as $s=\log(|S|)$ and we also define
$r$, $n$, $y$ and $z$ similarly.

In the signal model,\foreignlanguage{british}{ signal quantities with
capital letters, such as $S_t$, are complex numbers with magnitude
and phase values, $|S_t|$ and $\angle S_t$. }In the complex STFT
domain, using $a_t,b_t \in \mathbb{R}$, \foreignlanguage{british}{the
reverberation signal model} is given by
\begin{align}
R_{t} & =\sqrt{a_{t}}\,R_{t-1}\exp(j\theta_{t})+\sqrt{b_{t}}\,S_{t-1}\exp(j\psi_{t})\label{eq:1}\\
 & =\sum_{\tau=1}^{\infty}\left(\prod_{i=1}^{\tau-1}\left(\sqrt{a_{t-i+1}}\exp(j\theta_{t-i+1})\right)\right.\nonumber \\
 & \times\left.\sqrt{b_{t-\tau+1}}S_{t-\tau}\exp(j\psi_{t-\tau+1})\right)\text{,}\nonumber 
\end{align}
\begin{align}
 & a_{t}^{\frac{T_{60}}{L}}=10^{-6},\ \ \ \ \ \ \ \ b_{t}=\frac{1-a_{t}}{\text{DRR}}\label{eq:2}
\end{align}
where $L$ is the acoustic frame increment. In (\ref{eq:1}), the
factors $\exp(j \theta_t)$ and $\exp(j \psi_t)$, where $\theta_t$
and $\psi_t$ are uniformly distributed phases, are used. In (2),
the DRR is defined in the power spectral domain \cite{a179} \cite{a175}
and the $T_{60}$ and the DRR are both time and frequency dependent,
as described in \cite{a352}.

The expression in (\ref{eq:1}) is the convolution model for reverberation;
the most common reverberation model is this single-pole filter that
is described by the pole and zero positions that depend on the $T_{60}$
and the DRR \cite{a179} \cite{a333}. A convolution of infinite order
is used, with the two parameters of the $T_{60}$ and the DRR, to
describe reverberation \cite{a179} \cite{a175}. Models that describe
reverberation by a convolution are also discussed in \cite{a237}
\cite{a181}. \foreignlanguage{british}{The signal model is defined
by (1) and by}
\begin{align}
 & Y_{t}=S_{t}+Z_{t},\ \ \ \ \ \ Z_{t}=R_{t}+N_{t}\text{,}\label{eq:3}\\
 & \gamma_{t}=0.5\log\left(a_{t}\right),\ \ \delta_{t}=\gamma_{t}+r_{t-1}=\log\left(\sqrt{a_{t}}\left|R_{t-1}\right|\right)\text{,}\label{eq:5}\\
 & \beta_{t}=0.5\log\left(b_{t}\right),\ \ \epsilon_{t}=\beta_{t}+s_{t-1}=\log(\sqrt{b_{t}}\left|S_{t-1}\right|)\label{eq:6}
\end{align}
\foreignlanguage{british}{where} $b_t > 0$, $0 < a_t < 1$ and $\gamma_t < 0$.

Figure 1 shows graphs of $\beta$ against $T_{60}$ for a fixed DRR
and of $\beta$ against DRR for a fixed $T_{60}$. If $\text{DRR} = 0$
dB, then $b_t = 1 - a_t$. If $\beta_t = 0$, then $b_t=1$ and $\text{DRR} = 1-a_t$.

Figure 2 illustrates the flowchart of the signal model. The \foreignlanguage{british}{reverberation
signal model} in (\ref{eq:1}) uses $\sqrt{a_t}$ and $\sqrt{b_t}$
because the $a_t$ and $b_t$ reverberation parameters, in (\ref{eq:2}),
and the DRR are defined in the power spectral domain, as in \cite{a179}
\cite{a175}. The $a_t$ and $b_t$ parameters are mapped to $\gamma_t$
and $\beta_t$ using (\ref{eq:5}) and (\ref{eq:6}). The signals
$z_t$, $\delta_t$ and $\epsilon_t$ are the total distubance, the
old (decaying) reverberation and the new reverberation, respectively.
We note that $z_t$ is defined in the first paragraph of this section
and that $\delta_t$ and $\epsilon_t$ are defined in (\ref{eq:5})
and (\ref{eq:6}).

The signal model of how the reverberation parameters of $\gamma_t$
and $\beta_t$ change over time is a random walk model. This is used
in the algorithm's KF prediction step for $\gamma_t$ and $\beta_t$.

The signal model in Fig. 2 is directly linked to the alternating and
interacting KFs of the enhancement algorithm. The algorithm is
a collection of two KFs, the speech KF and the reverberation KF, that
estimate the speech and reverberation log-amplitude spectra and the
$\gamma_t$ and $\beta_t$ reverberation parameters. This KF algorithm
is described in detail in Sec. III.

\section{The speech enhancement algorithm}

The KF algorithm operates in the log-magnitude spectral domain, tracking
speech and reverberation. Figure \ref{fig:EDCs-1-2-1-2-1-1-1-2-1-1-1-2-1-1-1-1-1-1-1-1-2-5-001-1-3}
depicts the denoising and dereverberation algorithm that formulates
a model of reverberation as a first-order autoregressive process and
propagates the means and variances of the random variables. Almost
all the signals follow a Gaussian distribution and the distribution
of $s_t$ conditioned on observations up to time $\tau$ is given
by $p_{s_{t| \tau}}(s) \triangleq p(s_t | y_0, \dots, y_{\tau}) = N(s_{t| \tau}, \Sigma^{(s)}_{t| \tau})$.
In Fig. \ref{fig:EDCs-1-2-1-2-1-1-1-2-1-1-1-2-1-1-1-1-1-1-1-1-2-5-001-1-3},
a Gaussian distribution is denoted by its mean, $s_{t|\tau}$.

The core of the algorithm in Fig. \ref{fig:EDCs-1-2-1-2-1-1-1-2-1-1-1-2-1-1-1-1-1-1-1-1-2-5-001-1-3}
is the KF that is defined by the gray blocks in the flowchart diagram.
The non-linear KF estimates and tracks the posterior distributions
of the speech log-magnitude spectrum, $s_t$, the reverberation log-magnitude
spectrum, $r_t$, and the reverberation parameters, $\gamma_t$ and
$\beta_t$.

The input to the algorithm in Fig. \ref{fig:EDCs-1-2-1-2-1-1-1-2-1-1-1-2-1-1-1-1-1-1-1-1-2-5-001-1-3}
is the noisy reverberant speech in the time domain. The algorithm's
first step is to perform a STFT and obtain the signal in the complex
STFT domain. The algorithm does not alter the noisy reverberant phase,
$\angle Y$, and uses the noisy reverberant amplitude spectrum, $|Y|$,
in three ways: in the speech KF prediction step, in the KF update
step and in the noise power modelling. The main part of the algorithm
is the KF and the speech KF state, $\textbf{s}_{t} \in \mathbb{R}^{p}$,
is the speech log-spectrum from the previous $p$ frames,
\begin{align}
 & \textbf{s}_{t}=(s_{t}\ s_{t-1}\ \dots\ s_{t-p+1})^{T}\text{.}\label{eq:rr6}
\end{align}
The speech KF prediction step is based on autoregressive (AR) modelling
on the log-spectrum of pre-cleaned speech \cite{a268}. The reverberation
KF state is $r_t$ and the KF states of the reverberation parameters
are $\gamma_t$ and $\beta_t$. The KF observation is the noisy reverberant
speech log-spectrum, $y_t$, which is used in the KF update step to
compute the first two moments of the posterior of the speech log-spectrum.
The mean of the speech log-spectrum posterior is used together with
$\angle Y$ to create the enhanced speech signal using the inverse
STFT (ISTFT).

Apart from the speech log-spectrum, the non-linear KF also tracks
the reverberation log-spectrum, $r_t$, and the $\gamma_t$ and $\beta_t$
reverberation parameters. The KF, as defined by the gray blocks in
Fig. \ref{fig:EDCs-1-2-1-2-1-1-1-2-1-1-1-2-1-1-1-1-1-1-1-1-2-5-001-1-3},
has a speech KF prediction step, a reverberation KF prediction step
and a series of KF update steps. The reverberation KF is comprised
of the blocks ``Reverberation KF prediction'', ``KF Update'' and
``$\gamma_t$, $\beta_t$ KF Update''. These three blocks perform
joint denoising and dereverberation and estimate $s_{t|t}$ and $\Sigma_{t|t}^{\text{(s)}}$
to enhance noisy reverberant speech.

\begin{figure}[t]
\begin{minipage}[t]{0.48\columnwidth}%
\centering \includegraphics[bb=12bp 0bp 510bp 400bp,clip,width=1\columnwidth]{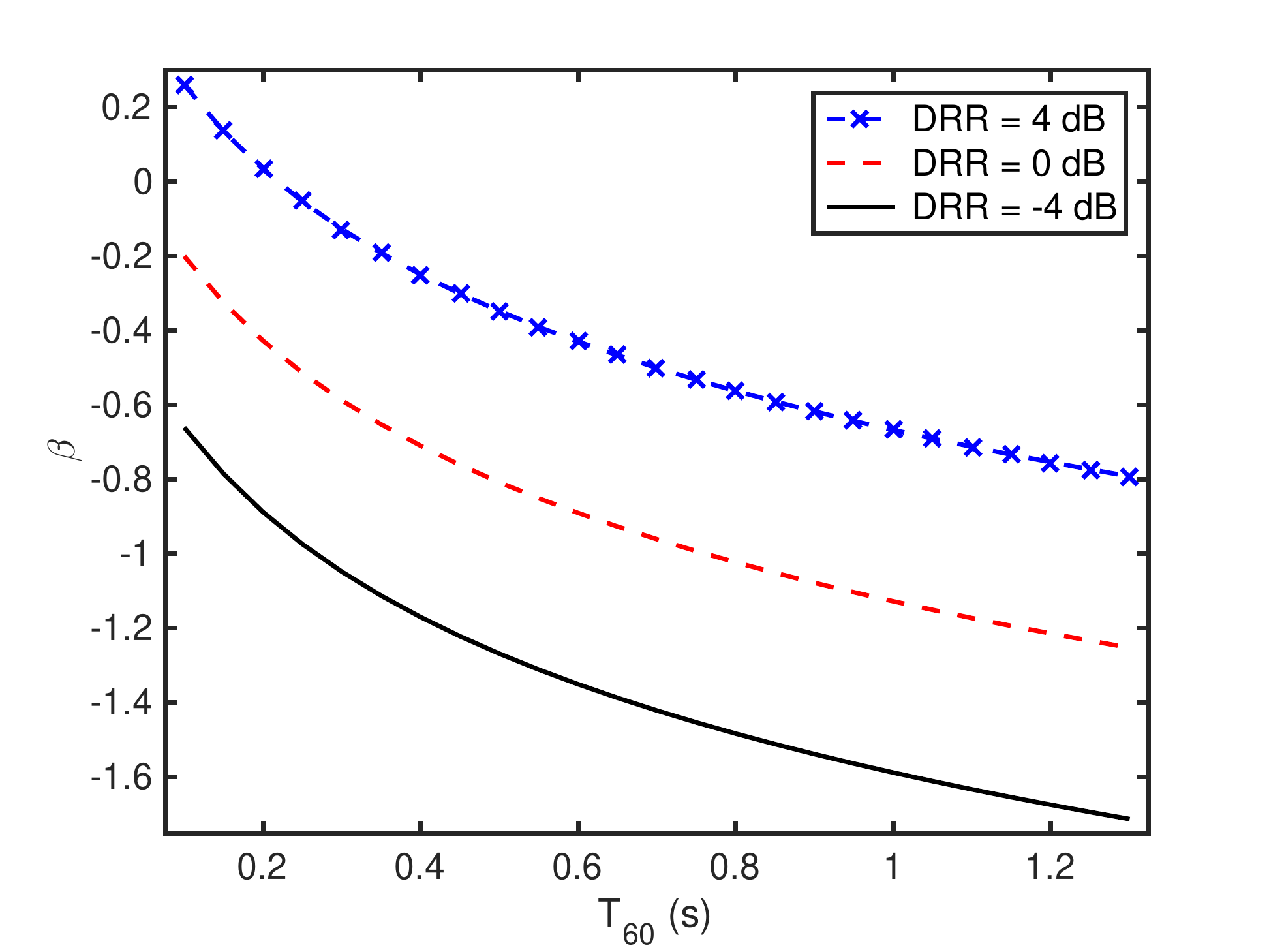}

\footnotesize{(a)}

\label{fig:EDCs-1-2-1-2-1-1-1-2-1-1-1-2-1-1-1-1-1-1-1-1-2-5-001-1-1-1-1-4-4}%
\end{minipage}\hfill{}%
\begin{minipage}[t]{0.48\columnwidth}%
\centering \includegraphics[bb=12bp 0bp 510bp 400bp,clip,width=1\columnwidth]{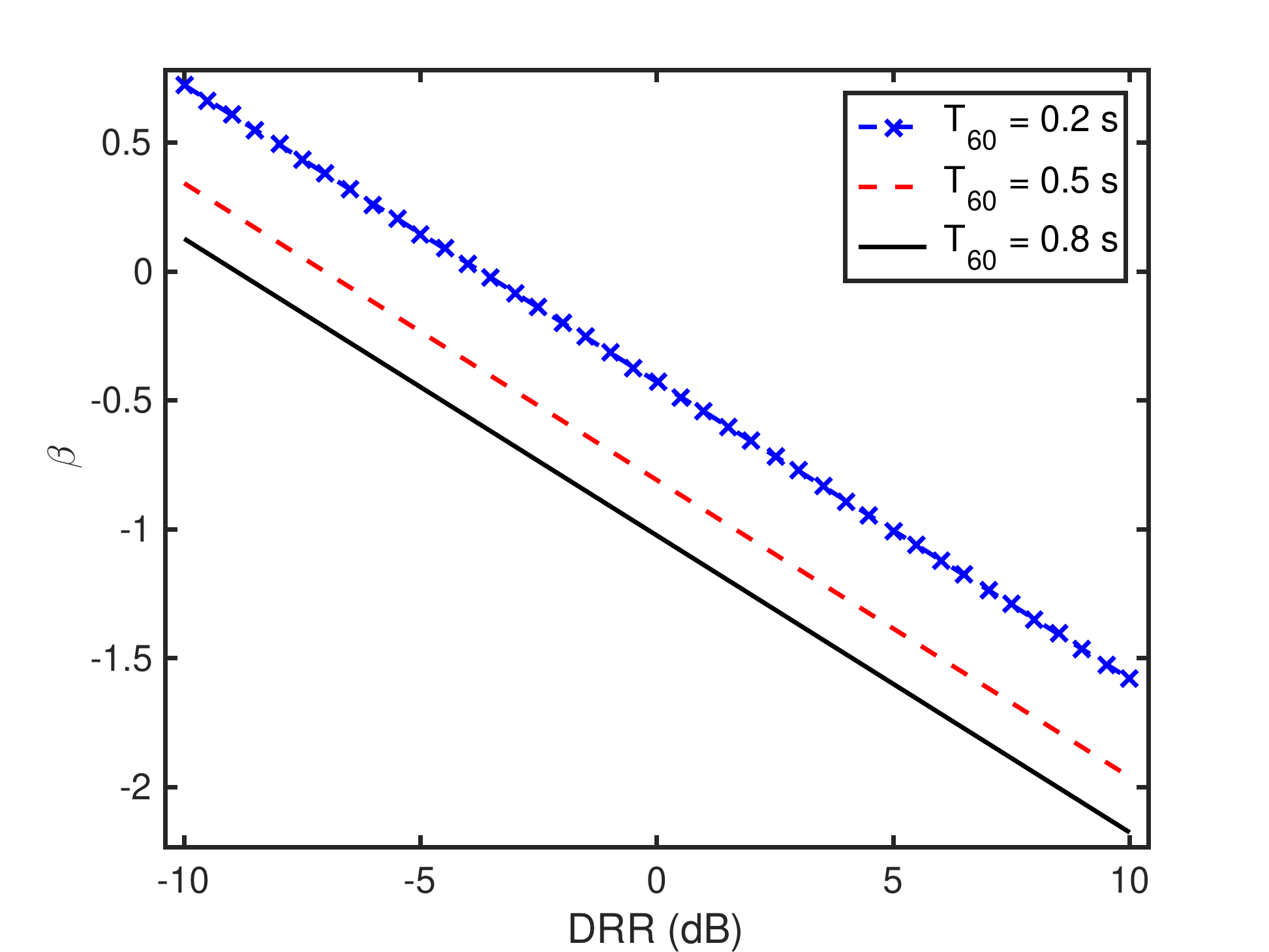}

\footnotesize{(b)}

\label{fig:EDCs-1-2-1-2-1-1-1-2-1-1-1-2-1-1-1-1-1-1-1-1-2-5-001-1-1-2-4-3-1}%
\end{minipage}

\caption{Plot of $\beta$, using $L=8$ ms in (\ref{eq:2}), against: (a) $T_{60}$
when the DRR is $4$, $0$ and $-4$ dB, and (b) DRR when the $T_{60}$
is $0.2$, $0.5$ and $0.8$ s.}
\end{figure}

\begin{figure}[t]
\centering\includegraphics[clip,width=0.97\columnwidth]{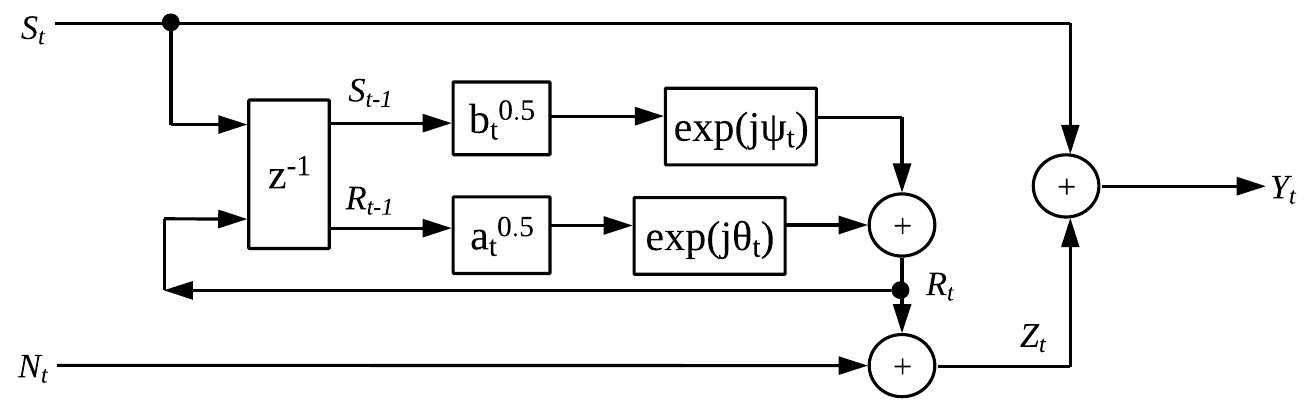}

\label{fig:EDCs-1-2-1-2-1-1-1-2-1-1-1-2-1-1-1-1-1-1-1-1-2-5-001-1-3-2-1}

\caption{The flowchart diagram of the proposed signal model of the speech enhancement
algorithm. The signal model is defined in (1) and (3).}
\end{figure}

\begin{figure}[t]
\centering\includegraphics[clip,width=1\columnwidth]{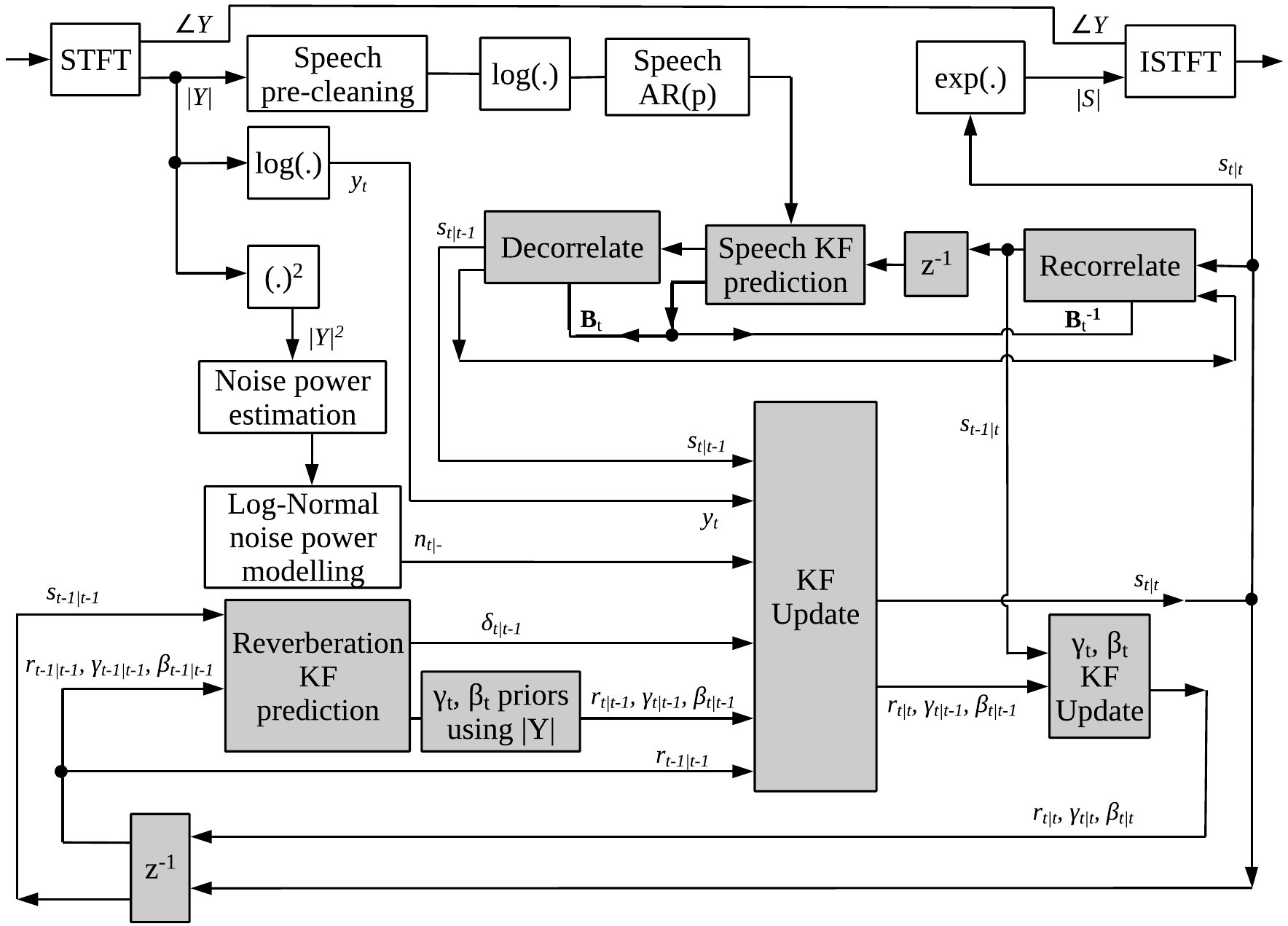}\caption{The flowchart of the algorithm to suppress noise and reverberation.
For clarity, the signal variances are not included. The gray blocks
constitute the KF that tracks the speech and reverberation log-spectra
and the reverberation parameters. The ``KF Update'' and ``$\gamma_t$,
$\beta_t$ KF Update'' blocks constitute the non-linear KF update
step that uses the signal model described in Sec. II.}

\label{fig:EDCs-1-2-1-2-1-1-1-2-1-1-1-2-1-1-1-1-1-1-1-1-2-5-001-1-3} 
\end{figure}

The structure of the rest of this algorithm description section is
as follows. Sections III.A and III.B present the speech and reverberation
KF prediction steps, respectively. Section III.C describes the KF
update step and Sec. III.D the priors for the $\gamma_t$ and $\beta_t$
parameters that are needed so that the KF (a) distinguishes between
speech and reverberation, and (b) does not diverge to non-realistic
$T_{60}$ and DRR estimates. Section III.E describes the unshaded
peripheral blocks in Fig. \ref{fig:EDCs-1-2-1-2-1-1-1-2-1-1-1-2-1-1-1-1-1-1-1-1-2-5-001-1-3}.

\subsection{The Speech KF Prediction Step}

The speech KF prediction step is linear and is related to the ``Speech
KF prediction'', ``Decorrelate'' and ``Recorrelate'' blocks in
Fig. \ref{fig:EDCs-1-2-1-2-1-1-1-2-1-1-1-2-1-1-1-1-1-1-1-1-2-5-001-1-3}.
The speech KF prediction step is described in \cite{a268} and in
\cite{a225} \cite{a236} and is based on conditional distributions
to model short-term dependencies. Decorrelation and recorrelation
of the speech KF state in (\ref{eq:rr6}) are performed after and
before the speech KF prediction step, respectively. The decorrelation
and recorrelation operations in Fig. \ref{fig:EDCs-1-2-1-2-1-1-1-2-1-1-1-2-1-1-1-1-1-1-1-1-2-5-001-1-3},
which are performed so that the non-linear KF update step can be applied,
perform vector-matrix and matrix-matrix multiplications for the speech
KF state mean and its covariance matrix, respectively, using $\textbf{B}_t \in \mathbb{R}^{p \times p}$
\cite{a123} \cite{a268}. The outputs of the ``Decorrelate'' block
are: (a) the first element of the speech KF state, and (b) the rest
elements of the speech KF state.

The KF prediction step propagates the first and second moments of
the speech KF state \cite{a268} \cite{a360}. Inter-frame linear
relationships are used for the speech KF prediction step that uses
AR modelling in the log-magnitude spectral domain. In the speech KF
prediction step, $\textbf{s}_t$ is predicted as a linear combination
of $\textbf{s}_{t-1}$ using the speech AR coefficients that are obtained
from the ``Speech AR(p)'' block in Fig. \ref{fig:EDCs-1-2-1-2-1-1-1-2-1-1-1-2-1-1-1-1-1-1-1-1-2-5-001-1-3},
which uses pre-cleaned speech as an input. After the speech KF prediction
step, $\textbf{s}_t$ is correlated; we decorrelate the speech KF
state with a linear transformation (using $\textbf{B}_t$) to simplify
the KF update step and impose the observation constraint \cite{a278}.
The KF update step changes only the first element of the speech KF
state and after the KF update step, recorrelation is applied with
a linear transformation (using $\textbf{B}_t^{-1}$) to continue the
KF recursion.

\subsection{The Reverberation KF Prediction Step}

The presented algorithm uses a KF prediction step for $\gamma_t$
and $\beta_t$ that assumes that the variance of $\gamma_t$ and $\beta_t$
increases over time, preserving their mean. The KF algorithm implements
a random-walk prediction step, performing the operations of
\begin{align}
 & \negthinspace\negthinspace\negthinspace\negthinspace\negthinspace\negthinspace\negthinspace\negthinspace\negthinspace\negthinspace\gamma_{t|t-1}^{\prime}=\gamma_{t-1|t-1},\ \ \ \ \ \Sigma_{t|t-1}^{(\gamma^{\prime})}=\Sigma_{t-1|t-1}^{(\gamma)}+Q_{\gamma},\label{eq:rr6-1}\\
 & \negthinspace\negthinspace\negthinspace\negthinspace\negthinspace\negthinspace\negthinspace\negthinspace\negthinspace\negthinspace\beta_{t|t-1}^{\prime}=\beta_{t-1|t-1},\ \ \ \ \ \Sigma_{t|t-1}^{(\beta^{\prime})}=\Sigma_{t-1|t-1}^{(\beta)}+Q_{\beta}
\end{align}
where $Q_{\gamma}$ is a fixed error variance for $\gamma$ and $Q_{\beta}$
is a fixed error variance for $\beta$. The values used for the prediction
error variances, $Q_{\gamma}$ and $Q_{\beta}$, depend on the rate
at which the $T_{60}$ and the DRR are likely to change in a real
situation.

After the reverberation KF prediction step, the algorithm computes
and imposes priors on $\gamma_t$ and $\beta_t$ using Gaussian-Gaussian
multiplication. The internally computed priors for $\gamma_t$ and
$\beta_t$ in the ``$\gamma_t$, $\beta_t$ priors'' block in Fig.
\ref{fig:EDCs-1-2-1-2-1-1-1-2-1-1-1-2-1-1-1-1-1-1-1-1-2-5-001-1-3}
are explained in Sec. III.D. After imposing the priors, the outputs
are $\gamma_{t|t-1}$, $\Sigma_{t|t-1}^{(\gamma)}$ and $\beta_{t|t-1}$,
$\Sigma_{t|t-1}^{(\beta)}$. We note that a prime diacritic, $\prime$,
is used in (7) and (8) to denote quantities before the priors.

\selectlanguage{british}%
The ``Reverberation KF prediction'' block in \foreignlanguage{english}{Fig.
\ref{fig:EDCs-1-2-1-2-1-1-1-2-1-1-1-2-1-1-1-1-1-1-1-1-2-5-001-1-3}}
estimates the first two moments of the prior distribution of the reverberation
spectral log-amplitude, i.e. $r_{t|t-1}$ and its variance.\foreignlanguage{english}{
The algorithm performs a reverberation KF prediction step based on
the previous posterior of both speech and reverberation using the
signal model in (\ref{eq:1}), where $a$ is less than unity and this
makes the reverberation KF prediction step stable.}

\selectlanguage{english}%
From (\ref{eq:1}) and Fig. 2, the STFT-domain reverberation is the
sum of two components arising, respectively, from the reverberation
and speech components of the previous frame. The old reverberation,
$\delta_t$, and the new reverberation, $\epsilon_t$, are defined
in (\ref{eq:5}) and (\ref{eq:6}), respectively. The KF algorithm
calculates the prior distributions of these two components in the
log-amplitude spectral domain using $\delta_{t|t-1} = \gamma_{t|t-1} + r_{t-1|t-1}$
and $\epsilon_{t|t-1} = \beta_{t|t-1} + s_{t-1|t-1}$. These equations
are based on (\ref{eq:5}) and (\ref{eq:6}) with a common condition
added to all terms. Assuming that $r_{t-1}$ and $s_{t-1}$ are uncorrelated
with $\gamma_t$ and $\beta_t$, respectively, \foreignlanguage{british}{the
means and variances of the two Gaussian distributions are added. The
variances therefore add,}
\begin{align}
 & \Sigma_{t|t-1}^{(\delta)}=\Sigma_{t|t-1}^{(\gamma)}+\Sigma_{t-1|t-1}^{\text{(r)}},\\
 & \Sigma_{t|t-1}^{(\epsilon)}=\Sigma_{t|t-1}^{(\beta)}+\Sigma_{t-1|t-1}^{\text{(s)}}\text{.}
\end{align}

As shown in Fig. \ref{fig:EDCs-1-2-1-2-1-1-1-2-1-1-1-2-1-1-1-1-1-1-1-1-2-5-001-1-3},
the final operation of t\foreignlanguage{british}{he ``Reverberation
KF prediction'' block is to compute the} prior distribution, $r_{t|t-1}$.
The addition in the c\foreignlanguage{british}{omplex STFT domain
of two random variables in the log-spectral domain is modelled. The
reverberation log-amplitude spectrum is estimated by modelling the
addition in the STFT domain of two random variables in the log-amplitude
spectral domain. }Given two disturbance sources, we combine them into
a single disturbance source.

From this point onwards in Sec. III.B, the time-frame subscript $t|t-1$
is omitted for clarity. For example, $p(\delta)$ is used instead
of $p_{\delta_{t|t-1}}(\delta)$, which is defined in Sec. III.

From (\ref{eq:1}), the reverberation component is the STFT-domain
sum of two elements arising, respectively, from the reverberation
and speech components in the previous frame. The log-amplitude spectral
domain distributions of these two elements, $\delta_{t|t-1}$ and
$\epsilon_{t|t-1}$, were calculated in the preceding paragraphs.
A two-dimensional Gaussian distribution is used for $p(\delta,\epsilon) = p(\delta) p(\epsilon)$
assuming independence between $\delta$ and $\epsilon$. We assume
that the phase difference, $\eta$, between the two disturbance sources,
$\delta$ and $\epsilon$, is uniformly distributed, i.e. $\eta \sim U(-\pi, \pi)$,
and independent of their magnitudes. We write $e^{2r}=e^{2\delta}+e^{2\epsilon}+2\cos(\eta)e^{\delta+\epsilon}$
and $r=0.5\log\left(e^{2\delta}+e^{2\epsilon}+2\cos(\eta)e^{\delta+\epsilon}\right)$,
which takes account of $\eta$ \cite{a225} \cite{a268}. Next, we
calculate
\begin{align}
\mathbb{E}\left\{ r\right\}  & =\int_{\eta=-\pi}^{\pi}p(\eta)\ \int_{\delta,\epsilon}r\ p(\delta,\epsilon)\ d\delta\ d\epsilon\ d\eta\nonumber \\
 & =\int_{\eta=0}^{\pi}\dfrac{1}{\pi}\ \int_{\delta,\epsilon}0.5\log\left(e^{2\delta}+e^{2\epsilon}+2\cos(\eta)e^{\delta+\epsilon}\right)\nonumber \\
 & \times p(\delta)\ p(\epsilon)\ d\delta\ d\epsilon\ d\eta\text{,}\label{eq:u11}\\
\mathbb{E}\left\{ r^{2}\right\}  & =\int_{\eta=0}^{\pi}\dfrac{1}{\pi}\ \int_{\delta,\epsilon}\left(0.5\log\left(e^{2\delta}+e^{2\epsilon}\right.\right.\nonumber \\
 & \left.\left.+2\cos(\eta)e^{\delta+\epsilon}\right)\right)^{2}\times p(\delta)\ p(\epsilon)\ d\delta\ d\epsilon\ d\eta\label{eq:7}
\end{align}
where $K_{(\delta,\epsilon)}$ sigma points are used to evaluate the
inner integral over $(\delta,\epsilon)$ \cite{a154} and $K_{\eta}$
the outer integral over $\eta$.

Equations (\ref{eq:u11}) and (\ref{eq:7}) estimate the first two
moments of the prior distribution of reverberation, $r_{t|t-1}$.
In this algorithm description section, we provide expressions for
either the second moment or the variance. We convert implicitly between
them using, for example, $\Sigma_{t}^{\text{(r)}} = \mathbb{E} \{ r_t^2 \}  - ( \mathbb{E} \{ r_t \} )^2$
for $r_t$.

\subsection{The Non-Linear KF Update Step}

The KF algorithm decomposes the noisy reverberant observation, $y_t$,
into its component parts using distributions in the log-magnitude
spectral domain. \foreignlanguage{british}{The decompositions are
based on }Fig. 2\foreignlanguage{british}{ and }the signal model in
(\ref{eq:3})\foreignlanguage{british}{ and }(\ref{eq:1}\foreignlanguage{british}{).
}The KF algorithm performs a series of low-dimensional operations
instead of a high-dimensional one in the KF update step.\foreignlanguage{british}{
The adaptive KF algorithm propagates backwards through Fig. 2 and
}decomposes\foreignlanguage{british}{: (a) }$y_t$\foreignlanguage{british}{
into speech, }$s_t$,\foreignlanguage{british}{ and into reverberation
and noise, }$z_t$\foreignlanguage{british}{, (b) the reverberation
and noise, }$z_t$\foreignlanguage{british}{, into reverberation,
}$r_t$\foreignlanguage{british}{, and into noise, }$n_t$\foreignlanguage{british}{,
and (c) the reverberation, }$r_t$\foreignlanguage{british}{, into
``old reverberation'', }$\delta_t$, and \foreignlanguage{british}{into}
\foreignlanguage{british}{``new reverberation'', }$\epsilon_t$\foreignlanguage{british}{.
}The reverberation and noise log-spectrum, $z_t$, is a variable of
the ``KF Update'' block in\foreignlanguage{british}{ }Fig. \ref{fig:EDCs-1-2-1-2-1-1-1-2-1-1-1-2-1-1-1-1-1-1-1-1-2-5-001-1-3}.

The KF algorithm in Fig. \ref{fig:EDCs-1-2-1-2-1-1-1-2-1-1-1-2-1-1-1-1-1-1-1-1-2-5-001-1-3}
uses the noisy reverberant observation, $y_t$, to first update $s_t$
and $z_t$ and then update $r_t$. The posterior $z_{t|t}$ is computed
in the ``KF Update'' block in Fig. \ref{fig:EDCs-1-2-1-2-1-1-1-2-1-1-1-2-1-1-1-1-1-1-1-1-2-5-001-1-3}.
In the proposed KF algorithm, the observed $y_t$ affects $z_{t|t}$
directly and, in turn, $z_{t|t}$ affects $r_{t|t}$. We hence divide
the observation update into two steps: (a) we use the log-spectrum
observation, $y_t$, to estimate the posterior distributions $s_{t|t}$
and $z_{t|t}$ because $Y_t = S_t + Z_t$ in (\ref{eq:3}), and (b)
we use $z_{t|t}$ as an \textquotedblleft observation\textquotedblright{}
to obtain the posterior distributions $r_{t|t}$ and $n_{t|t}$ because
$Z_t = R_t + N_t$ in (\ref{eq:3}). In (b), we calculate an updated
version of $r_{t}$ by using the posterior $z_{t|t}$ as a KF observation
constraint. Hence, according to (a) and (b), the log-spectrum observation,
$y_t$, provides new information about $s_t$ and $r_t$.

The sequence of operations involved in \foreignlanguage{british}{the
``Reverberation KF prediction'', ``KF update'' and ``$\gamma_t$,
$\beta_t$ KF update'' blocks} are listed in Table \ref{tab:ppssaaa}.\foreignlanguage{british}{
The ``Reverberation KF prediction'' block} \foreignlanguage{british}{in}
Fig. \ref{fig:EDCs-1-2-1-2-1-1-1-2-1-1-1-2-1-1-1-1-1-1-1-1-2-5-001-1-3},
which was \foreignlanguage{british}{presented in Sec. III.B, performs
the first five operations }in Table \ref{tab:ppssaaa}. \foreignlanguage{british}{The
``KF update'' and ``$\gamma_t$, $\beta_t$ KF update'' blocks
in }Fig. \ref{fig:EDCs-1-2-1-2-1-1-1-2-1-1-1-2-1-1-1-1-1-1-1-1-2-5-001-1-3}\foreignlanguage{british}{
perform the next seven operations }in Table \ref{tab:ppssaaa}\foreignlanguage{british}{,
i.e. steps 6-12. The bottom ``$z^{-1}$'' block }in Fig. \ref{fig:EDCs-1-2-1-2-1-1-1-2-1-1-1-2-1-1-1-1-1-1-1-1-2-5-001-1-3}
\foreignlanguage{british}{performs step 13. These 13 steps constitute
the dereverberation KF update step. The non-linear dereverberation
KF update step computes the first two moments of the posterior distributions
for most signal quantities and, moreover, includes the prediction
step as well for some quantities.} Both means and variances are computed
for the tracked Gaussian signals; for clarity, the variances, such
as $\Sigma_{t|t}^{\text{(r)}}$ for $r_t$, are not included in Table
\ref{tab:ppssaaa}.

{
\small

\ctable[caption = \textrm{\normalfont The operations performed in the ``Reverberation KF prediction'', ``KF Update'' and ``$\gamma_t$, $\beta_t$ KF Update'' blocks in Fig. \ref{fig:EDCs-1-2-1-2-1-1-1-2-1-1-1-2-1-1-1-1-1-1-1-1-2-5-001-1-3}. Steps 7-12 perform specific signal decompositions propagating backwards through Fig. 2}.,   
label = tab:ppssaaa,   
pos = hp,   
doinside=\hspace*{0pt}, width=.492\textwidth ]{p{0.2cm} p{7.8cm}}{}{ 
\FL & \textbf{Inputs:} (a) $s_{t|t-1}$ from the speech KF prediction step, (b) $n_{t|-}$ from external noise estimation, (c) $y_t$ from observation, and (d) $s_{t-1|t}$ from $s_{t|t}$ in step 7 and $\textbf{B}_t^{-1}$ from Sec. III.A.
\ML 1: & $\gamma_{t|t-1} \ \leftarrow \ \gamma_{t-1|t-1}$ from step 13 \smallskip 
\NN 2: & $\beta_{t|t-1} \ \leftarrow \ \beta_{t-1|t-1}$ from step 13 \smallskip 
\NN 3: & $\delta_{t|t-1} \ \leftarrow \ \gamma_{t|t-1}, r_{t-1|t-1}$ from steps 1, 13 \smallskip 
\NN 4: & $\epsilon_{t|t-1} \ \leftarrow \ \beta_{t|t-1}, s_{t-1|t-1}$ from steps 2, 13 \smallskip 
\NN 5: & $r_{t|t-1} \ \leftarrow \ \delta_{t|t-1}, \epsilon_{t|t-1}$ from steps 3, 4 \smallskip 
\NN 6: & $z_{t|t-1} \ \leftarrow \ r_{t|t-1}, n_{t|-}$ from step 5 and input (b) \smallskip 
\NN 7: & $s_{t|t}, z_{t|t} \ \leftarrow \ s_{t|t-1}, z_{t|t-1}, y_t$ from 6 and inputs (a), (c) \smallskip 
\NN 8: & $r_{t|t}, [n_{t|t}] \ \leftarrow \ r_{t|t-1}, n_{t|-}, z_{t|t}$ from 5, 7 and input (b) \smallskip 
\NN 9: & $\epsilon_{t|t}^{\prime} \ \leftarrow \ \beta_{t|t-1}, s_{t-1|t}$ from step 2 and input (d) \smallskip 
\NN 10: & $\delta_{t|t}, \epsilon_{t|t} \ \leftarrow \ \delta_{t|t-1}, \epsilon_{t|t}^{\prime}, r_{t|t}$ from steps 3, 8, 9 \smallskip 
\NN 11: & $\gamma_{t|t}, [r_{t-1|t}] \ \leftarrow \ \gamma_{t|t-1}, r_{t-1|t-1}, \delta_{t|t}$ from steps 1, 10, 13 \smallskip 
\NN 12: & $\beta_{t|t}, [s_{t-1|t}] \ \leftarrow \ \beta_{t|t-1}, s_{t-1|t}, \epsilon_{t|t}$ from 2, 10 and input (d)\smallskip 
\NN 13: & $r_{t-1|t-1}, \gamma_{t-1|t-1}, \beta_{t-1|t-1} \ \leftarrow \ r_{t|t}, \gamma_{t|t}, \beta_{t|t}$
\LL } 
}

For clarity, from this point onwards in Sec. III.C, the time subscript
is included only if it differs from $t|t-1$. Table \ref{tab:ppssaaa}
shows the time subscripts. We also denote $n_{t|-}$ by $n_{t|t-1}$.

The KF algorithm computes the total distubance, $z_t$, from (\ref{eq:3})
in step 6 using similar equations to step 5. A two-dimensional Gaussian
distribution is used for $p(r,n)$ and independence is assumed between
$r$ and $n$. Hence, $p(r,n) = p(r) p(n)$. The phase-sensitive KF
algorithm assumes that the phase difference, $\zeta$, between the
two disturbance sources, $r$ and $n$, is uniformly distributed and
independent of their magnitudes. From (\ref{eq:3}), we write $e^{2z}=e^{2r}+e^{2n}+2\cos(\zeta)e^{r+n}$
and thus $z=0.5\log\left(e^{2r}+e^{2n}+2\cos(\zeta)e^{r+n}\right)$.
Next, using $p(\zeta)=\frac{1}{2\pi}$, the first two moments of $z$
are given by
\begin{align}
\mathbb{E}\left\{ z\right\}  & =\int_{\zeta=0}^{\pi}\dfrac{1}{\pi}\int_{r,n}0.5\log\left(e^{2r}+e^{2n}+2\cos(\zeta)e^{r+n}\right)\nonumber \\
 & \times p(r)\ p(n)\ dr\ dn\ d\zeta\text{,}\label{eq:v13}\\
\mathbb{E}\left\{ z^{2}\right\}  & =\int_{\zeta=0}^{\pi}\dfrac{1}{\pi}\int_{r,n}\left(0.5\log\left(e^{2r}+e^{2n}\right.\right.\nonumber \\
 & \left.\left.+2\cos(\zeta)e^{r+n}\right)\right)^{2}\times p(r)\ p(n)\ dr\ dn\ d\zeta\label{eq:v14}
\end{align}
where $K_{(r,n)}$ sigma points are used to evaluate the inner integral
over $(r,n)$ \cite{a154} and $K_{\zeta}$ the outer integral over
$\zeta$.

The KF algorithm performs noise suppression with steps 6 and 7. Step
7 decomposes $y_t$ into $s_t$ and $z_t$, as shown in the signal
model in Fig. 2, estimating both $s_{t|t}$ and $z_{t|t}$.

Step 7 performs the first signal decomposition, $y_t$ into $s_t$
and $z_t$, when propagating backwards through the signal model in
Fig. 2. Step 7 applies the observation constraint, $y_t$, according
to (\ref{eq:3}). As in \cite{a225} \cite{a236}, the variables are
first transformed according to $(s,z,\lambda) \Rightarrow (u,y,\lambda)$
where $u = z-s$ and $\lambda = \angle S - \angle Z$. This variable
transformation is performed to allow the imposition of the scalar
KF observation, $y_t$.

The noisy reverberant log-amplitude spectrum, $y$, is given by $y = s + 0.5 \log(1 + \exp(2(z-s)) + 2 \cos(\lambda) \exp(z-s))$.
The KF update step assumes that $\angle Z$ is uniformly distributed,
$\angle Z \sim U(-\pi,\pi)$. Therefore, $\lambda \sim U(-\pi, \pi)$.
The first two moments of the posteriors of $s_t$ and $z_t$ are computed
using
\begin{align}
 & \mathbb{E}\left\{ s_{t}^{m_{1}}z_{t}^{m_{2}}\,|\,y_{0},y_{1},\dots,y_{t}\right\} \nonumber \\
 & =\int_{\lambda=-\pi}^{\pi}\int_{u=-\infty}^{\infty}s^{m_{1}}z^{m_{2}}\ p(u,\lambda|y_{t})\ du\ d\lambda\nonumber \\
 & \propto\dfrac{1}{|\Delta|}\int_{\lambda=-\pi}^{\pi}p(\lambda)\int_{u=-\infty}^{\infty}s^{m_{1}}z^{m_{2}}\ p(s,z)\ du\ d\lambda\nonumber \\
 & \propto\int_{\lambda=0}^{\pi}\int_{u=-\infty}^{\infty}s^{m_{1}}z^{m_{2}}\ p(s)\ p(z)\ du\ d\lambda\label{eq:12}
\end{align}
where the Jacobian determinant is $\Delta = -1$ and the moment indexes,
$m_1$ and $m_2$, are integers, $0 \leq m_1, m_2 \leq 2$. We denote
the variables for two moment indexes by $m_1$ and $m_2$.

The first two moments of the posterior distributions of $s_t$ and
$z_t$ are estimated. In (\ref{eq:12}), the priors of the speech
and of the noise and late reverberation are assumed to be independent,
i.e. $p(s,z) = p(s) p(z)$. In addition, in (\ref{eq:12}), $K_{\lambda}$
weighted sigma points are used to evaluate the outer integral over
$\lambda$ \cite{a268}.

Step 8 performs the second signal decomposition, $z_t$ into $r_t$
and $n_t$, when propagating backwards through the proposed signal
model in Fig. 2. Step 8 decomposes $z_{t|t}$ into $r_t$ and $n_t$,
according to (\ref{eq:3}), estimating both $r_{t|t}$ and $n_{t|t}$.
\foreignlanguage{british}{Step 8} performs an integral over \foreignlanguage{british}{$z_{t}$}
where the integrand is similar to step 7 and (\ref{eq:12}) and to
the KF update step in \cite{a225}\foreignlanguage{british}{ }\cite{a236}\foreignlanguage{british}{.
Instead of a scalar observation, as in step 7, the observation in
step 8 is a distribution; step 8 performs an outer integral over the
observation distribution and the integrand is similar to step 7. Step
8 uses $\mathbb{E}\left\{ r^{m} \right\} = \int_{z_{t}} \mathbb{E} \left\{ r^{m} | z_{t} \right\} dz_{t}$
where }$m \in \mathbb{N}_0$ and $0 \leq m \leq 2$\foreignlanguage{british}{.}
Step 7 computes a two-dimensional integral over the variables of $(z-s)$
and of the phase difference between $S$ and $Z$, using $y_t$ that
reduces the probability space from three to two dimensions. In step
8, the KF algorithm calculates a three-dimensional integral over:
(a) $(r-n)$, (b) the phase difference between $R$ and $N$, i.e.
$\iota = \angle R - \angle N$, and (c) the posterior of $z_{t}$.
Assuming $\angle N \sim U(-\pi,\pi)$, step 8 computes
\begin{align}
 & \mathbb{E}\left\{ r_{t}^{m_{1}}n_{t}^{m_{2}}\,|\,y_{0},y_{1},\dots,y_{t}\right\} \propto\int_{z_{t}=-\infty}^{\infty}p_{z_{t|t}}(z)\nonumber \\
 & \times\int_{\iota=0}^{\pi}\ \int_{u'=-\infty}^{\infty}r^{m_{1}}n^{m_{2}}\ p(r)\ p(n)\ du'\ d\iota\ dz_{t}\label{eq:u16}
\end{align}
where $u' = (r-n)$ and $\iota \sim U(-\pi,\pi)$. In (\ref{eq:u16}),
$K_{\iota}$ weighted sigma points are used to evaluate the integral
over $\iota$ and $K_{z_{t|t}}$ sigma points are used to evaluate
the integral over $z_{t}$ \cite{a154}.

\selectlanguage{british}%
I\foreignlanguage{english}{n Table \ref{tab:ppssaaa}}, the signals
in square brackets are calculated but are not used in the KF recursion.
In step 8, the posterior of $n_t$ is estimated using (\ref{eq:u16})
but is not used in the KF recursion.

\selectlanguage{english}%
Step 9 determines a preliminary estimate of the posterior distribution
of new reverberation component, $\epsilon_t$ in (\ref{eq:6}). This
preliminary estimate, denoted $\epsilon_{t|t}^{\prime}$, combines
an updated estimate of the previous frame's speech, $s_{t-1}$ with
the prior estimate of the reverberation parameter, $\beta_{t}$. In
step 9,\foreignlanguage{british}{ two random variables in the log-amplitude
spectral domain are added; the means and variances of the two Gaussian
variables are}: $\epsilon_{t|t}^{\prime} = \beta_{t|t-1} + s_{t-1|t}$\foreignlanguage{british}{
and }$\Sigma_{t|t}^{(\epsilon^{\prime})} = \Sigma_{t|t-1}^{(\beta)} + \Sigma_{t-1|t}^{\text{(s)}}$.
In this addition, $\beta_t$ and $s_{t-1}$ are assumed to be independent.

Step 10 decomposes the reverberation, $r_t$, into a new reverberation
component, $\epsilon_t$ and an old decaying reverberation component,
$\delta_t$, using (1), (4) and (5). Step 10 uses the prior distributions
for these two components from steps 3 and 9. \foreignlanguage{british}{Step
10 performs the same operation as step 8.} In analogy to step 7, the
variable transformations in steps 8 and 10 are $(r,n,\iota) \Rightarrow (u',z,\iota)$
and $(\delta,\epsilon,\xi) \Rightarrow (u'',r,\xi)$, respectively.
In step 10, the KF algorithm performs an integral over \foreignlanguage{british}{$r$}
where the integrand is similar to the KF update step in \cite{a225}\foreignlanguage{british}{.}
Step 10 estimates the posterior distributions of $\delta$ and $\epsilon$;
it estimates both $\delta_{t|t}$ and $\epsilon_{t|t}$ using $\delta_{t|t-1}$
and $\epsilon_{t|t}'$. Step 10 computes
\begin{align}
 & \mathbb{E}\left\{ \delta_{t}^{m_{1}}\epsilon_{t}^{m_{2}}\,|\,y_{0},y_{1},\dots,y_{t}\right\} \propto\int_{r_{t}=-\infty}^{\infty}p_{r_{t|t}}(r)\nonumber \\
 & \times\int_{\xi=0}^{\pi}\ \int_{u''=-\infty}^{\infty}\delta^{m_{1}}\,\epsilon^{m_{2}}\ p(\delta)\ p_{\epsilon_{t|t}^{\prime}}(\epsilon)\ du''\ d\xi\ dr_{t}\label{eq:u17}
\end{align}
where $u'' = (\delta - \epsilon)$ and $\xi \sim U(-\pi, \pi)$ is
the phase difference of the STFT-domain signals of $\delta$ and $\epsilon$.
Using (\ref{eq:u17}), we decompose $r_{t|t}$ into old reverberation
and new reverberation, estimating $\delta_{t|t}$ and $\epsilon_{t|t}$.
In (\ref{eq:u17}), $K_{\xi}$ sigma points are used to evaluate the
integral over $\xi$ and $K_{r_{t|t}}$ the integral over $r_{t}$.

Steps 10-12 perform the final signal decomposition, $r_t$ into $\gamma_t$
and $\beta_t$, when propagating backwards through the signal model
in Fig. 2.\foreignlanguage{british}{ In steps 11 and 12, the KF algorithm
computes the first two moments of the posterior distributions of $\gamma_t$
and $\beta_t$. In step 11, the algorithm }performs an integral over
\foreignlanguage{british}{$\delta_t$} using weighted sigma points
where the integrand models the \foreignlanguage{british}{addition
of two random variables in the log-amplitude spectral domain, $\gamma_t$
and $r_{t-1}$. Likewise, in step 12, the algorithm }performs an integral
over \foreignlanguage{british}{$\epsilon_t$ }using sigma points where
the integrand models the \foreignlanguage{british}{addition of two
random variables in the log-amplitude spectral domain, $\beta_t$
and $s_{t-1}$. }Steps 11 and 12 perform a linear KF update and impose
a straight line observation constraint because the signal model is
that $\gamma_t$ and $r_{t-1}$ are additive and that $\beta_t$ and
$s_{t-1}$ are additive, according to (\ref{eq:3}) and (\ref{eq:1}).

In step 11, the KF algorithm decomposes the old reverberation component,
$\delta_t$, according to (\ref{eq:6}) into the sum of a reverberation
parameter, $\gamma_t$, and the previous frame's reverberation, $r_{t-1}$.
We define $\textbf{x}_1 \sim \text{N}(\textbf{x}_1; \, \pmb{\mu}_1, \textbf{S}_1)$
where $\textbf{x}_1 = (\gamma_t, \, r_{t-1}, \, w_{t})^T \in \mathbb{R}^3$,
$w_{t}$ is zero-mean Gaussian with variance $\Sigma_{t}^{(\delta)}$,
$\pmb{\mu}_1 = (\gamma_{t|t-1}, \, r_{t-1|t-1}, \, 0)^T \in \mathbb{R}^3$
and $\textbf{S}_1 = \text{diag}((\Sigma_{t|t-1}^{(\gamma)}, \, \Sigma_{t-1|t-1}^{\text{(r)}}, \, \Sigma_{t|t}^{(\delta)})^T) \in \mathbb{R}^{3 \times 3}$
where $\text{diag}(\textbf{j})$ is a diagonal matrix with the elements
of $\textbf{j}$ on the main diagonal of the matrix. If $\textbf{x}_1 \sim \text{N}(\textbf{x}_1; \pmb{\mu}_1, \textbf{S}_1)$,
then
\begin{align}
(\textbf{x}_{1}|\textbf{g}^{T}\textbf{x}_{1}=\delta_{t|t}) & \sim\text{N}\left(\textbf{x}_{1};\,(\textbf{I}-\textbf{H}_{1}\textbf{g}^{T})\pmb{\mu}_{1}+\textbf{H}_{1}\delta_{t|t},\right.\nonumber \\
 & \ \ \ \left.(\textbf{I}-\textbf{H}_{1}\textbf{g}^{T})\textbf{S}_{1}\right)
\end{align}
where $\textbf{H}_1 = \textbf{S}_1 \textbf{g} (\textbf{g}^T \textbf{S}_1 \textbf{g})^{-1}$
and $\textbf{g} = (1, \, 1, \, -1)^T \in \mathbb{R}^3$ \cite{a337}.
We compute $(\textbf{I} - \textbf{H}_1\textbf{g}^T)\pmb{\mu}_1+\textbf{H}_1 \delta_{t|t} \in \mathbb{R}^3$
and $(\textbf{I}-\textbf{H}_1\textbf{g}^T)\textbf{S}_1 \in \mathbb{R}^{3 \times 3}$
and set $\gamma_{t|t}$ equal to the first element of this computed
mean and $\Sigma_{t|t}^{(\gamma)}$ equal to the first element of
this variance.

Likewise, for step 12, we define $\textbf{x}_2 \sim \text{N}(\textbf{x}_2; \, \pmb{\mu}_2, \textbf{S}_2)$
where $\textbf{x}_2 = (\beta_t, \, s_{t-1}, \, q_{t})^T \in \mathbb{R}^3$,
$q_{t}$ is zero-mean Gaussian with variance $\Sigma_{t}^{(\epsilon)}$,
$\pmb{\mu}_2 = (\beta_{t|t-1}, \, s_{t-1|t}, \, 0)^T \in \mathbb{R}^3$
and, moreover, $\textbf{S}_2 = \text{diag}((\Sigma_{t|t-1}^{(\beta)}, \, \Sigma_{t-1|t}^{\text{(s)}}, \, \Sigma_{t|t}^{(\epsilon)})^T) \in \mathbb{R}^{3 \times 3}$.
Next, we use
\begin{align}
(\textbf{x}_{2}|\textbf{g}^{T}\textbf{x}_{2}=\epsilon_{t|t}) & \sim\text{N}\left(\textbf{x}_{2};\,(\textbf{I}-\textbf{H}_{2}\textbf{g}^{T})\pmb{\mu}_{2}+\textbf{H}_{2}\epsilon_{t|t},\right.\nonumber \\
 & \ \ \ \left.(\textbf{I}-\textbf{H}_{2}\textbf{g}^{T})\textbf{S}_{2}\right)
\end{align}
where $\textbf{H}_2 = \textbf{S}_2 \textbf{g} (\textbf{g}^T \textbf{S}_2 \textbf{g})^{-1}$
\cite{a337}. The KF algorithm computes $(\textbf{I} - \textbf{H}_2\textbf{g}^T)\pmb{\mu}_2+\textbf{H}_2 \epsilon_{t|t} \in \mathbb{R}^3$
and $(\textbf{I}-\textbf{H}_2\textbf{g}^T)\textbf{S}_2 \in \mathbb{R}^{3 \times 3}$
and sets $\beta_{t|t}$ equal to the first element of this computed
mean and $\Sigma_{t|t}^{(\beta)}$ equal to the first element of this
computed variance.\smallskip

In steps 11 and 12, the presented KF algorithm computes $\gamma_{t|t}$,
$\Sigma_{t|t}^{(\gamma)}$ and $\beta_{t|t}$, $\Sigma_{t|t}^{(\beta)}$,
respectively. Steps 11 and 12 use $p(\gamma_t, r_{t-1} | \delta_{t|t} + w_t)$
and $p(\beta_t, s_{t-1} | \epsilon_{t|t} + q_t)$, respectively.\smallskip

Step 13 applies a one-frame delay, i.e. \foreignlanguage{british}{$z^{-1}$,
}to continue the KF recursion \foreignlanguage{british}{as shown in}
Fig. \ref{fig:EDCs-1-2-1-2-1-1-1-2-1-1-1-2-1-1-1-1-1-1-1-1-2-5-001-1-3}.
In summary, the 13 steps have the four main operations of: (a) step
5, (b) step 7, (c) step 8, and (d) step 11. The operations performed
in the other steps are either simple or identical to these operations.

\selectlanguage{british}%
According to the 13 steps\foreignlanguage{english}{ in Table \ref{tab:ppssaaa},
the $\gamma_t$ and $\beta_t$ reverberation parameters are affected
by the observed $y_t$ through a series of operations that estimate
the first two moments of the posterior distributions. The estimate
of $\gamma_t$ is affected by the previous estimate of $\beta_t$
because of the sequence of operations: $\beta_{t-1|t-1} \Rightarrow \beta_{t|t-1} \Rightarrow \epsilon_{t|t}^{\prime} \Rightarrow \delta_{t|t} \Rightarrow \gamma_{t|t}$.
The estimate of $\beta_t$ depends on the previous estimate of $\gamma_t$
because of the sequence of operations: $\gamma_{t-1|t-1} \Rightarrow \gamma_{t|t-1} \Rightarrow \delta_{t|t-1} \Rightarrow \epsilon_{t|t} \Rightarrow \beta_{t|t}$.}

The proposed 13 steps do not include the speech KF prediction step,
which is shown in \foreignlanguage{english}{Fig. \ref{fig:EDCs-1-2-1-2-1-1-1-2-1-1-1-2-1-1-1-1-1-1-1-1-2-5-001-1-3}},
that is used to calculate $s_{t-1|t}$ that is needed in steps 9 and
12. \foreignlanguage{english}{After step 7, according to} \foreignlanguage{english}{Fig.
\ref{fig:EDCs-1-2-1-2-1-1-1-2-1-1-1-2-1-1-1-1-1-1-1-1-2-5-001-1-3}
and Sec. III.B, recorrelation of the speech KF state is performed
with $\textbf{B}_t^{-1}$. Using $s_{t|t}$, from the recorrelation
operation, $s_{t-1|t}$ is obtained. }In steps 9 and 12, $s_{t-1|t}$
is used as a better estimate of $s_{t-1}$ than $s_{t-1|t-1}$.

In step 9, we note that two sub-indexes of $t|t-1$ and $t-1|t$ give
rise to a sub-index of $t|t$: $\epsilon_{t|t}^{\prime} = f(\beta_{t|t-1}, s_{t-1|t})$.
In step 9, we introduce\foreignlanguage{english}{ the notation $\epsilon_{t|t}^{\prime}$
to }avoid using the same symbol for two different posterior distributions,
\foreignlanguage{english}{$\epsilon_{t}$} and \foreignlanguage{english}{$\epsilon_{t}^{\prime}$}.
\selectlanguage{english}%

\subsection{The Priors for the Reverberation Parameters \label{sec:991}}

This section describes the priors for the $\gamma_t$ and $\beta_t$
reverberation parameters, which are based on \cite{a258} \cite{a349}
and \cite{a256}. The priors for $\gamma_t$ and $\beta_t$ are imposed
using Gaussian-Gaussian multiplication; $\gamma_t$ is modelled with
a Gaussian distribution and its internal prior is also a Gaussian.
Likewise, $\beta_t$ is modelled with a Gaussian and its internal
prior is also a Gaussian.

The priors for $\gamma_t$ and $\beta_t$ are estimated from spectral
log-amplitude observations in the free decay region (FDR), which is
comprised of $M_t$ consecutive frames with decreasing energy. We
define the look-ahead factor $C$ and the frame index $l=t-M_t+C+1,t-M_t+C+2,...,t+C$.
The least squares (LS) fit to the FDR is found using $r_l = \theta_1 x_l + \theta_2$
where $x_l$ is the time index in seconds and depends on $L$. The
parameters of the straight line are the slope, $\theta_1$, and the
y-intercept, $\theta_2$. For clarity, the frame subscript $t$ is
omitted from $\theta_1$ and $\theta_2$. We define $\bm{\theta}=(\theta_{1}\ \theta_{2})^{T} \in \mathbb{R}^{2}$.
Its Gaussian distribution is
\begin{align}
 & N(\bm{\theta};\bm{\mu}_{\theta},\bm{\Sigma}_{\theta})\,\propto\,|\bm{\Sigma}_{\theta}|^{-0.5}\nonumber \\
 & \times\exp\left(-0.5(\bm{\theta}-\bm{\mu}_{\theta})^{T}\bm{\Sigma}_{\theta}^{-1}(\bm{\theta}-\bm{\mu}_{\theta})\right)\label{eq:n20}
\end{align}
where $\bm{\mu}_{\theta} \in \mathbb{R}^{2}$ and $\bm{\Sigma}_{\theta} \in \mathbb{R}^{2 \times 2}$.

The log-likelihood of $\bm{\theta}$ is given by $l(\bm{\theta}) = \text{c}_1 - 0.5 (\bm{\theta} - \bm{\mu}_{\theta})^T \bm{\Sigma}_{\theta}^{-1} (\bm{\theta} - \bm{\mu}_{\theta}) = \text{c}_2 - 0.5 \sum_{\tau=t-M_t+C+1}^{t+C} ((\theta_1 x_{\tau} + \theta_2 - r_{\tau})^T (\Sigma_{\tau}^{(r)})^{-1} (\theta_1 x_{\tau} + \theta_2 - r_{\tau}))$
where $\text{c}_1$ and $\text{c}_2$ are constants.\smallskip

\begin{figure}[t]
\begin{minipage}[t]{0.48\columnwidth}%
\centering \includegraphics[bb=2bp 0bp 545bp 411bp,clip,width=1.02\columnwidth]{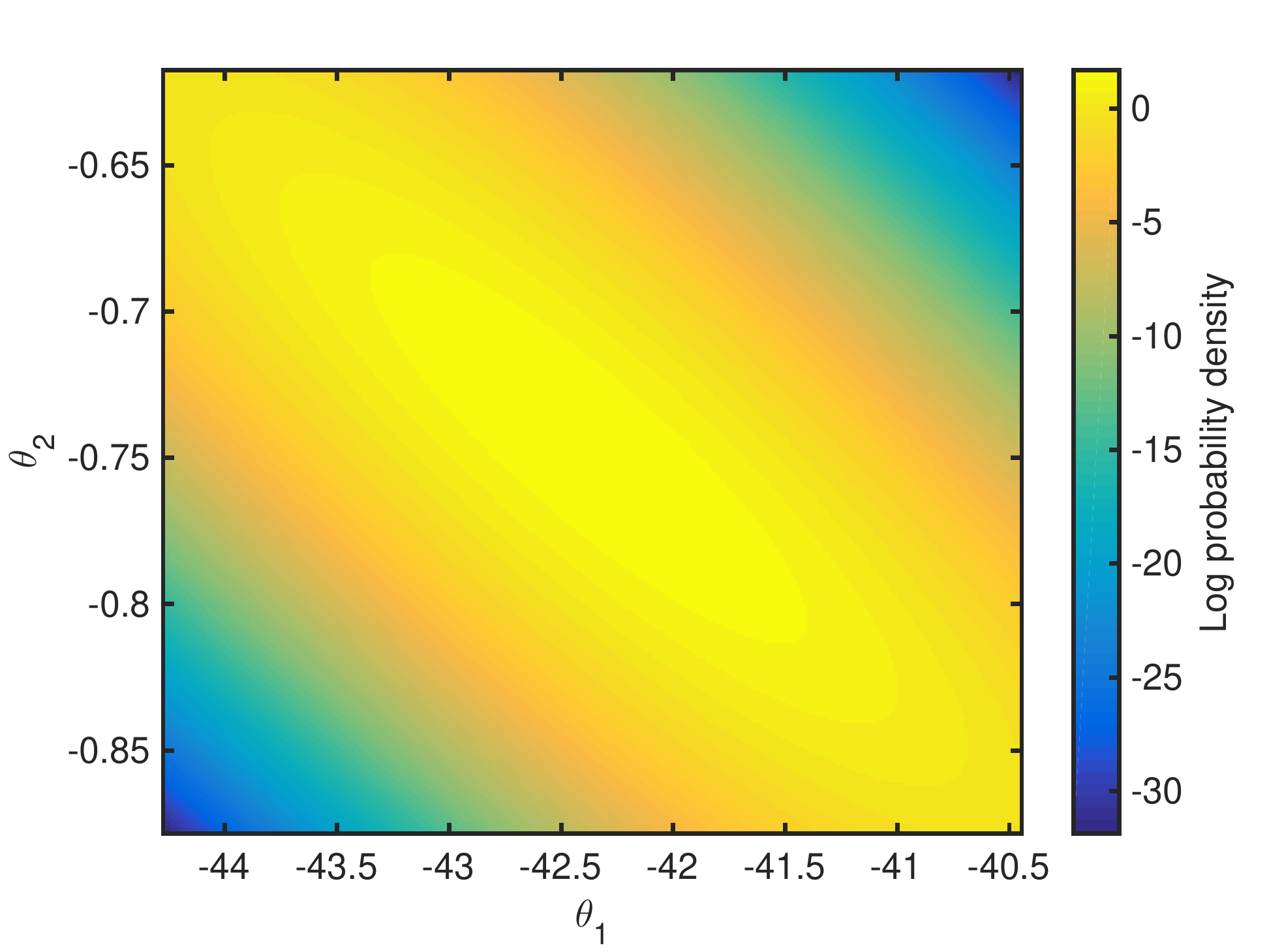}

\caption{Plot of the log-likelihood of the Gaussian distribution of $\bm{\theta}$
in (\ref{eq:n20}). $\bm{\Sigma}_{\theta} = \textbf{A}^{-1}$ is non-diagonal.
$\textbf{A}$ is computed in (\ref{eq:n21}) from the FDR.}

\label{fig:y71}%
\end{minipage}\hfill{}%
\begin{minipage}[t]{0.48\columnwidth}%
\centering \includegraphics[bb=12bp 0bp 510bp 400bp,clip,width=0.92\columnwidth]{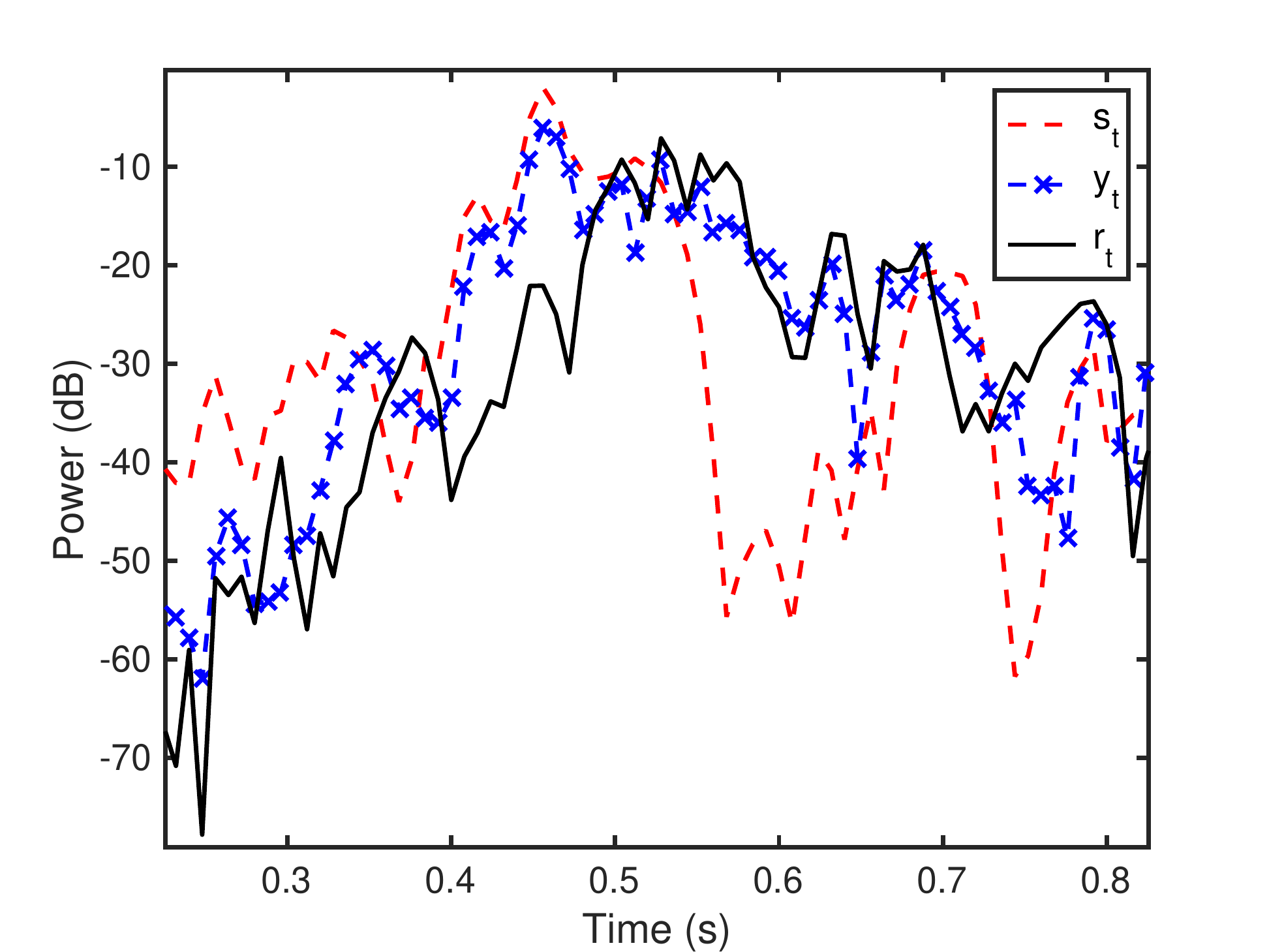}

\caption{Plot of the true $s_t$, $y_t$ and the true $r_t$ at $1$ kHz when
the noise type is white, $\text{SNR} = 20$ dB, $T_{60} = 0.68$ s
and $\text{DRR} = -2.52$ dB.}

\label{fig:y72}%
\end{minipage}
\end{figure}

We define the vectors $\textbf{x} \in \mathbb{R}^{M_t}$ and $\textbf{r} \in \mathbb{R}^{M_t}$
from $x_l$ and $r_l$, respectively. We define $\bm{\Sigma}_{r} \in \mathbb{R}^{M_t \times M_t}$
as the covariance matrix of $\textbf{r}$. The regression coefficients
as a Gaussian distribution are $N(\bm{\theta};\bm{\mu}_{\theta},\bm{\Sigma}_{\theta})$
where $\bm{\mu}_{\theta}=\textbf{A}^{-1}\textbf{b}$ and $\bm{\Sigma}_{\theta} = \textbf{A}^{-1}$,
\begin{align}
 & \textbf{A}=\left(\begin{array}{c}
\textbf{x}^{T}\bm{\Sigma}_{r}^{-1}\textbf{x}\\
\textbf{1}^{T}\bm{\Sigma}_{r}^{-1}\textbf{x}
\end{array}\ \begin{array}{c}
\textbf{x}^{T}\bm{\Sigma}_{r}^{-1}\textbf{1}\\
\textbf{1}^{T}\bm{\Sigma}_{r}^{-1}\textbf{1}
\end{array}\right)\text{,}\label{eq:n21}\\
 & \textbf{b}=\left(\begin{array}{c}
\textbf{x}^{T}\bm{\Sigma}_{r}^{-1}\textbf{r}\\
\textbf{1}^{T}\bm{\Sigma}_{r}^{-1}\textbf{r}
\end{array}\right)
\end{align}
where $\textbf{1} \in \mathbb{R}^{M_t}$ is a vector of ones. Figure
\ref{fig:y71} shows the log-likelihood of the Gaussian distribution
of $\bm{\theta}$ with mean $\bm{\mu}_{\theta}$ and variance $\bm{\Sigma}_{\theta}$
when the correlation between $\theta_1$ and $\theta_2$ is strong.
We note that the maximum likelihood solution for the mean of $\theta_1$,
when $\theta_2$ is not considered, can be found in \cite{a350}.

The noisy reverberation, $z_t$, is the observation and, in this case,
our aim is to find $r_t$ and $\Sigma_t^{(r)}$. We denote the noise
floor in the power spectral domain by $|N|^2$. In a FDR, at high
RNRs and when $|N|^2$ is low, we have $r_t \approx z_t$ because:
$z = r + 0.5\log(1 + \text{RNR}^{-1} + 2 \cos(\zeta) \, \text{RNR}^{-0.5})$,
which was also used in (\ref{eq:v13}) and (\ref{eq:v14}), where
$\text{RNR} = \exp(2(r-n))$.

To solve the problem of finding $r_t$ and $\Sigma_t^{(r)}$ given
the observed $z_t$, we first employ a minimum MSE (MMSE) algorithm,
such as the traditional Log-MMSE estimator \cite{Ephraim1985}, to
remove the noise and estimate the signal's spectral amplitude, as
in the end of Sec. 3.1 in \cite{a256}, and then perform a KF update
step and use $z_t$ as the KF observation. For the linear KF \cite{a171}
\cite{a125}, when the observation noise covariance matrix tends to
zero at high RNRs, then the mean of the KF state goes to the KF observation
and the variance of the KF state goes to zero. Therefore, a solution
to finding $r_t$ and $\Sigma_t^{(r)}$ is to set $r_t = z_t$ and
$\Sigma_t^{(r)}$ equal to a small value, such as $1$ $\text{dB}^2$.

The $T_{60}$ can be estimated from $\theta_1$ using $T_{60} = - \frac{3 \log(10)}{\theta_1}$,
where $\theta_1$ is computed from FDR data in the log-magnitude spectral
domain \cite{a349}. The KF algorithm, which models $\gamma_t$ with
a Gaussian distribution, uses a Gaussian prior for $\gamma$ with
mean $L \mu_{\theta_1}$ and variance $L^2 \Sigma_{\theta_1}$ because
$\gamma = L \theta_1$. Using $\gamma = L \theta_1$, $\theta_1$
is directly mapped to $\gamma_t$, without going via the $T_{60}$.

The algorithm estimates a prior for $\beta$ directly, at the same
time as the prior for $\gamma$, rather than first estimating the
$T_{60}$ and the DRR. Figure \ref{fig:y72} depicts $y_t$, $s_t$
and $r_t$ over time, at $1$ kHz, when the $T_{60}$ is $0.68$ s,
the DRR is $-2.52$ dB, the noise type is white and the SNR is $20$
dB. In Fig. \ref{fig:y72}, $s_t$ and $r_t$ are the true speech
and reverberation powers; the latter is computed by convolving the
RIR without its first $30$ ms with the speech signal \cite{a333}.
If we choose an appropriate FDR, then we can see that (\ref{eq:1})
is related to the FDR and the first frame of the FDR to the DRR. The
$\beta$ reverberation parameter is computed from the difference between
the log-amplitude of the first frame and the y-intercept of the LS
fit, i.e. $\theta_2$. For the LS fit to the FDR, we choose to not
use the first frame of the FDR. For the subtraction of two independent
random variables, their means subtract and their variances add; $\theta_2$
has a mean and a variance and the log-amplitude of the first frame
of the FDR has a mean and a fixed small variance. The prior for $\beta$
is given by
\begin{align}
 & \beta=r_{t-M_{t}+C+1}-\theta_{2}\text{.}
\end{align}

We denote the mean of the internally computed Gaussian $\gamma_t$
prior by $\gamma_{t|t-1}^{\prime\prime}$ and its variance by $\Sigma_{t|t-1}^{(\gamma^{\prime\prime})}$.
We use
\begin{align}
 & \gamma_{t|t-1}=\dfrac{\gamma_{t|t-1}^{\prime}\Sigma_{t|t-1}^{(\gamma^{\prime\prime})}+\gamma_{t|t-1}^{\prime\prime}\Sigma_{t|t-1}^{(\gamma^{\prime})}}{\Sigma_{t|t-1}^{(\gamma^{\prime})}+\Sigma_{t|t-1}^{(\gamma^{\prime\prime})}}\text{,}\\
 & \Sigma_{t|t-1}^{(\gamma)}=\dfrac{\Sigma_{t|t-1}^{(\gamma^{\prime})}\Sigma_{t|t-1}^{(\gamma^{\prime\prime})}}{\Sigma_{t|t-1}^{(\gamma^{\prime})}+\Sigma_{t|t-1}^{(\gamma^{\prime\prime})}}\text{.}
\end{align}
Likewise, we denote the mean of the Gaussian $\beta_t$ prior by $\beta_{t|t-1}^{\prime\prime}$
and its variance by $\Sigma_{t|t-1}^{(\beta^{\prime\prime})}$. For
$\beta_t$, we use
\begin{align}
 & \beta_{t|t-1}=\dfrac{\beta_{t|t-1}^{\prime}\Sigma_{t|t-1}^{(\beta^{\prime\prime})}+\beta_{t|t-1}^{\prime\prime}\Sigma_{t|t-1}^{(\beta^{\prime})}}{\Sigma_{t|t-1}^{(\beta^{\prime})}+\Sigma_{t|t-1}^{(\beta^{\prime\prime})}}\text{,}\\
 & \Sigma_{t|t-1}^{(\beta)}=\dfrac{\Sigma_{t|t-1}^{(\beta^{\prime})}\Sigma_{t|t-1}^{(\beta^{\prime\prime})}}{\Sigma_{t|t-1}^{(\beta^{\prime})}+\Sigma_{t|t-1}^{(\beta^{\prime\prime})}}\text{.}
\end{align}

Assuming that the RIR is known, it is straightforward to compute the
$T_{60}$: compute the energy decay curve, plot it on a log scale
and estimate the $T_{60}$ from its slope. On the contrary, estimating
the $T_{60}$ blindly is not trivial. The KF algorithm estimates the
$T_{60}$, applying internally estimated priors using the decay rate
of the LS fit to the FDR \cite{a349}, so that the KF does not diverge
and does not treat reverberation as speech.

\subsection{The Peripheral Blocks of the Algorithm}

This section describes the unshaded peripheral blocks of the KF algorithm
in Fig. \ref{fig:EDCs-1-2-1-2-1-1-1-2-1-1-1-2-1-1-1-1-1-1-1-1-2-5-001-1-3}.
The algorithm uses pre-cleaning before performing speech AR modelling,
as in \cite{a268} and \cite{a125}. Pre-cleaning has also been used
in \cite{a203} \cite{a233}, in \cite{a126} \cite{a123} and in
\cite{a278} \cite{a204}. The ``Speech pre-cleaning'' block in
Fig. \ref{fig:EDCs-1-2-1-2-1-1-1-2-1-1-1-2-1-1-1-1-1-1-1-1-2-5-001-1-3}
affects the ``Speech KF prediction'' block and not the observation
of the non-linear ``KF Update'' block, i.e. $y_t$, used in step
7.

For the ``Noise power estimation'' block, an external noise power
estimator, such as \cite{a76}, is used. External noise estimation
and log-normal noise power modelling, as in \cite{a307} \cite{a313},
are used for $n_{t|-}$, which is then used in step 6 of Table I.

In summary, the KF algorithm tracks speech and reverberation in the
spectral log-magnitude domain along with $\gamma_t$ and $\beta_t$,
as described in Fig. \ref{fig:EDCs-1-2-1-2-1-1-1-2-1-1-1-2-1-1-1-1-1-1-1-1-2-5-001-1-3}
and in the signal model in Fig. 2.

\section{Implementation, Testing and Validation}

We use acoustic frames of length $32$ ms and an acoustic frame increment
of $L=8$ ms in (2). We also use modulation frames of $64$ ms and
a modulation frame increment of $8$ ms \cite{a268} \cite{a123}.
In Fig. \ref{fig:EDCs-1-2-1-2-1-1-1-2-1-1-1-2-1-1-1-1-1-1-1-1-2-5-001-1-3},
for speech amplitude spectrum pre-cleaning, we use the Log-MMSE estimator
\cite{Ephraim1985} followed by the WPE dereverberation algorithm
\cite{a334} \cite{a327}. The dimension of the speech KF state, $\textbf{s}_t$,
in (\ref{eq:rr6}) is $p=2$. In the ``Noise power estimation''
block of Fig. \ref{fig:EDCs-1-2-1-2-1-1-1-2-1-1-1-2-1-1-1-1-1-1-1-1-2-5-001-1-3},
we use external noise power estimation from \cite{a76} using the
implementation in \cite{a335}.

The outer integrals in Secs. III.B and III.C are performed using sigma
points, as in \cite{a268}. The numbers of sigma points used in (\ref{eq:u11})-(\ref{eq:u17})
are $K_{(\delta,\epsilon)} = K_{(r,n)} = K_{z_{t|t}} = K_{r_{t|t}} = 3$
and $K_{\eta} = K_{\zeta} = K_{\lambda} = K_{\iota} = K_{\xi} = 6$.
For the latter cases, for $\xi$, the sigma points are at $\xi = ((1:K_{\xi})-0.5) \frac{\pi}{K_{\xi}}$
and the weights are all equal to $\frac{1}{K_{\xi}}$ \cite{a268}
\cite{a225}. With this choice of sigma points for $\xi$, the integral
will be exact for an integrand that is a sum of terms of the form
$\cos(n \xi)$ for $0 \leq n \leq 2K_{\xi}+1$.

In Sec. III.D, for the $\gamma_t$ and $\beta_t$ priors, the look-ahead
factor is $C=3$ frames. For the FDR that is comprised of frames with
decreasing energy, $M_t$ is computed in every frame.

\begin{figure}[t]
\begin{minipage}[t]{0.48\columnwidth}%
\centering \includegraphics[bb=12bp 0bp 510bp 400bp,clip,width=1\columnwidth]{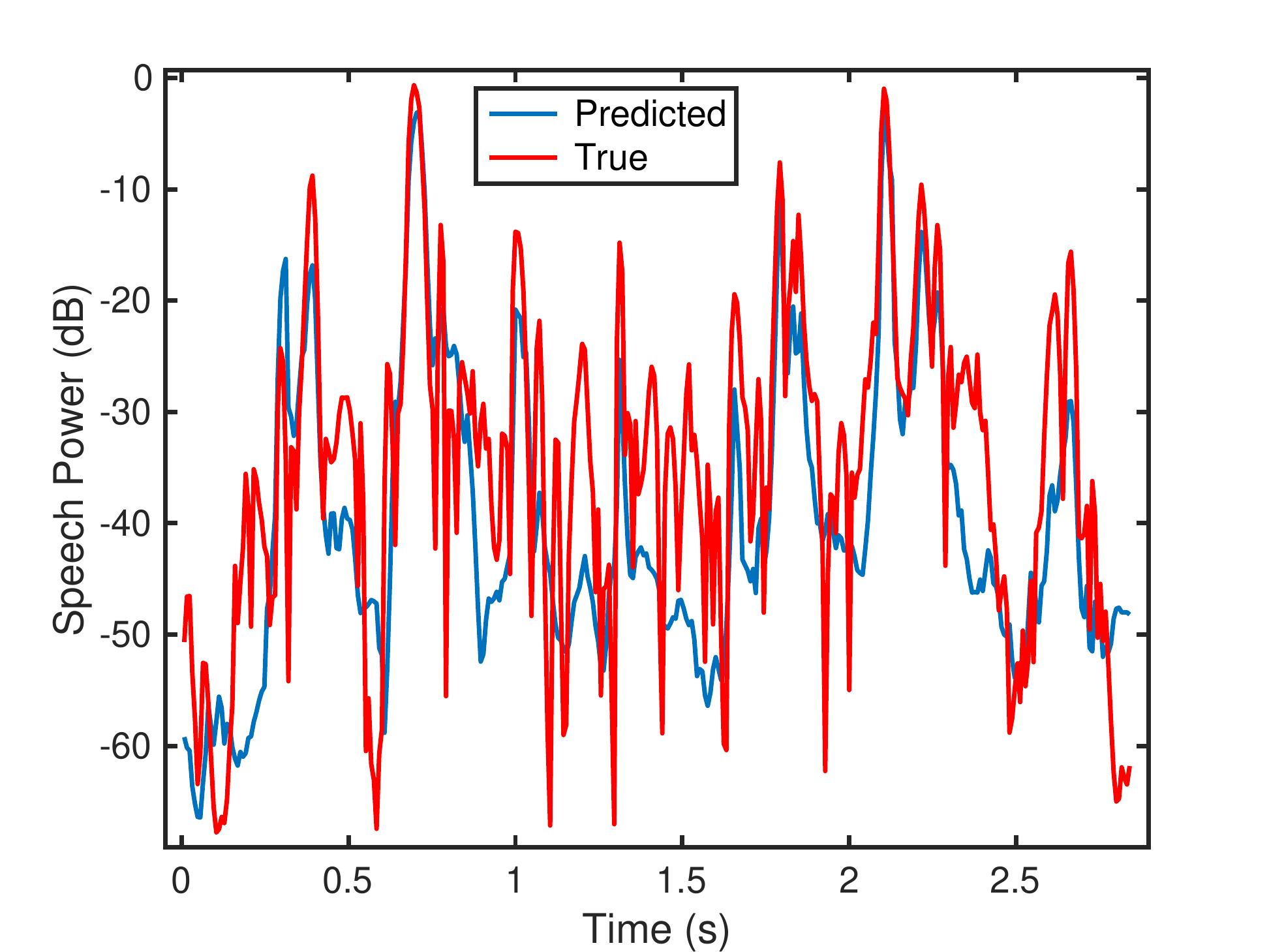}

{\footnotesize (a)}%
\end{minipage}\hfill{}%
\begin{minipage}[t]{0.48\columnwidth}%
\centering \includegraphics[bb=12bp 202bp 530bp 604bp,clip,width=1\columnwidth]{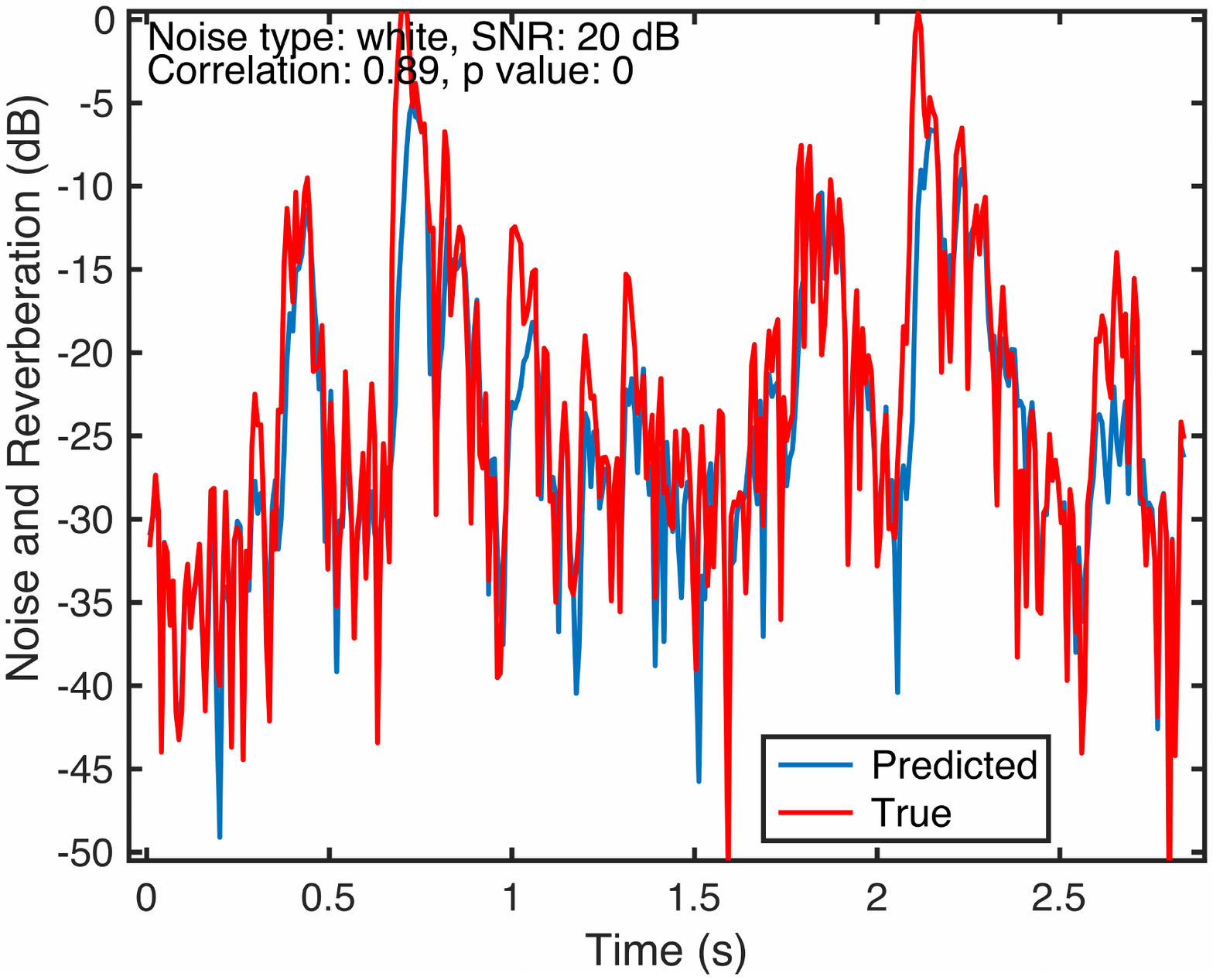}

{\footnotesize (b)}%
\end{minipage}

\vspace{1.10mm}

\begin{minipage}[t]{0.48\columnwidth}%
\centering \includegraphics[bb=12bp 202bp 530bp 604bp,clip,width=1\columnwidth]{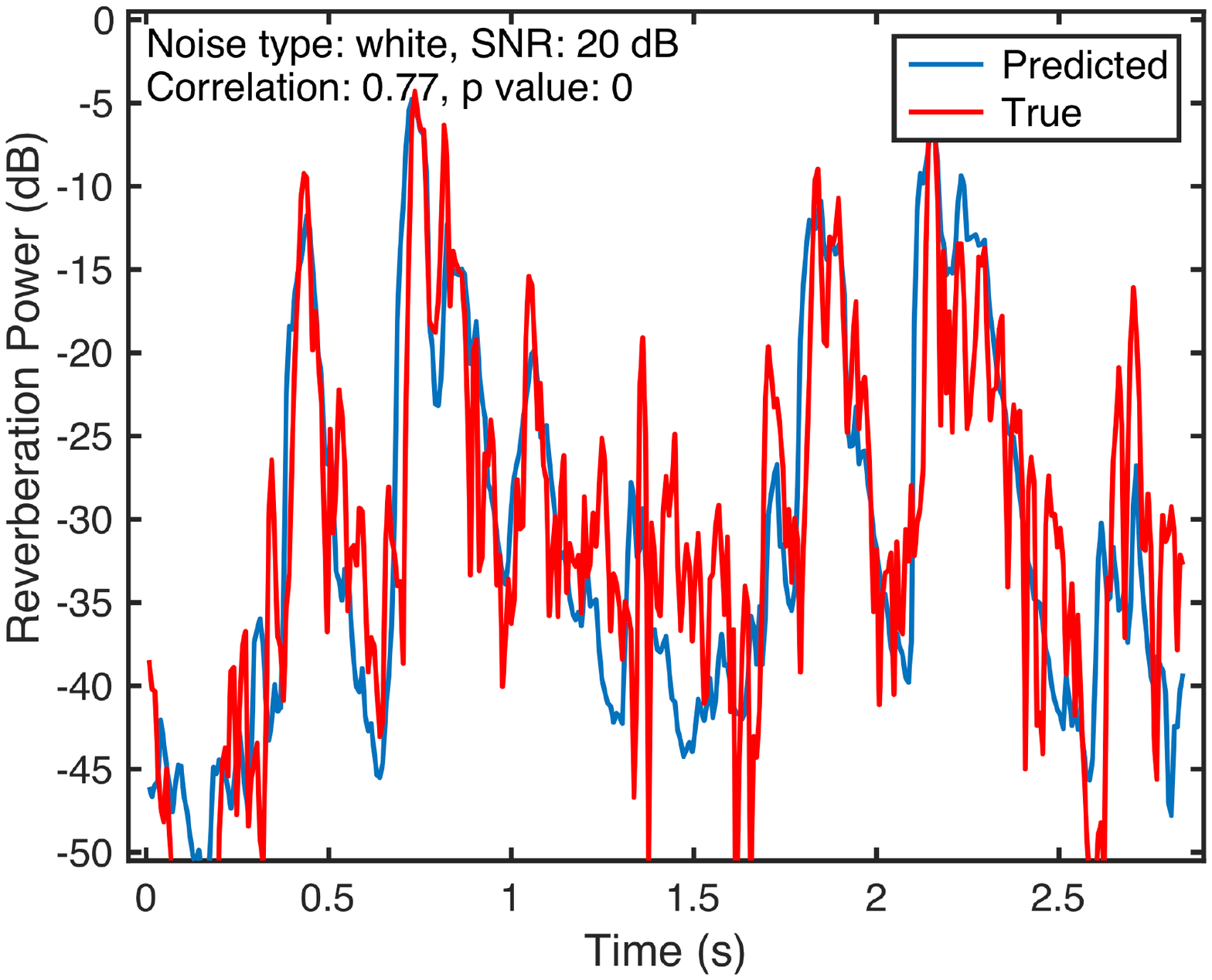}

{\footnotesize (c)}%
\end{minipage}\hfill{}%
\begin{minipage}[t]{0.48\columnwidth}%
\centering \includegraphics[bb=10bp 0bp 510bp 400bp,clip,width=1\columnwidth]{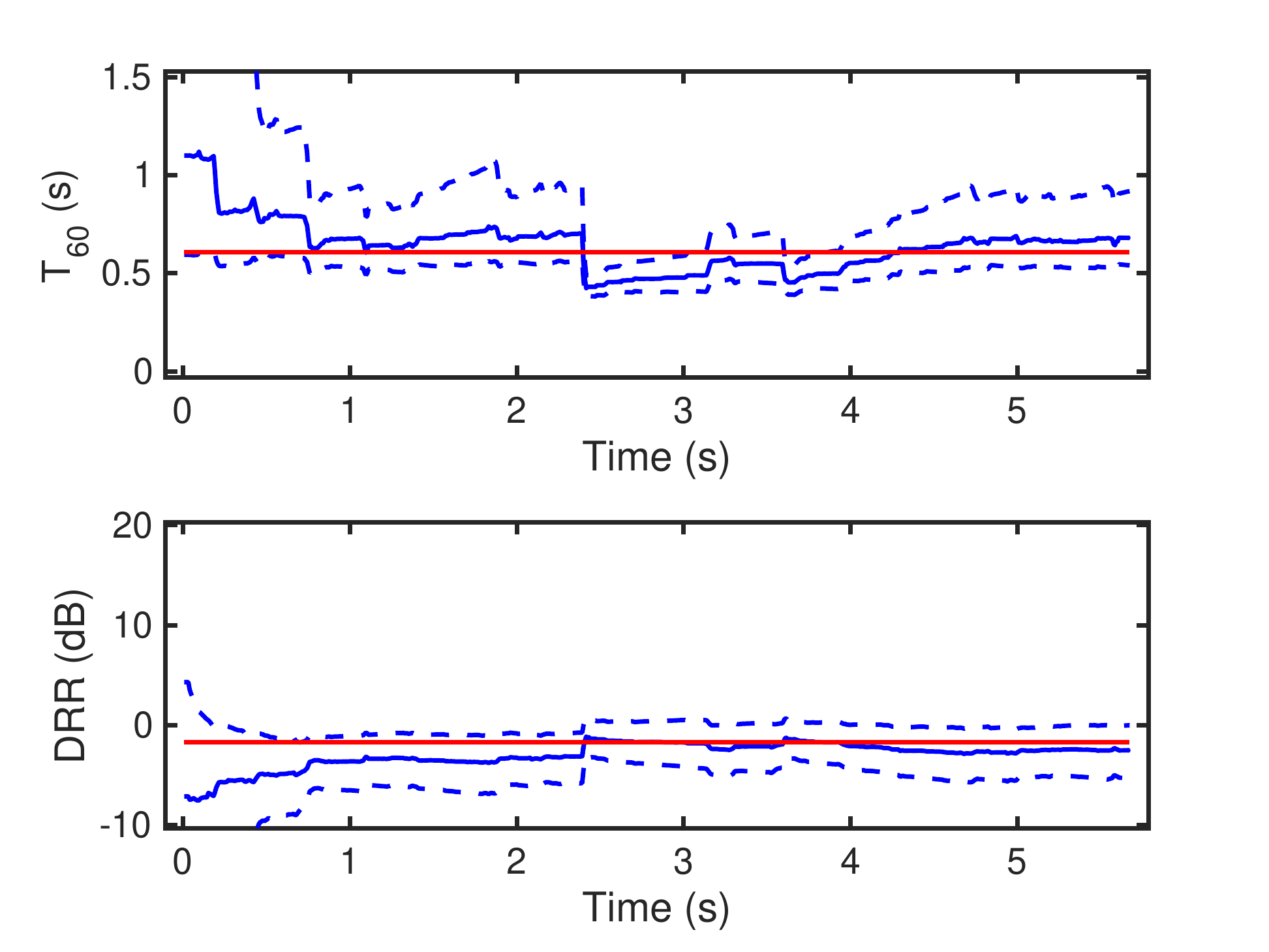}

{\footnotesize (d)}%
\end{minipage}

\caption{Plot of the predicted and true: (a) speech power, i.e. $s_{t|t}$
and the true $s_t$, at $1$ kHz, (b) noise and reverberation power,
i.e. $z_{t|t}$ and the true $z_t$, at $1$ kHz, (c) reverberation
power, i.e. $r_{t|t}$ and the true $r_t$, at $1$ kHz, and (d) $T_{60}$
and DRR reverberation parameters. In (a)-(d), the $T_{60}$ is $0.61$
s and the DRR is $-1.74$ dB, while the noise type is white and the
SNR is $20$ dB.}
\end{figure}

Figures 6(a)-(c) illustrate $s_t$, $z_t$ and $r_t$ at $1$ kHz
over time, as Fig. \ref{fig:EDCs-1-2-1-2-1-1-1-2-1-1-1-2-1-1-1-1-1-1-1-1-2-5-001-1-3}
in \cite{a125}. Figure 6(a) shows the predicted and true speech power,
i.e. $s_{t|t}$ and the true $s_t$, and Fig. 6(b) the predicted and
true noise and reverberation power, i.e. $z_{t|t}$ and the true $z_t$.
The correlation coefficient between the true $z_t$ and $z_{t|t}$
is $0.89$. Figure 6(c) shows the predicted and true reverberation
power, i.e. $r_{t|t}$ and the true $r_t$. The correlation coefficient
between the true $r_t$ and $r_{t|t}$ is $0.77$ $(<0.89)$. The
noise type is white, the SNR is $20$ dB, $T_{60} = 0.61$ s and $\text{DRR} = -1.74$
dB.

The ordering of the graphs in Figs. 6(a)-(c) matches the ordering
of the signal decompositions in the KF algorithm, in Table \ref{tab:ppssaaa}.
The noisy reverberant observation is first decomposed into $s_t$
and $z_t$ with step 7 in Table \ref{tab:ppssaaa}; Figs. 6(a) and
(b) illustrate $s_{t|t}$ and $z_{t|t}$, respectively, over time.
Then, $z_{t|t}$ is decomposed into $r_t$ and $n_t$ with step 8;
Fig. 6(c) depicts $r_{t|t}$ over time.

\begin{table}[t]
\caption{{\normalfont List of the environments sorted with respect to the $T_{60}$ from A to I and from J to V. The room dimensions are $5 \times 4 \times 4$ or $10 \times 7 \times 3$ m. The source-microphone distance is $1.5$ m. The wall reflection coefficient is adjusted to vary the $T_{60}$ and the DRR. For $a$ and $b$ in (\ref{eq:2}), $L=8$ ms.}}
\centering%
\begin{tabular}{|c|c|c|c|c|c|}
\hline 
\textbf{Index} & \textbf{$T_{60}$ (s)} & \textbf{DRR (dB)} & \textbf{Room} \textbf{(m$\times$m$\times$m)} & \textbf{$a$} & \textbf{$b$}\tabularnewline
\hline 
A  & $0.18$ & $8.43$ & $5 \times 4 \times 4$ & $0.54$ & $0.07$\tabularnewline
\hline 
B & $0.25$ & $5.78$ & $5 \times 4 \times 4$ & $0.64$ & $0.10$\tabularnewline
\hline 
C  & $0.33$ & $3.13$ & $5 \times 4 \times 4$ & $0.72$ & $0.14$\tabularnewline
\hline 
D  & $0.40$ & $1.69$ & $5 \times 4 \times 4$ & $0.76$ & $0.16$\tabularnewline
\hline 
E & $0.47$ & $0.25$ & $5 \times 4 \times 4$ & $0.79$ & $0.20$\tabularnewline
\hline 
F & $0.54$ & $-0.74$ & $5 \times 4 \times 4$ & $0.81$ & $0.23$\tabularnewline
\hline 
G & $0.61$ & $-1.74$ & $5 \times 4 \times 4$ & $0.83$ & $0.25$\tabularnewline
\hline 
H & $0.64$ & $-2.13$ & $5 \times 4 \times 4$ & $0.84$ & $0.26$\tabularnewline
\hline 
I & $0.68$ & $-2.52$ & $5 \times 4 \times 4$ & $0.85$ & $0.27$\tabularnewline
\hline 
J & $0.21$ & $8.07$ & $10 \times 7 \times 3$ & $0.59$ & $0.06$\tabularnewline
\hline 
K  & $0.31$ & $2.74$ & $10 \times 7 \times 3$ & $0.70$ & $0.16$\tabularnewline
\hline 
L & $0.40$ & $0.17$ & $10 \times 7 \times 3$ & $0.76$ & $0.23$\tabularnewline
\hline 
M & $0.50$ & $0.11$ & $10 \times 7 \times 3$ & $0.80$ & $0.19$\tabularnewline
\hline 
N & $0.59$ & $-0.73$ & $10 \times 7 \times 3$ & $0.83$ & $0.20$\tabularnewline
\hline 
O & $0.64$ & $-0.95$ & $10 \times 7 \times 3$ & $0.84$ & $0.20$\tabularnewline
\hline 
P & $0.69$ & $-1.12$ & $10 \times 7 \times 3$ & $0.85$ & $0.19$\tabularnewline
\hline 
Q & $0.71$ & $-1.68$ & $10 \times 7 \times 3$ & $0.86$ & $0.21$\tabularnewline
\hline 
R & $0.73$ & $-2.01$ & $10 \times 7 \times 3$ & $0.86$ & $0.22$\tabularnewline
\hline 
S & $0.85$ & $-2.09$ & $10 \times 7 \times 3$ & $0.88$ & $0.19$\tabularnewline
\hline 
T & $0.97$ & $-2.95$ & $10 \times 7 \times 3$ & $0.89$ & $0.22$\tabularnewline
\hline 
U & $1.01$ & $-3.11$ & $10 \times 7 \times 3$ & $0.90$ & $0.20$\tabularnewline
\hline 
V & $1.05$ & $-3.33$ & $10 \times 7 \times 3$ & $0.90$ & $0.22$\tabularnewline
\hline 
\end{tabular}

\label{tab:abchhghahhaa}
\end{table}

The $T_{60}$ and DRR estimates should converge to their true constant
values when the talker and the microphone are not moving and the frequency
variations in the reflection coefficients are not modelled \cite{Allen1979}
\cite{a359}. Internally estimated priors for $\gamma_t$ and $\beta_t$
make the reverberation parameters converge over time. Figure 6(d)
shows the predicted and true $T_{60}$ and DRR reverberation parameters
over time. The dashed lines are the signals' standard deviations,
computed from $\gamma_{t}$ and $\beta_{t}$.

\section{Evaluation of the Algorithm}

The proposed KF algorithm is evaluated in terms of the perceptual
evaluation of speech quality (PESQ) \cite{ITU_T_P862}, the cepstral
distance (CD) spectral divergence metric \cite{Kitawaki1988}, the
reverberation decay tail (RDT) dereverberation metric \cite{a325}
and the STOI speech intelligibility metric \cite{a308} \cite{a309}.
The ideal values of CD and RDT, which have been also used in \cite{a179},
are zero. For evaluation, the TIMIT database \cite{a187}, sampled
at $16$ kHz, and the RSG-10 noise database \cite{a58} are used.
From the TIMIT database, $52$ clean speech utterances are chosen.
We use artificially-created reverberation with the image method \cite{Allen1979}
using the implementation in \cite{a330}. The wall reflection coefficient
is adjusted to vary the $T_{60}$ and hence also the DRR. The KF algorithm
is evaluated with noisy reverberant speech signals at various SNRs,
from $5$ to $20$ dB, using random noise segments and the noise types
of white, babble and factory.

\begin{figure}[t]
\begin{minipage}[t]{0.48\columnwidth}%
\centering \includegraphics[bb=10bp 0bp 525bp 411bp,clip,width=1\columnwidth]{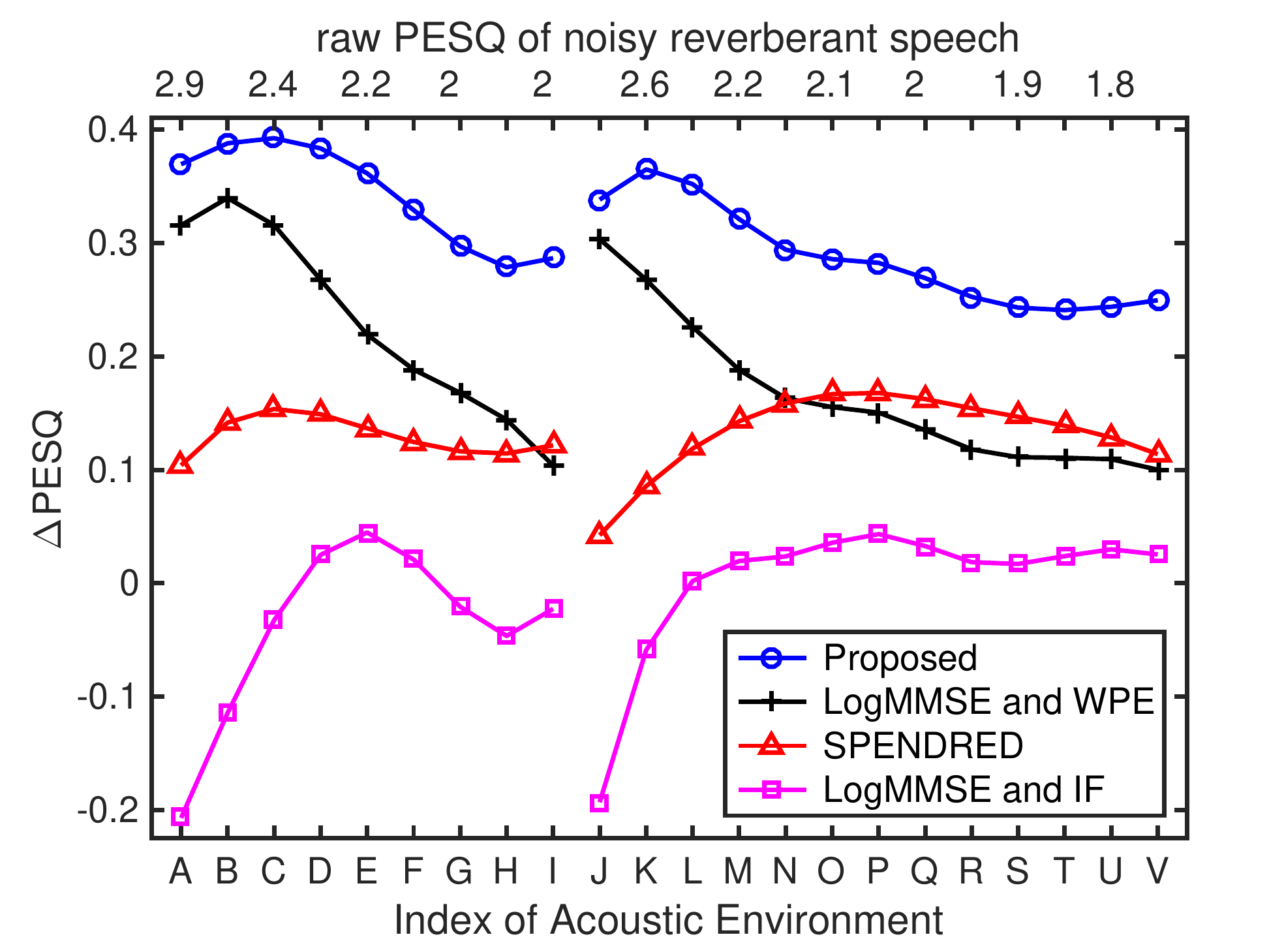}

{\footnotesize (a)}%
\end{minipage}\hfill{}%
\begin{minipage}[t]{0.48\columnwidth}%
\centering \includegraphics[bb=10bp 0bp 525bp 411bp,clip,width=1\columnwidth]{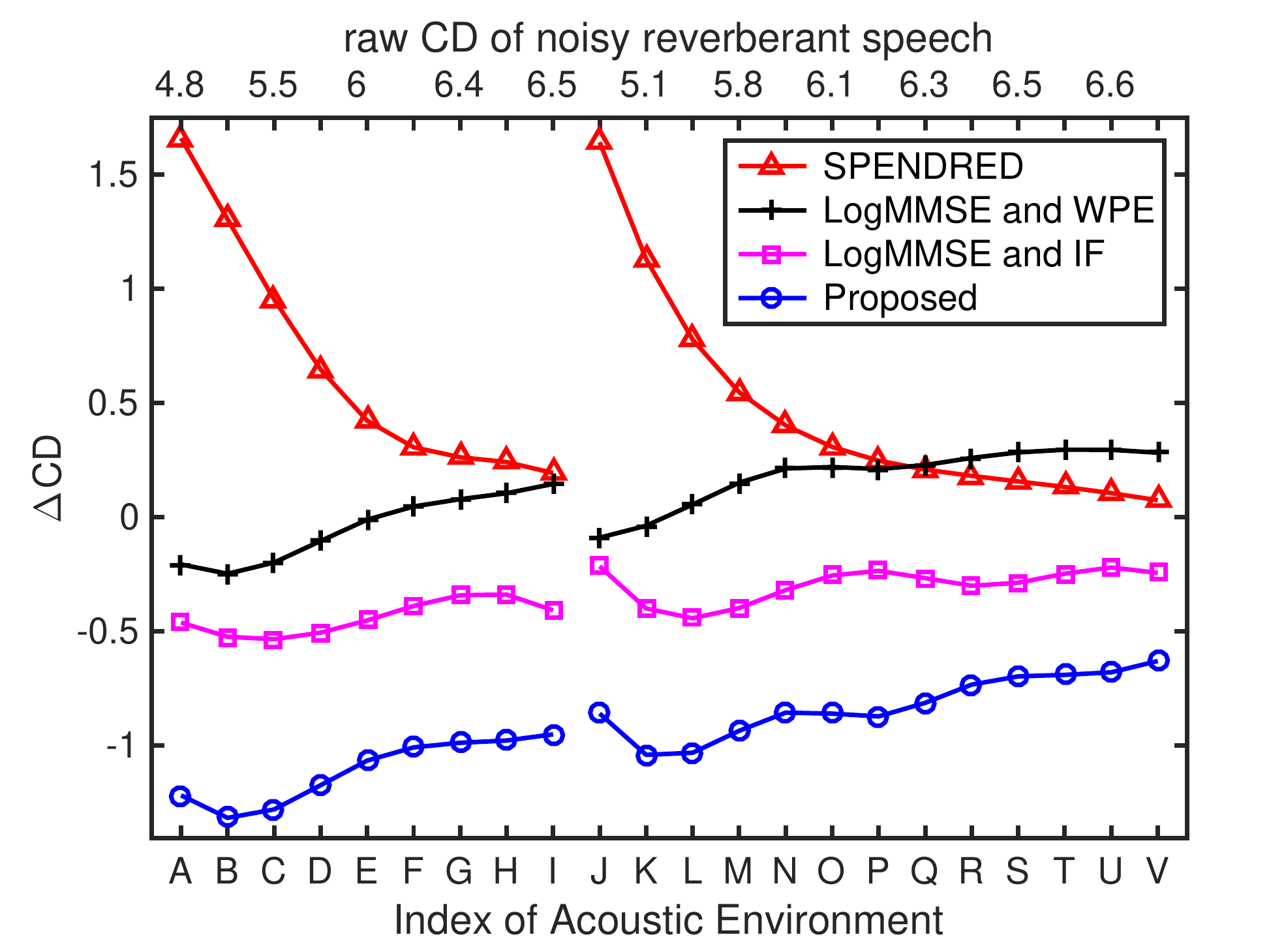}

{\footnotesize (b)}%
\end{minipage}

\vspace{1.10mm}

\begin{minipage}[t]{0.48\columnwidth}%
\centering \includegraphics[bb=10bp 0bp 525bp 411bp,clip,width=1\columnwidth]{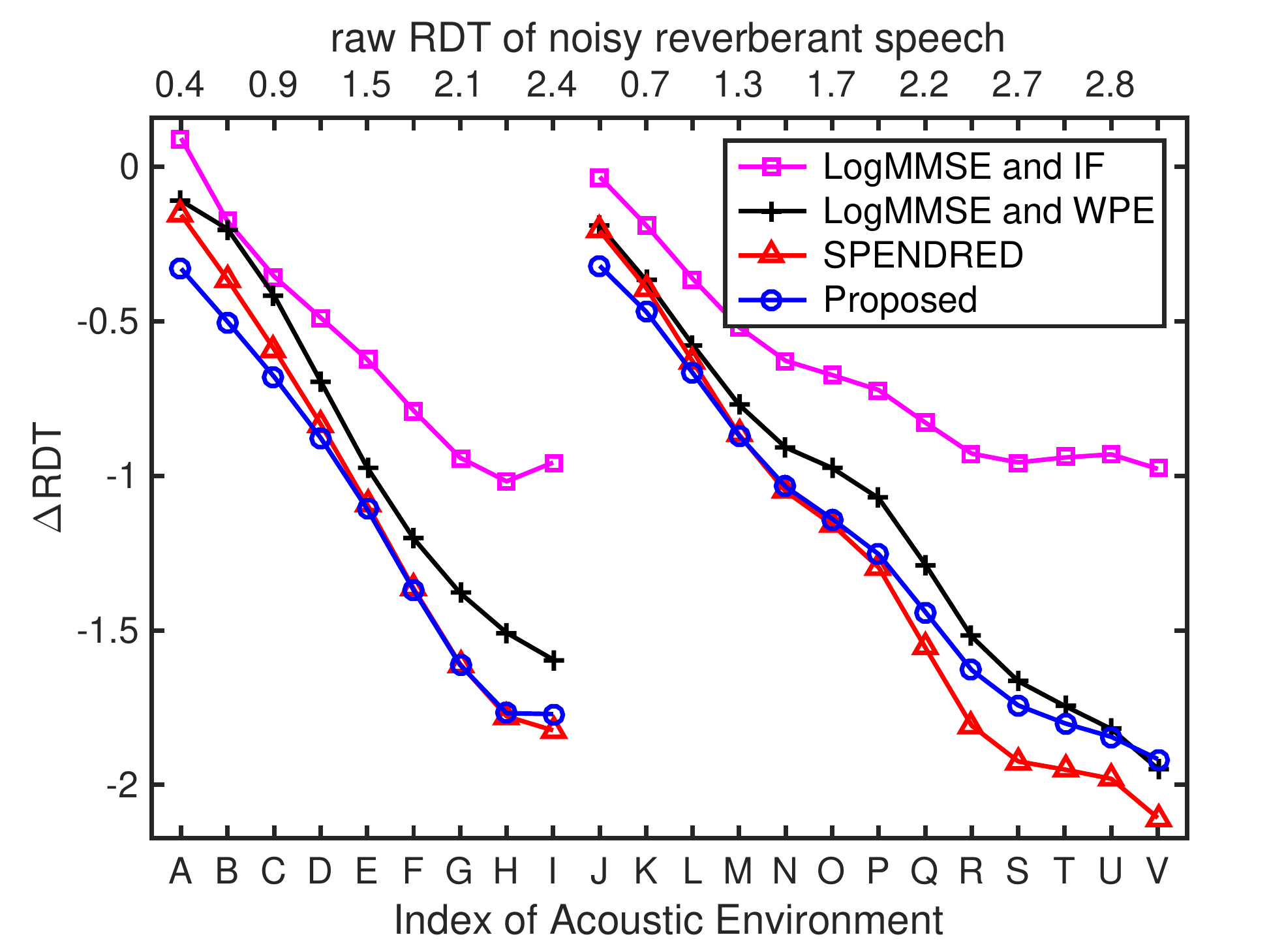}

{\footnotesize (c)}%
\end{minipage}\hfill{}%
\begin{minipage}[t]{0.48\columnwidth}%
\centering \includegraphics[bb=6bp 0bp 525bp 411bp,clip,width=1\columnwidth]{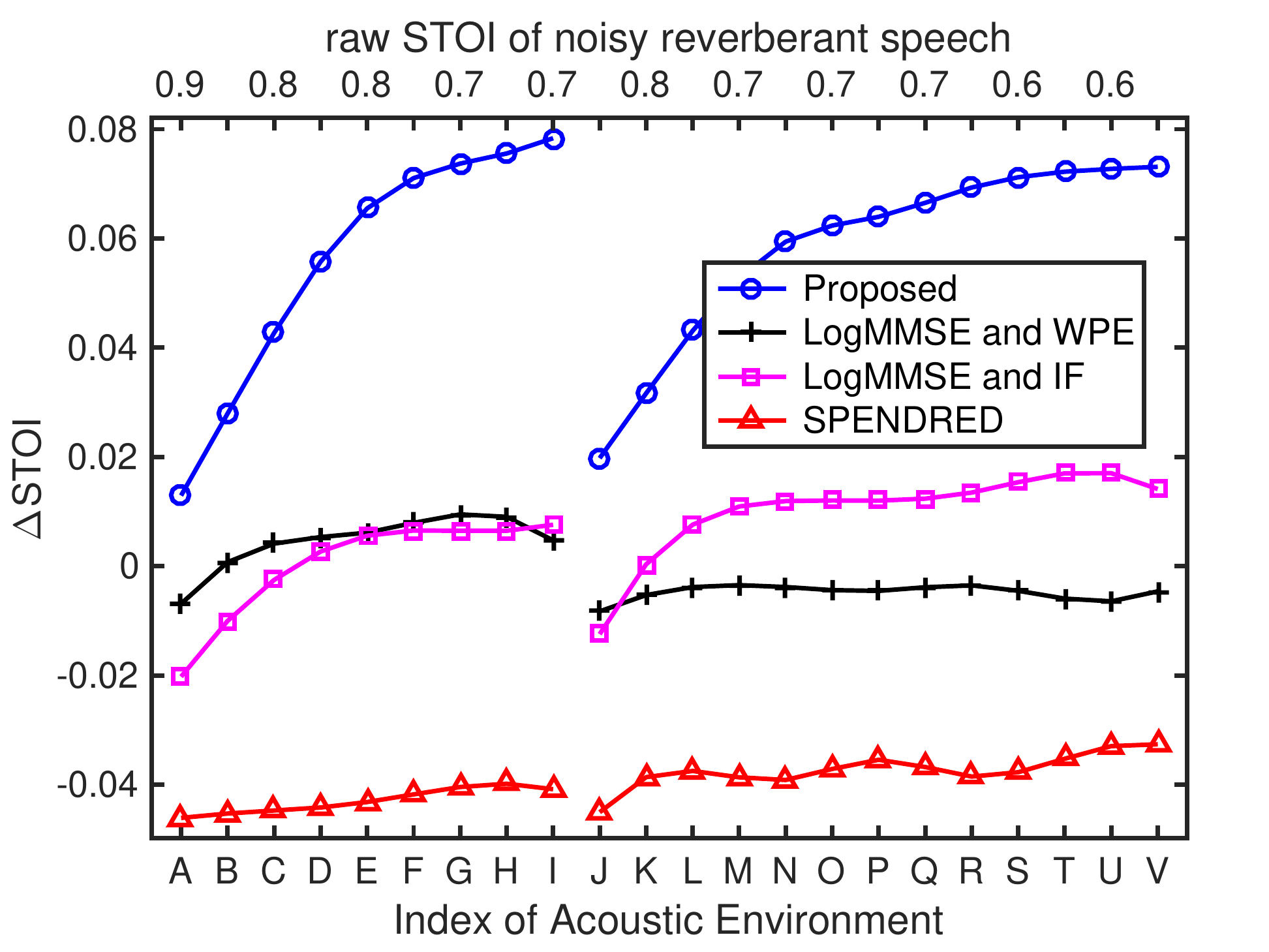}

{\footnotesize (d)}%
\end{minipage}

\caption{Plot of: (a) $\Delta$PESQ where higher scores signify better speech
quality, (b) $\Delta$CD where lower values signify better speech
quality, (c) $\Delta$RDT where lower values signify better dereverberation,
and (d) $\Delta$STOI where higher scores signify better intelligibility.
The graphs are against the index of the acoustic environment and,
hence, against the $T_{60}$ and the DRR. The mean over the noises
of white, babble and factory at $20$ dB SNR is shown.}
\end{figure}

\begin{figure}[t]
\begin{minipage}[t]{0.48\columnwidth}%
\centering \includegraphics[bb=10bp 0bp 525bp 411bp,clip,width=1\columnwidth]{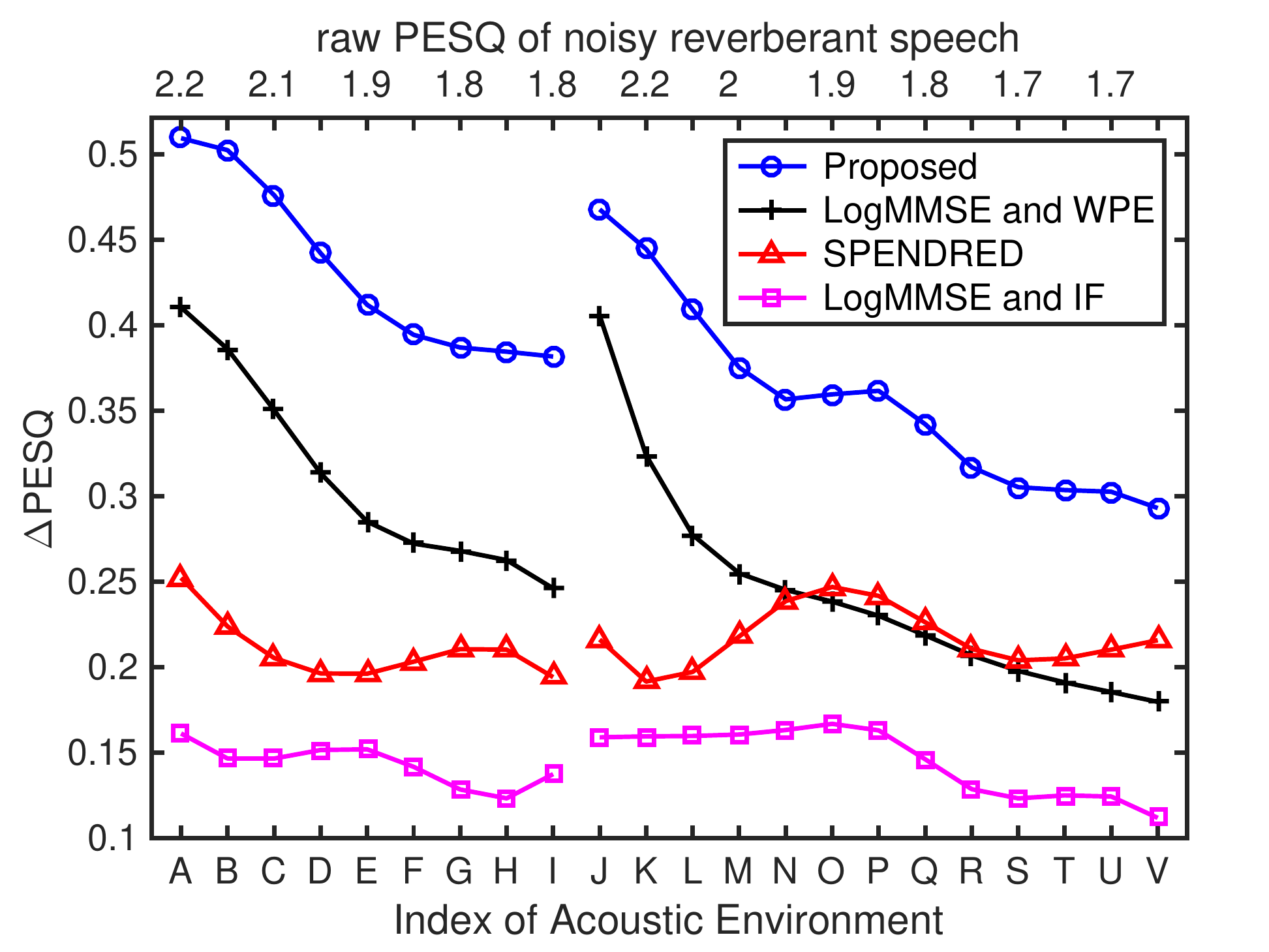}

{\footnotesize (a)}%
\end{minipage}\hfill{}%
\begin{minipage}[t]{0.48\columnwidth}%
\centering \includegraphics[bb=10bp 0bp 525bp 411bp,clip,width=1\columnwidth]{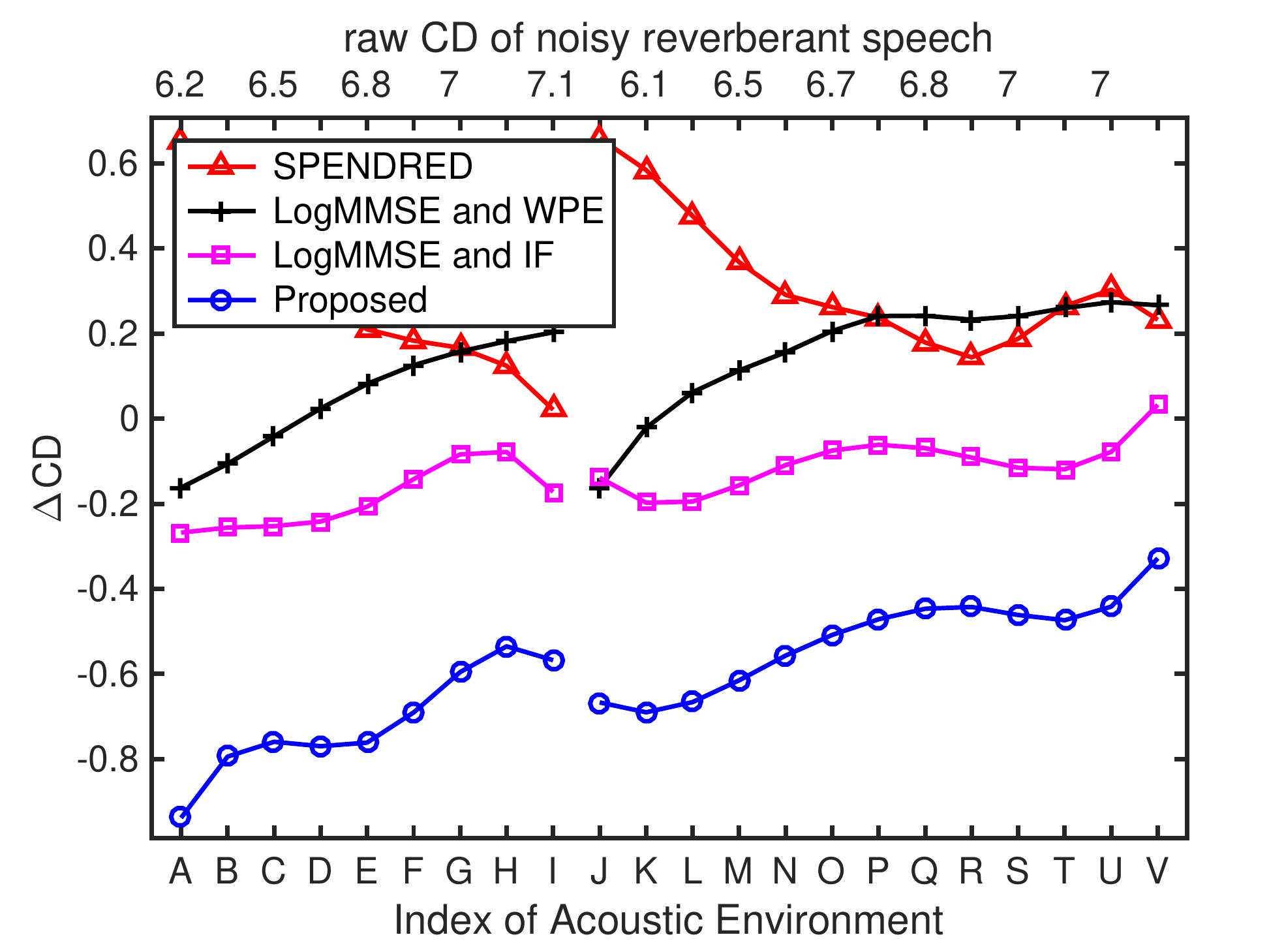}

{\footnotesize (b)}%
\end{minipage}

\vspace{1.10mm}

\begin{minipage}[t]{0.48\columnwidth}%
\centering \includegraphics[bb=10bp 0bp 525bp 411bp,clip,width=1\columnwidth]{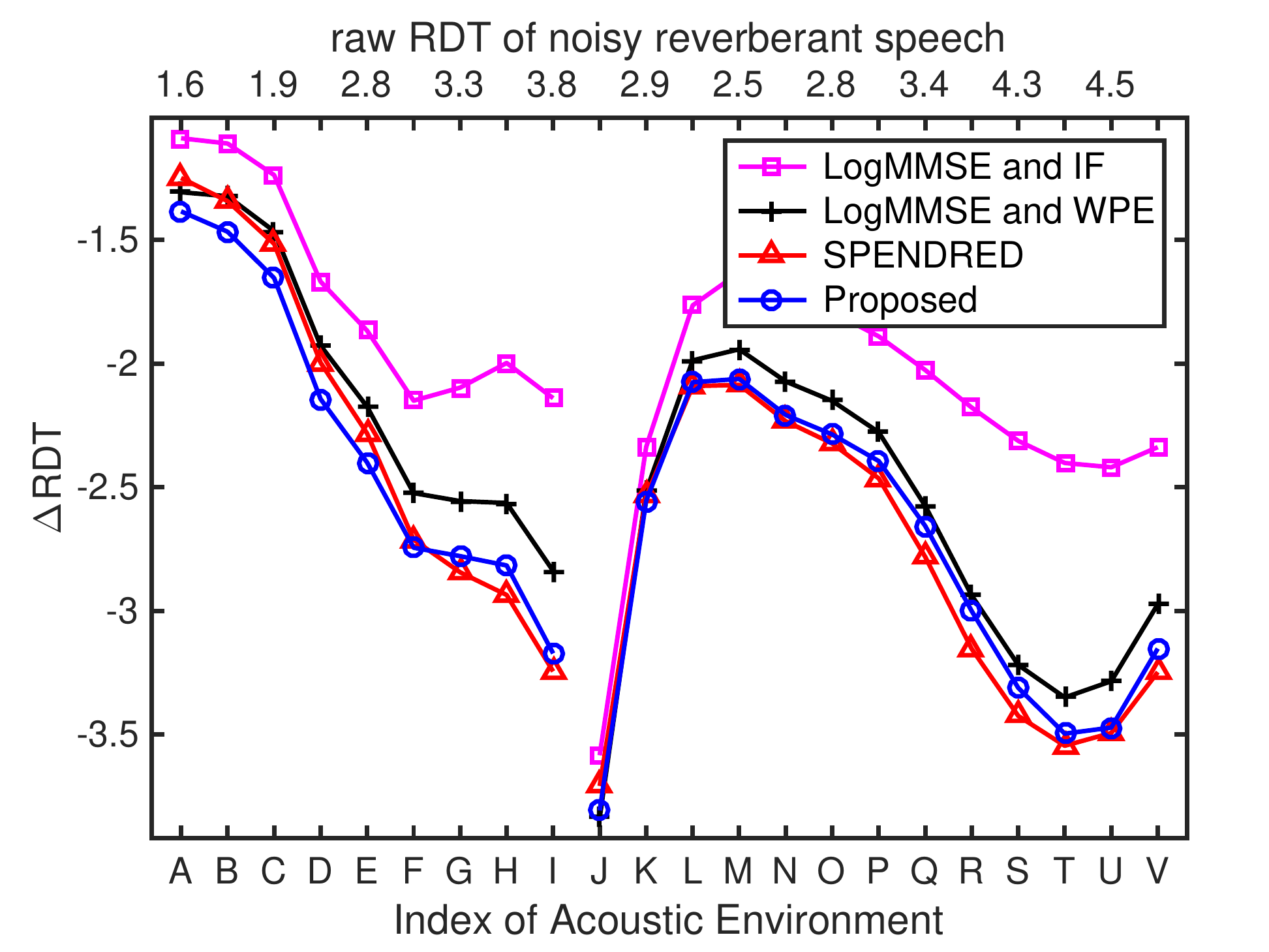}

{\footnotesize (c)}%
\end{minipage}\hfill{}%
\begin{minipage}[t]{0.48\columnwidth}%
\centering \includegraphics[bb=6bp 0bp 525bp 411bp,clip,width=1\columnwidth]{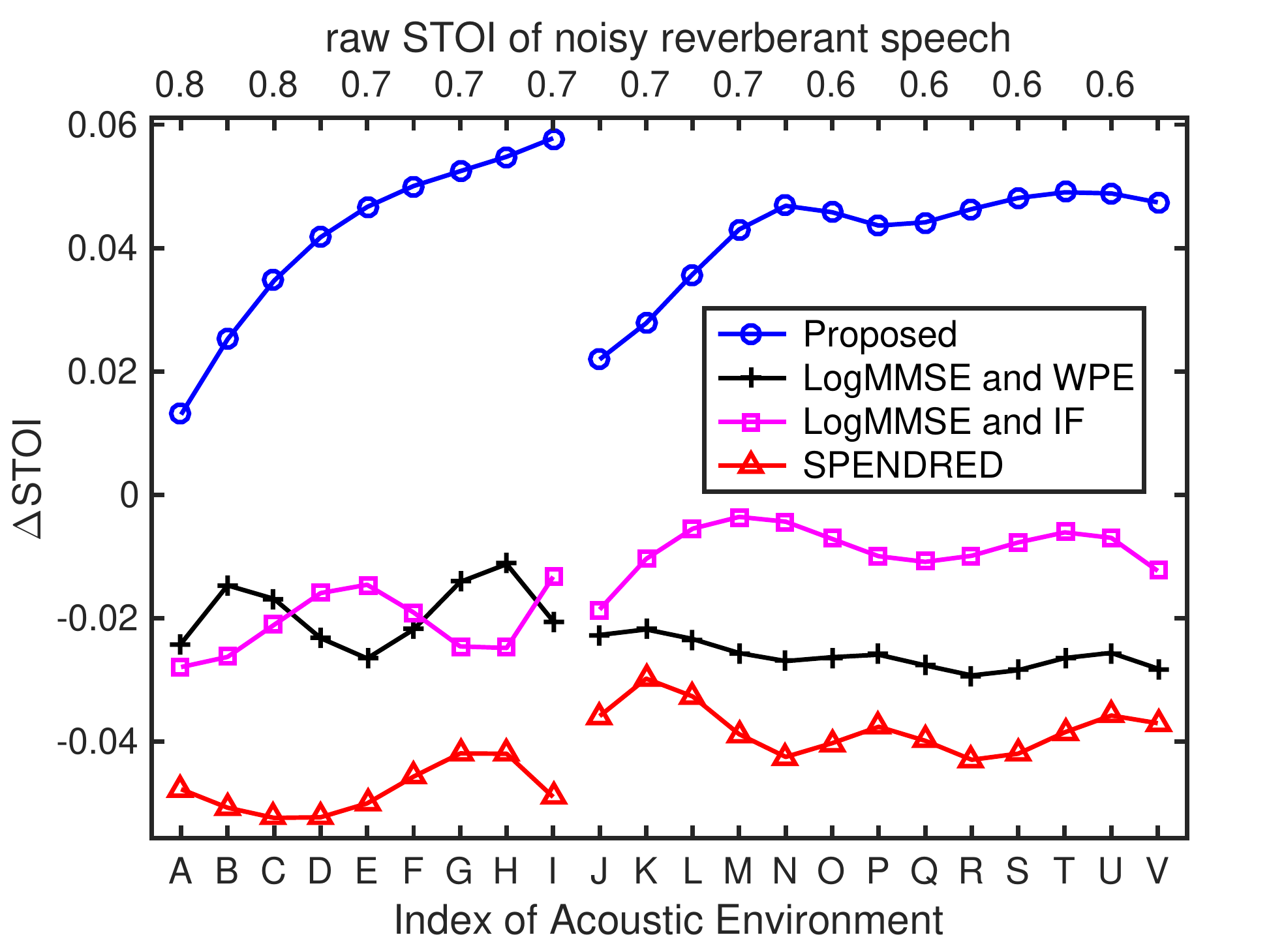}

{\footnotesize (d)}%
\end{minipage}

\caption{Plot of: (a) $\Delta$PESQ, (b) $\Delta$CD, (c) $\Delta$RDT, and
(d) $\Delta$STOI. The graphs are against the index of the acoustic
environment. The average over the noise types of white, babble and
factory at $10$ dB SNR is shown.}
\end{figure}

The KF algorithm is compared to the SPENDRED algorithm \cite{a179}
\cite{a175}, which jointly performs blind denoising and dereverberation,
and to algorithms that sequentially perform denoising and dereverberation,
specifically to the Log-MMSE estimator \cite{Ephraim1985} followed
by the WPE algorithm \cite{a327} \cite{a249} and to the Log-MMSE
estimator followed by an inverse-filter (IF) dereverberation method
that is based on $T_{60}$ and DRR estimates \cite{a256}. For the
IF, we blindly estimate the $T_{60}$ and the DRR for the entire speech
utterance using the algorithm in \cite{a256} \cite{a342}, whose
implementation was generously provided by the author. As found in
the Acoustic Characterisation of Environments (ACE) challenge \cite{a257},
the $T_{60}$ estimator in \cite{a256} had the best performance of
the examined $T_{60}$ estimators.

For evaluation, the examined acoustic conditions are shown in Table
\ref{tab:abchhghahhaa}. Two different rooms with dimensions $5 \times 4 \times 4$
m and $10 \times 7 \times 3$ m are used. The distance between the
microphone and the talker is $1.5$ m. Table \ref{tab:abchhghahhaa}
is sorted with respect to the $T_{60}$ and the DRR from A to I and
from J to V. We plot the improvement in PESQ, $\Delta$PESQ, against
the index of the acoustic environment to evaluate the KF algorithm.
The top axis shows the raw PESQ of the noisy reverberant speech and
is monotonic from A to I and from J to V. Similar graphs are also
plotted for the other metrics. The ordering of the legends matches
that of the algorithms at low $T_{60}$ values.

Table \ref{tab:abchhghahhaa} also shows the $a$ and $b$ reverberation
values of the RIRs and we note that one of the baselines that we use,
i.e. the SPENDRED algorithm \cite{a179} \cite{a175}, assumes that
$b \leq 1$. This is valid in Table \ref{tab:abchhghahhaa} because
$b << 1$ and its range is small.

\begin{figure}[t]
\begin{minipage}[t]{0.48\columnwidth}%
\centering \includegraphics[bb=10bp 0bp 525bp 411bp,clip,width=1\columnwidth]{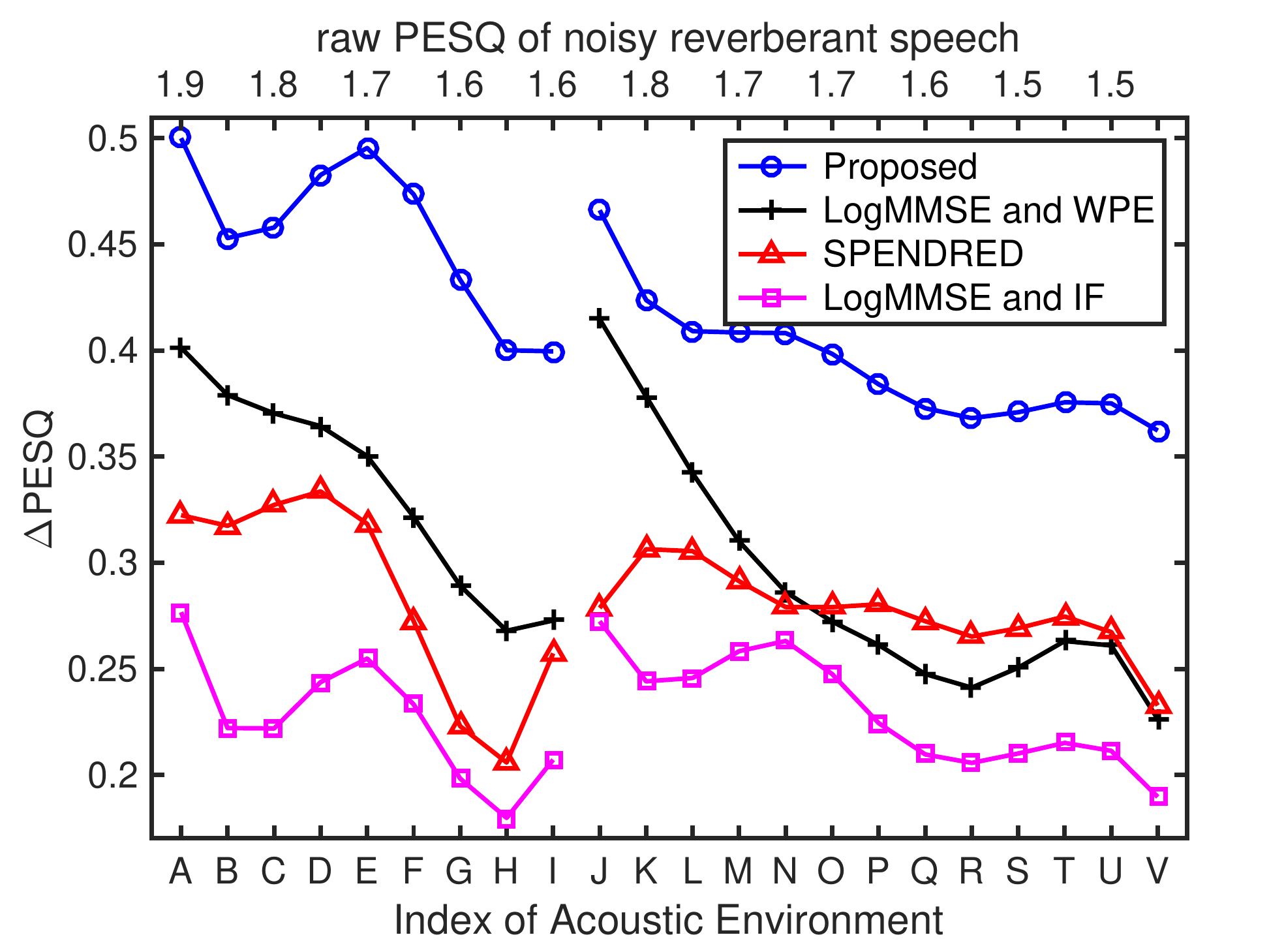}

{\footnotesize (a)}%
\end{minipage}\hfill{}%
\begin{minipage}[t]{0.48\columnwidth}%
\centering \includegraphics[bb=10bp 0bp 525bp 411bp,clip,width=1\columnwidth]{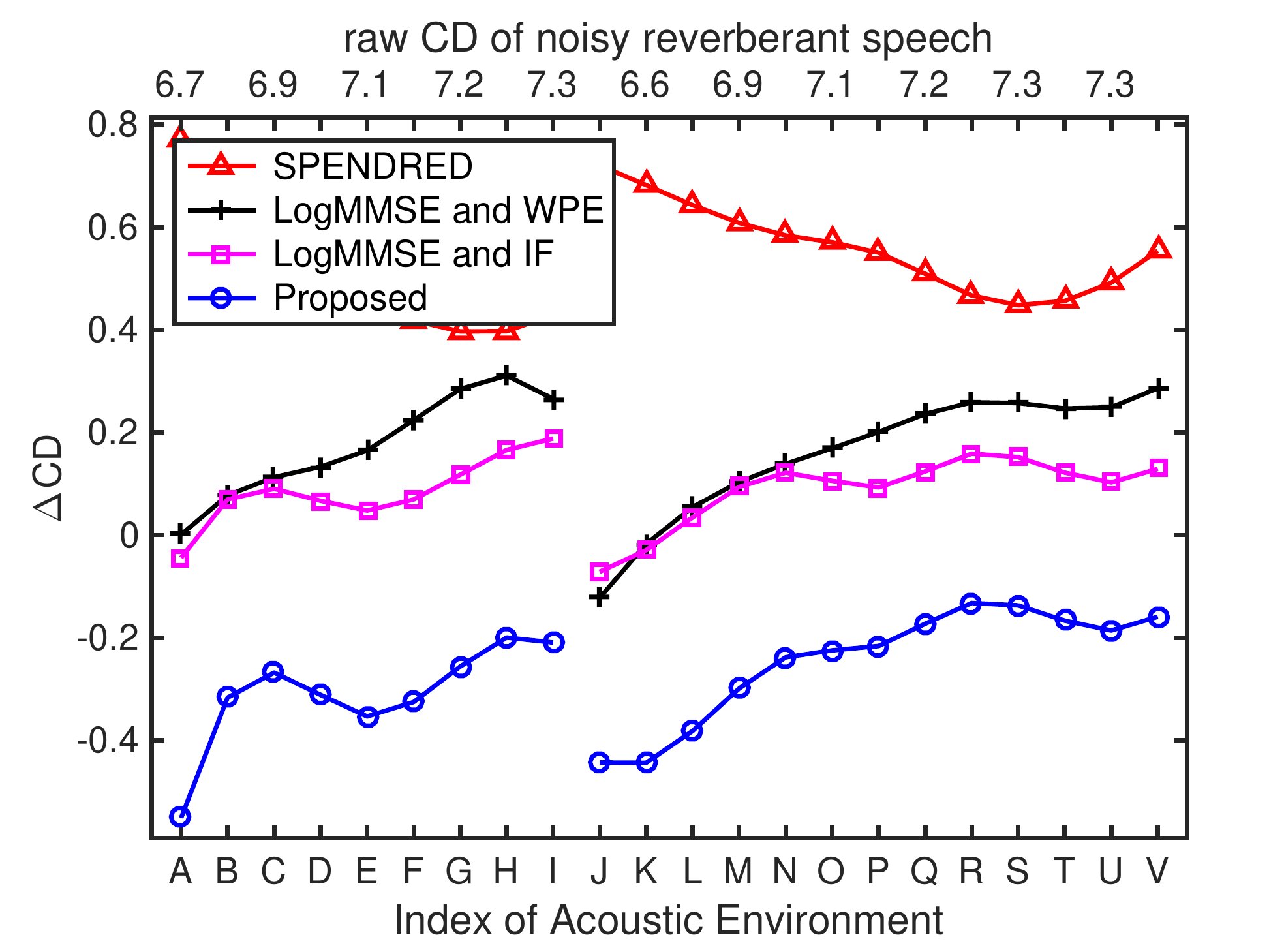}

{\footnotesize (b)}%
\end{minipage}

\vspace{1.10mm}

\begin{minipage}[t]{0.48\columnwidth}%
\centering \includegraphics[bb=10bp 0bp 525bp 411bp,clip,width=1\columnwidth]{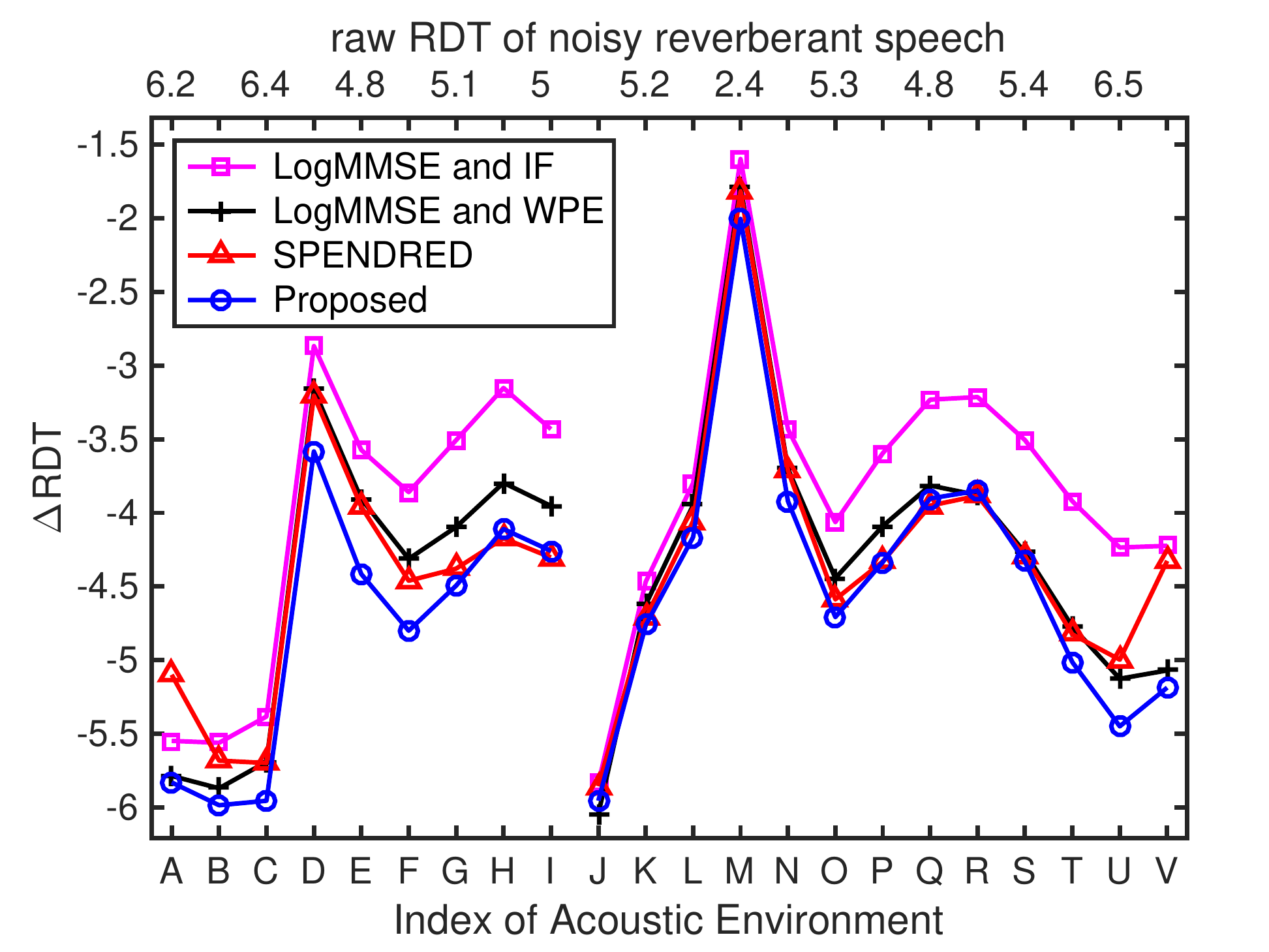}

{\footnotesize (c)}%
\end{minipage}\hfill{}%
\begin{minipage}[t]{0.48\columnwidth}%
\centering \includegraphics[bb=6bp 0bp 525bp 411bp,clip,width=1\columnwidth]{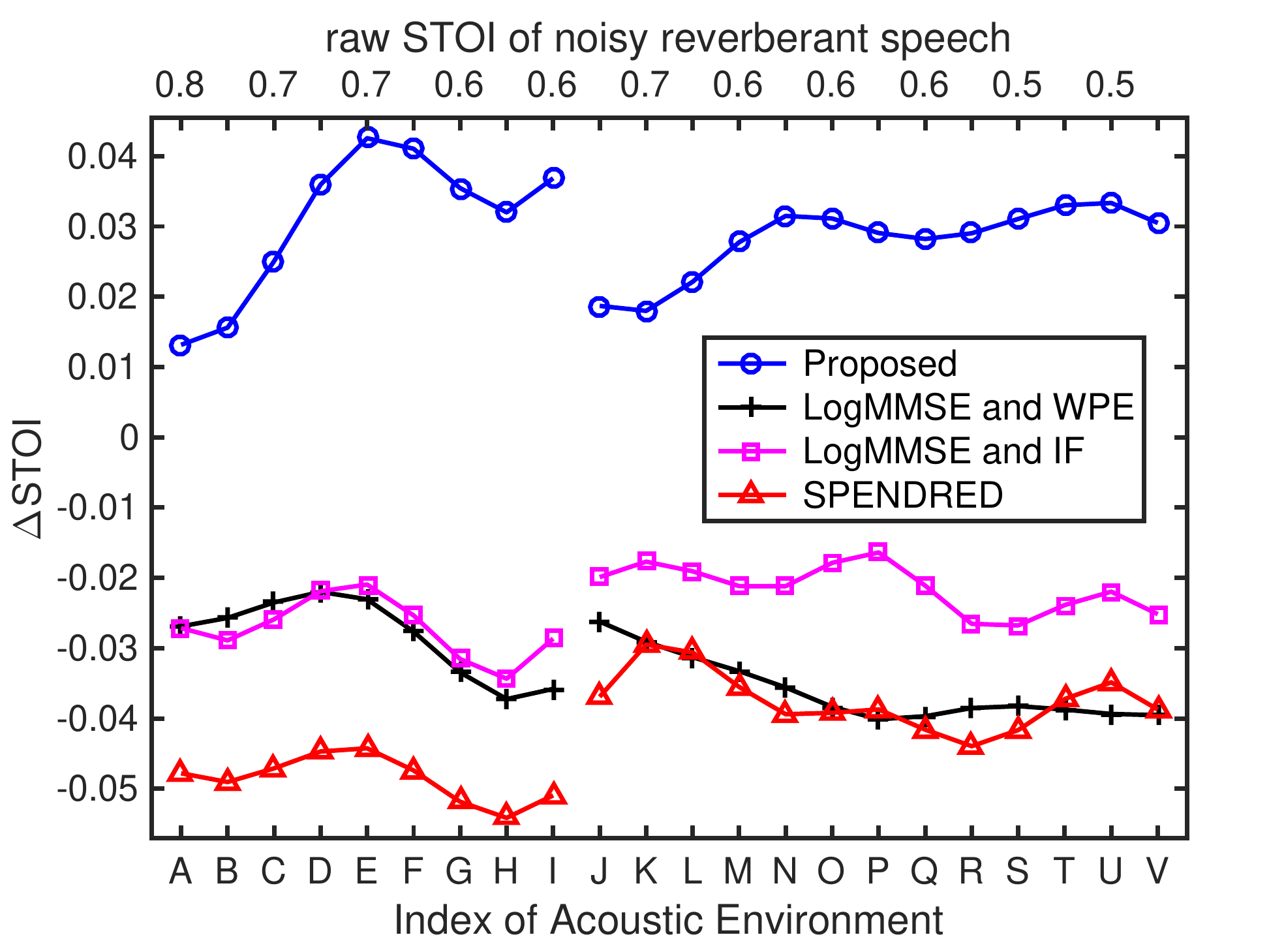}

{\footnotesize (d)}%
\end{minipage}

\caption{Plot of: (a) $\Delta$PESQ, (b) $\Delta$CD, (c) $\Delta$RDT, and
(d) $\Delta$STOI. The graphs are against the index of the acoustic
environment. The average over the noise types of white, babble and
factory at $5$ dB SNR is shown.}
\end{figure}

\subsection{Overall Performance Against the $T_{60}$ and the DRR}

This section investigates the overall performance of the KF algorithm
against the $T_{60}$ and the DRR. Figure 7 examines: (a) the $\Delta$PESQ,
where higher scores signify better speech quality, (b) the $\Delta$CD,
where lower values signify better speech quality, (c) the $\Delta$RDT,
where lower values signify better dereverberation, and (d) the $\Delta$STOI,
where higher scores signify better intelligibility. Figure 7 first
presents the results that are related to speech quality, in (a) and
(b). Then, it shows the dereverberation results, in (c), and the speech
intelligibility results, in (d). The graphs are against the index
of the acoustic environment and, hence, against the $T_{60}$ and
the DRR. The average over the noise types of white, babble and factory
at $20$ dB SNR is shown. We note that noisy reverberant speech with
$T_{60} \approx 0.6$ s has a raw PESQ score of $2$ in Fig. 7(a).

In Fig. 7(a), the top axis is monotonic from A to I and from J to
V and there is no transition from index I to index J.

\begin{figure}[t]
\begin{minipage}[t]{0.48\columnwidth}%
\centering \includegraphics[bb=10bp 0bp 525bp 411bp,clip,width=1\columnwidth]{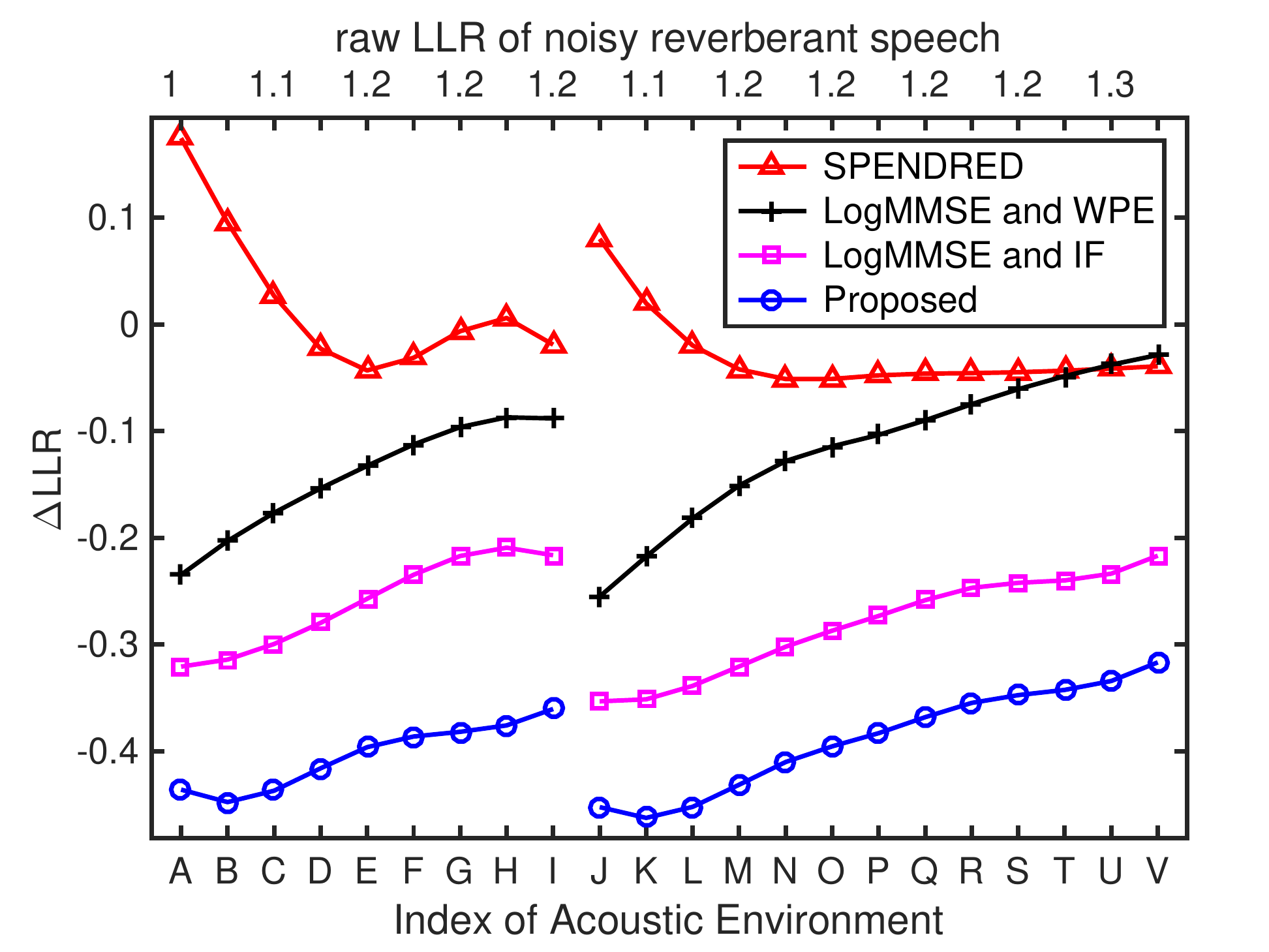}

{\footnotesize (a)}%
\end{minipage}\hfill{}%
\begin{minipage}[t]{0.48\columnwidth}%
\centering \includegraphics[bb=6bp 0bp 525bp 411bp,clip,width=1\columnwidth]{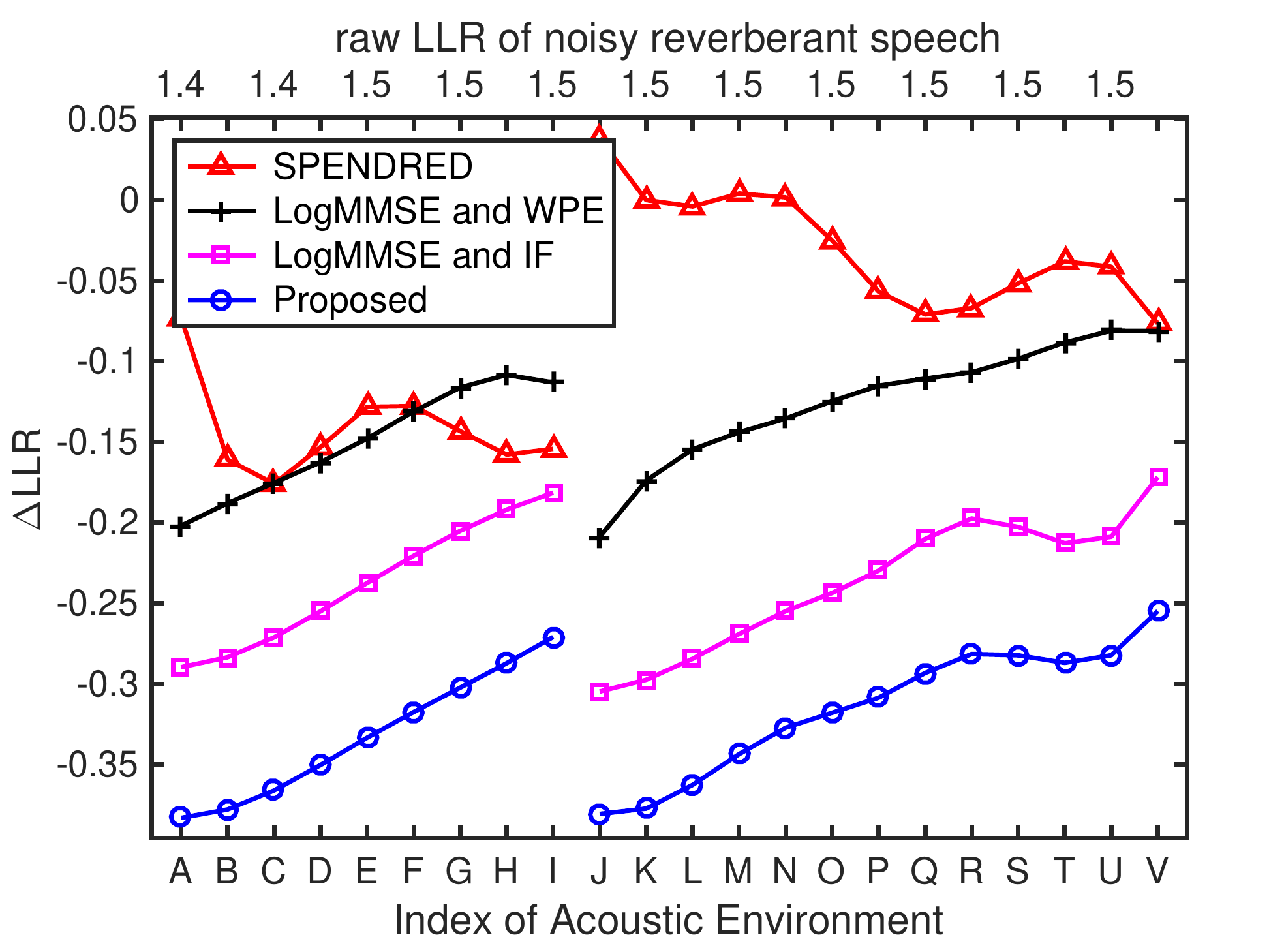}

{\footnotesize (b)}%
\end{minipage}\caption{Plot of $\Delta$LLR where lower scores signify better speech quality
when white noise is used: (a) at $20$ dB SNR, and (b) at $10$ dB
SNR.}
\end{figure}

\begin{figure}[t]
\begin{minipage}[t]{0.48\columnwidth}%
\centering \includegraphics[bb=10bp 0bp 525bp 411bp,clip,width=1\columnwidth]{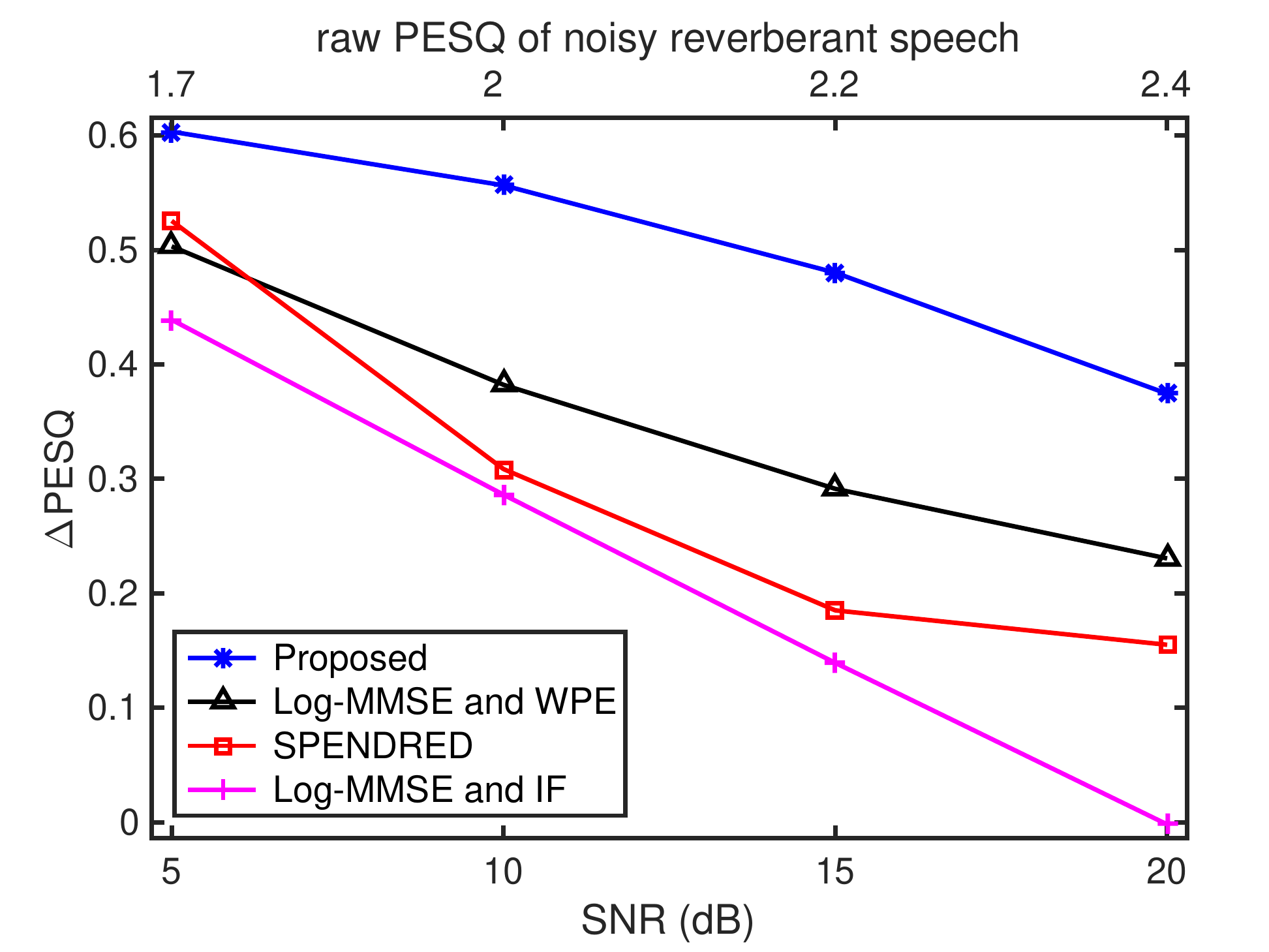}

{\footnotesize (a)}%
\end{minipage}\hfill{}%
\begin{minipage}[t]{0.48\columnwidth}%
\centering \includegraphics[bb=6bp 0bp 525bp 411bp,clip,width=1\columnwidth]{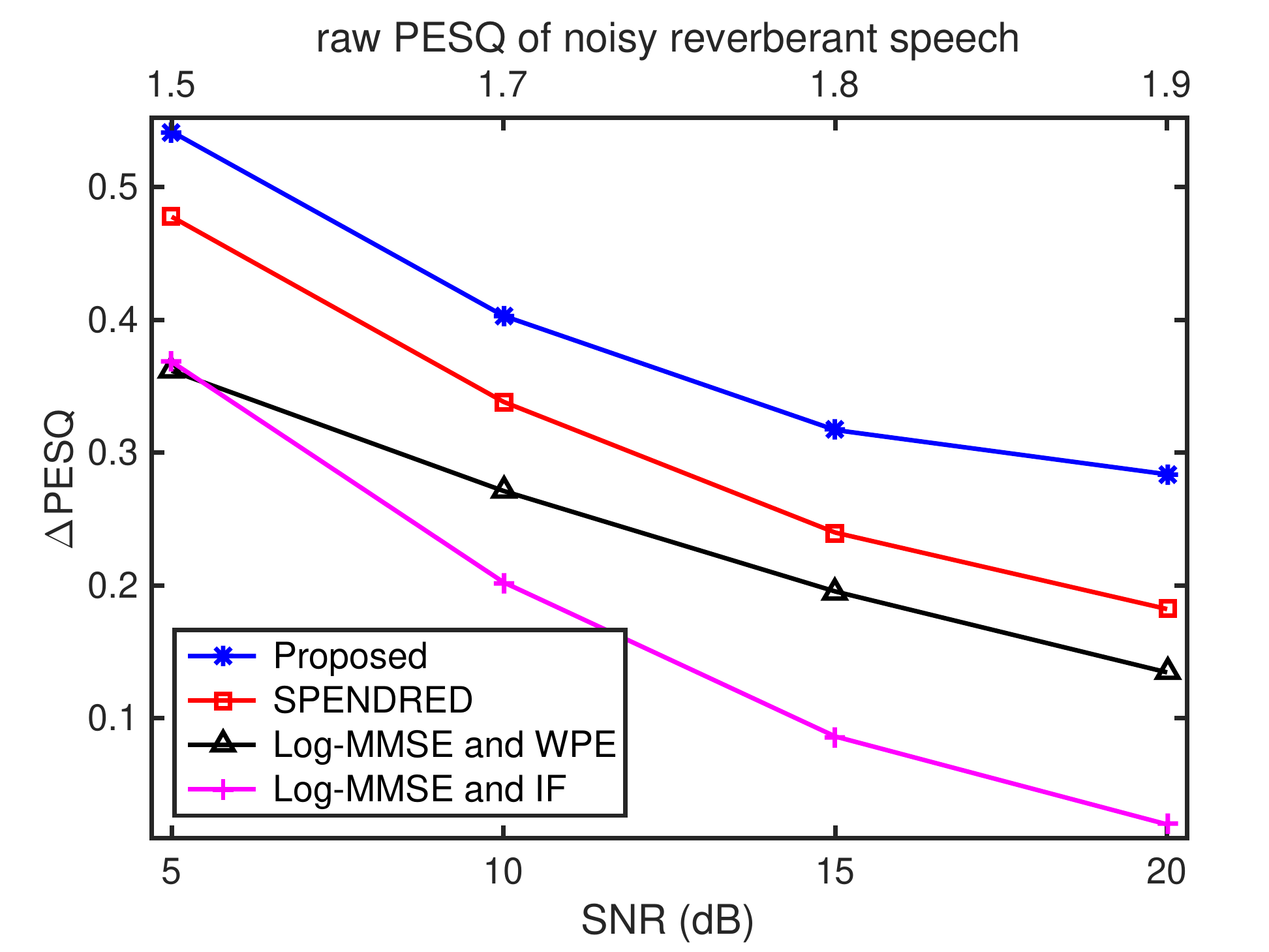}

{\footnotesize (b)}%
\end{minipage}

\vspace{1.10mm}

\begin{minipage}[t]{0.48\columnwidth}%
\centering \includegraphics[bb=10bp 0bp 525bp 411bp,clip,width=1\columnwidth]{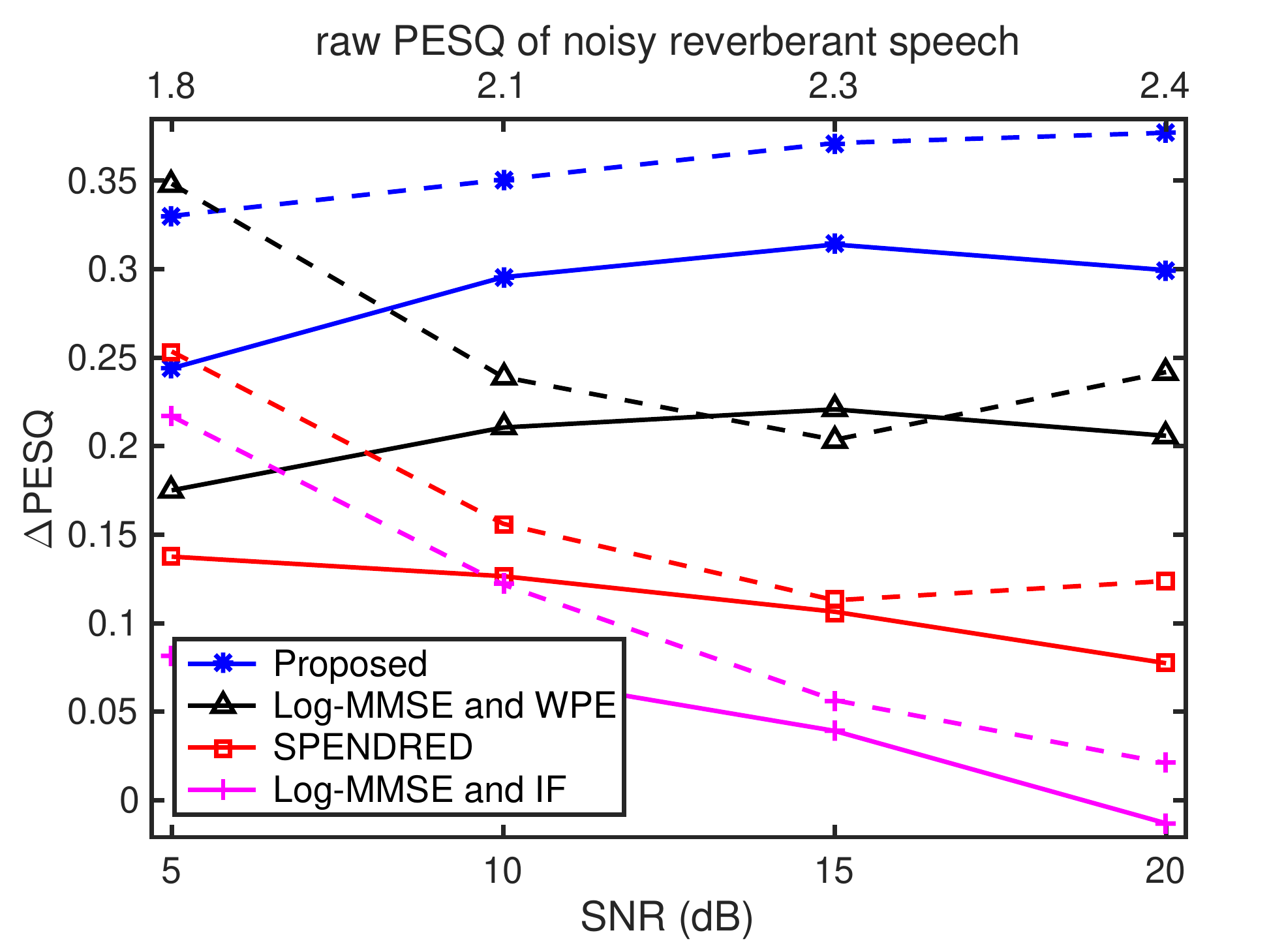}

{\footnotesize (c)}%
\end{minipage}\hfill{}%
\begin{minipage}[t]{0.48\columnwidth}%
\centering \includegraphics[bb=10bp 0bp 525bp 411bp,clip,width=1\columnwidth]{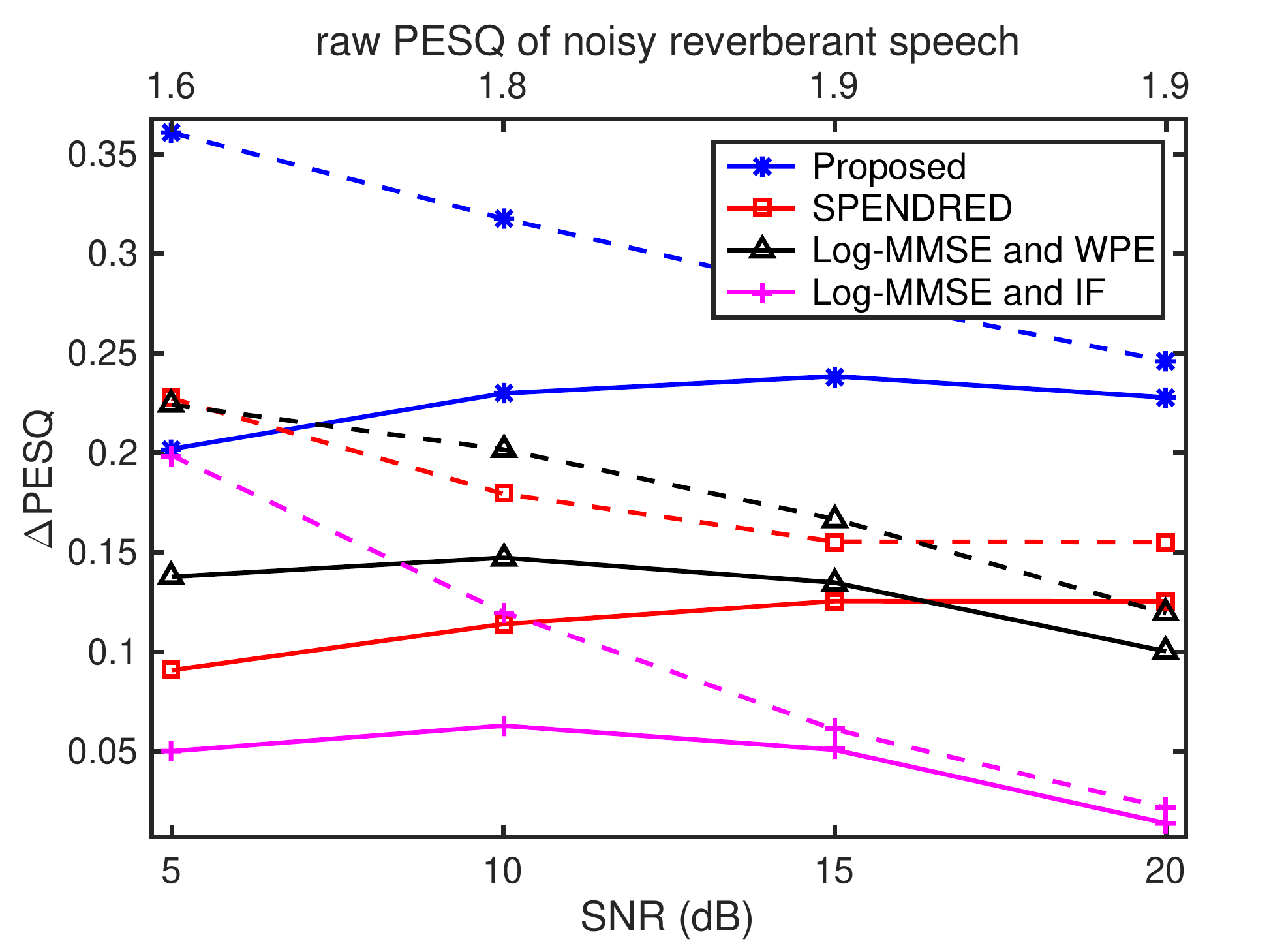}

{\footnotesize (d)}%
\end{minipage}

\caption{Plot of $\Delta$PESQ against SNR for: (a), (b) white noise, and (c),
(d) babble noise (solid lines) and factory noise (dashed lines). In
(a) and (c), $T_{60} = 0.40$ s and \text{DRR} = 0.17 dB (i.e. index
L in Table \ref{tab:abchhghahhaa}). In (b) and (d), $T_{60} = 0.73$
s and \text{DRR} = -2.1 dB (i.e. index R). In (c) and (d), babble
noise is used for the top axis and for the legends.}
\end{figure}

From Fig. 7(a), in terms of PESQ, the proposed algorithm consistently
yields improved speech quality in challenging environments compared
to the examined baselines. Compared to the unprocessed noisy reverberant
speech, the algorithm shows improved performance for all the examined
$T_{60}$ range from $0.18$ to $1.05$ s and DRR range from $8.43$
to $-3.33$ dB. Compared to the unprocessed speech, for a $T_{60}$
of $0.3$ and $1$ s, the algorithm has a $\Delta$PESQ of $0.35$
and $0.25$, respectively, for the SNR of $20$ dB averaged over the
examined noises.

From Fig. 7(b), in terms of CD, the KF algorithm yields improved speech
quality in acoustic environments with a $T_{60}$ from $0.18$ to
$1.05$ s. The KF algorithm shows a deteriorating CD improvement with
increasing $T_{60}$. Compared to the unprocessed signal, for a $T_{60}$
of $0.3$ and $1$ s, the algorithm has a CD improvement of approximately
$-1.1$ and $-0.6$, respectively, for the SNR of $20$ dB averaged
over the tested noise types.

From Fig. 7(c), we observe that the KF algorithm yields improved speech
dereverberation in challenging acoustic conditions and that the $\Delta$RDT
improves with increasing $T_{60}$ values. From the raw RDT of noisy
reverberant speech in Fig. 7(c), we observe that the indexes A and
J have very low reverberation. The indexes A and J have raw RDT scores
$0.4$ and $0.7$, respectively. The indexes G-I and T-V have high
reverberation with a high raw RDT score. The proposed algorithm in
these high reverberation cases achieves a large RDT improvement decreasing
the RDT metric from approximately $2.6$ to $0.8$.

From Fig. 7(d), the KF algorithm shows marginally improved STOI performance
for the examined $T_{60}$ range compared to the unprocessed speech
and to the examined baselines. For a $T_{60}$ of $0.3$ and $1$
s, the algorithm has a $\Delta$STOI of $0.03$ and $0.07$, respectively.
With increasing $T_{60}$, the $\Delta$STOI scores slightly increase.
The baselines have negative $\Delta$STOI.

\begin{figure}[t]
\begin{minipage}[t]{0.48\columnwidth}%
\centering \includegraphics[bb=10bp 0bp 525bp 411bp,clip,width=1\columnwidth]{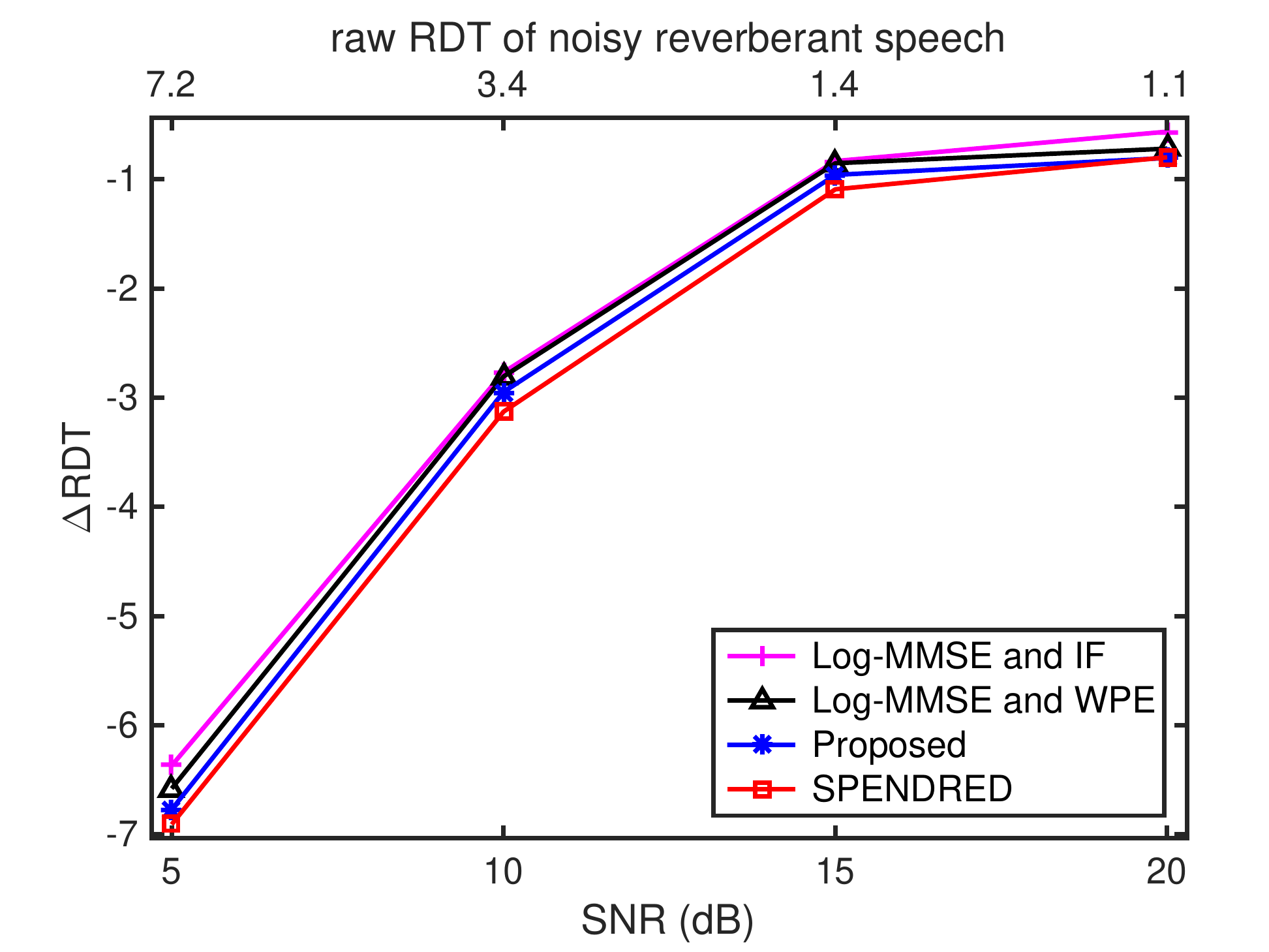}

{\footnotesize (a)}%
\end{minipage}\hfill{}%
\begin{minipage}[t]{0.48\columnwidth}%
\centering \includegraphics[bb=6bp 0bp 525bp 411bp,clip,width=1\columnwidth]{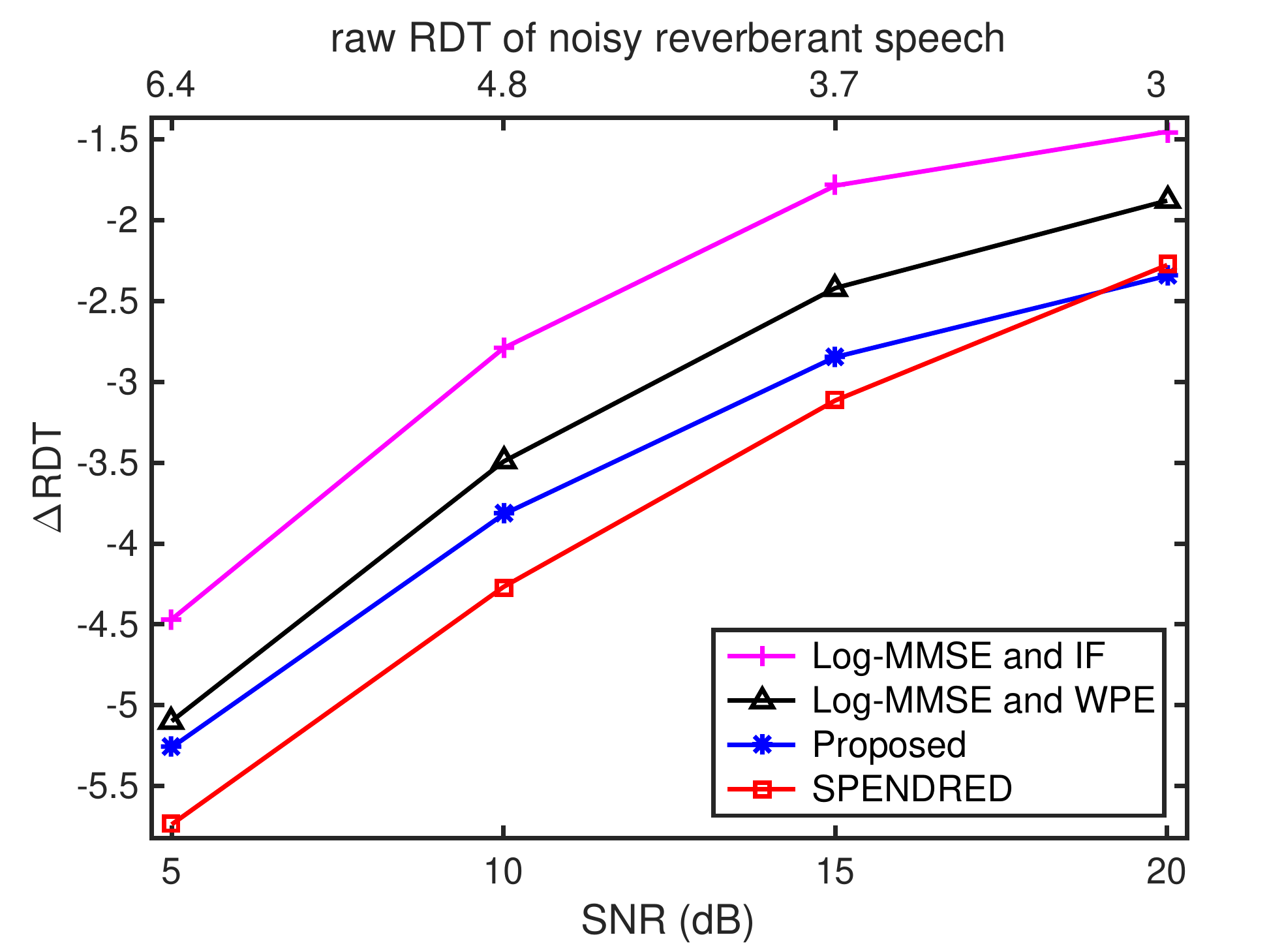}

{\footnotesize (b)}%
\end{minipage}

\vspace{1.10mm}

\begin{minipage}[t]{0.48\columnwidth}%
\centering \includegraphics[bb=10bp 0bp 525bp 411bp,clip,width=1\columnwidth]{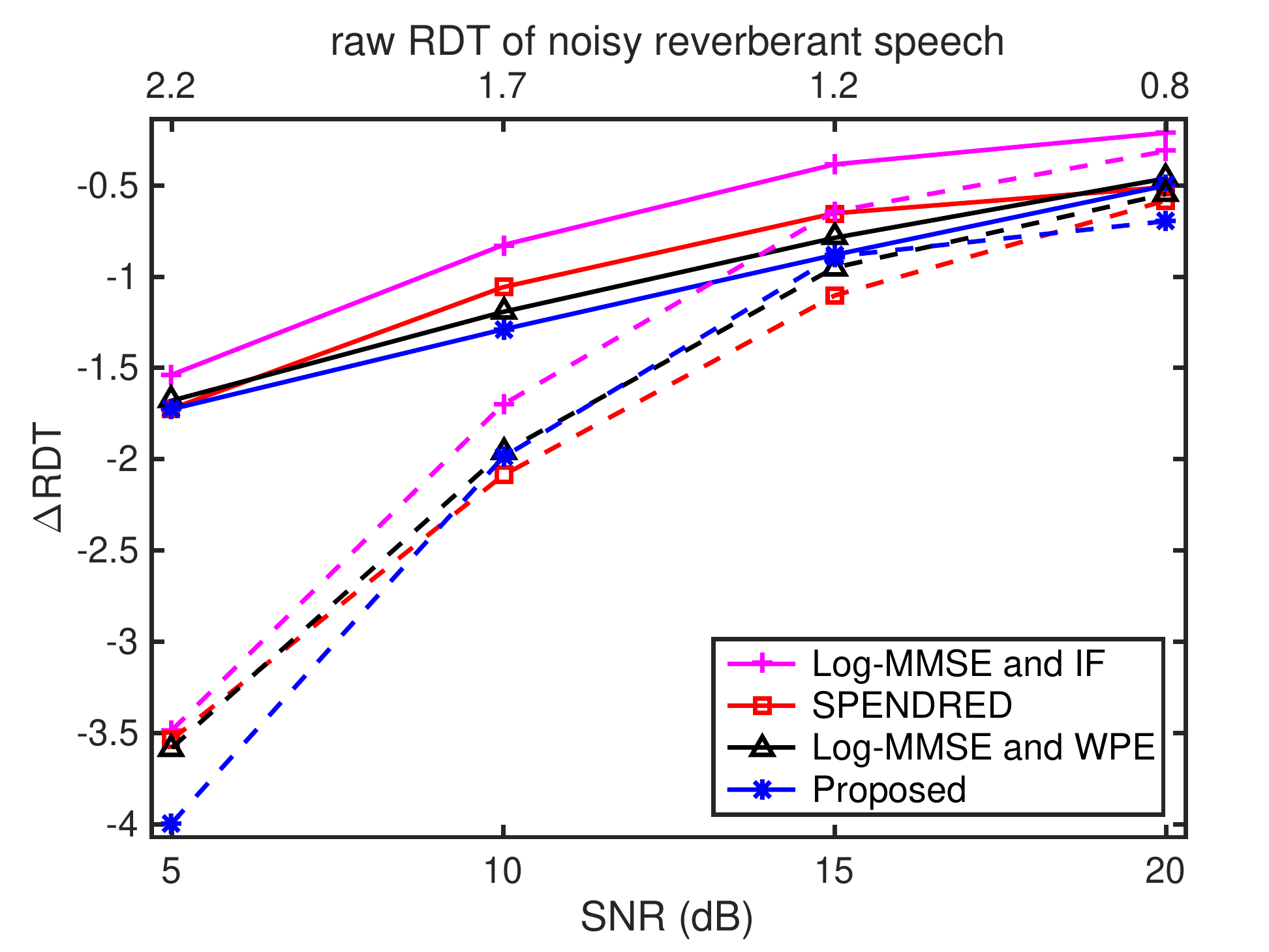}

{\footnotesize (c)}%
\end{minipage}\hfill{}%
\begin{minipage}[t]{0.48\columnwidth}%
\centering \includegraphics[bb=6bp 0bp 525bp 411bp,clip,width=1\columnwidth]{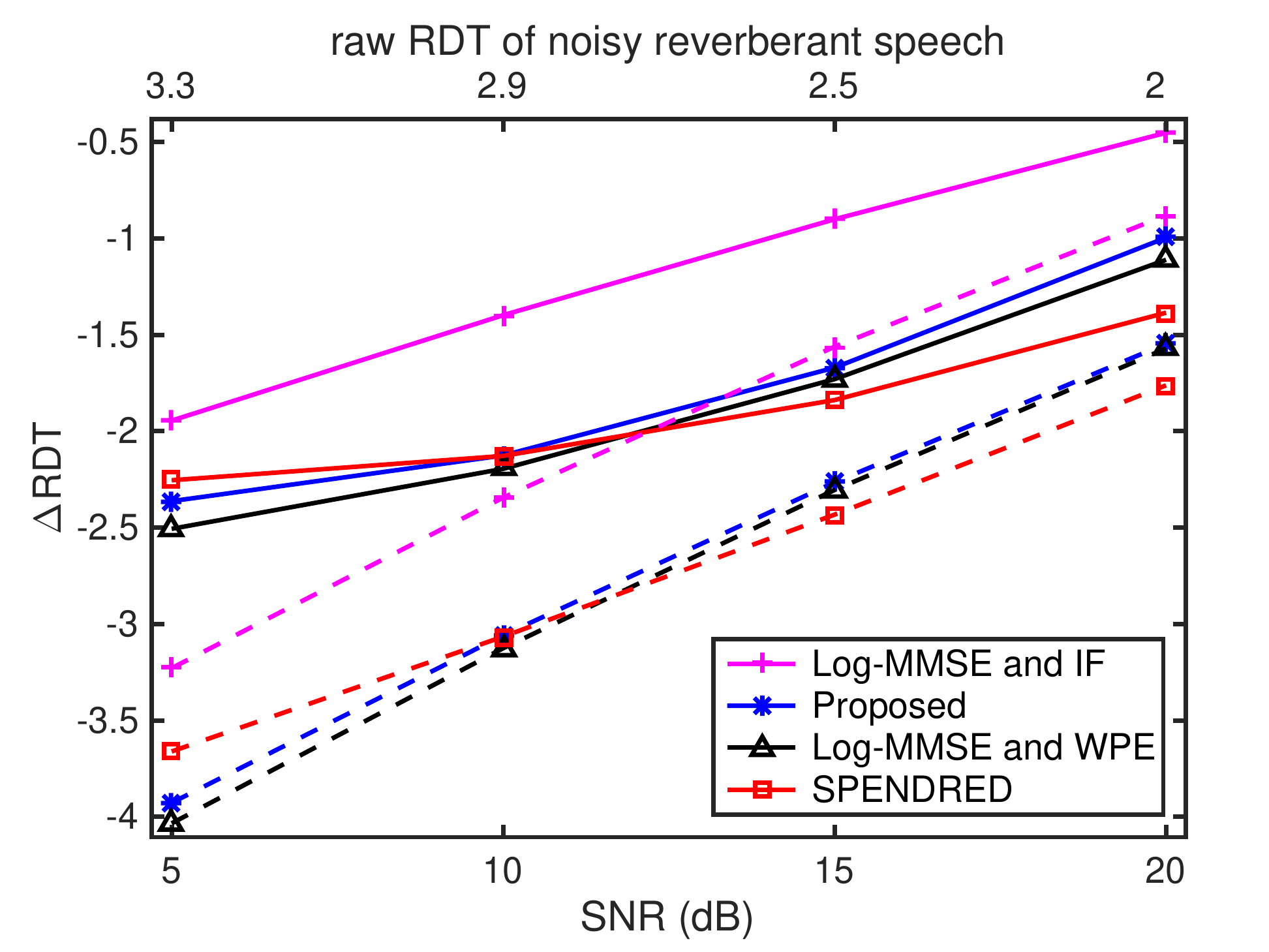}

{\footnotesize (d)}%
\end{minipage}

\caption{Plot of $\Delta$RDT against SNR for: (a), (b) white noise, and (c),
(d) babble noise (solid lines) and factory noise (dashed lines). In
(a) and (c), $T_{60} = 0.40$ s and \text{DRR} = 0.17 dB (i.e. index
L in Table \ref{tab:abchhghahhaa}). In (b) and (d), $T_{60} = 0.73$
s and \text{DRR} = -2.1 dB (i.e. index R). In (c) and (d), babble
noise is used for the top axis and for the legends.}
\end{figure}

We use the noise types of white, babble and factory at $10$ dB SNR
and we obtain Fig. 8. Figure 8 shows similar graphs to Fig. 7 but
for a SNR of $10$ dB. For all the examined $T_{60}$ values, the
KF algorithm shows a consistent improvement in the evaluation metrics
of PESQ, CD, RDT and STOI.

We now use the noise types of white, babble and factory at $5$ dB
SNR and we obtain Fig. 9. Figure 9 shows similar graphs to Figs. 7
and 8. For all the examined SNRs, the KF algorithm shows a consistent
improvement in the examined metrics, depending on the $T_{60}$. From
Fig. 9(a), we observe that the $\Delta$PESQ scores of the KF algorithm
at $5$ dB SNR are similar to the $\Delta$PESQ scores at $10$ dB
SNR in Fig. 8(a).

The KF algorithm shows improved PESQ performance for the examined
SNRs from $5$ dB to $20$ dB. The algorithm has better speech quality
performance compared to the baselines for all the examined $T_{60}$
values. Compared to the unprocessed noisy reverberant speech, for
the SNR of $5$ dB, the algorithm has a $\Delta$PESQ of about $0.45$
for a $T_{60}$ of $0.6$ s and a raw PESQ of $1.6$. For the SNR
of $20$ dB, the algorithm has a $\Delta$PESQ of approximately $0.3$
for a $T_{60}$ of $0.6$ s and a raw PESQ of $2$. Comparing Figs.
7-9, we observe that the raw PESQ decreases with decreasing SNR while
the $\Delta$PESQ increases.

The KF algorithm yields improved dereverberation performance in terms
of RDT in adverse conditions in Fig. 9(c). The algorithm's performance
improves with decreasing SNR. We observe that the indexes A-C and
J-K have a high raw RDT despite that they have low reverberation at
$5$ dB SNR, when the effect of noise is higher than that of reverberation.

The results of the log-likelihood ratio (LLR) speech quality metric
\cite{Quackenbush1988} \cite{a357}, which is used as the main evaluation
metric in the dereverberation algorithm in \cite{a319}, resemble
the results of CD in Fig. 10. Both CD and LLR have been used in the
REVERB challenge \cite{a261} \cite{a262}. It was found in \cite{a357}
that LLR correlates well with speech quality although slightly less
well than PESQ. Figure 10 examines $\Delta$LLR when white noise is
used at $20$ and $10$ dB SNR. From Fig. 10, we observe that decreasing
the SNR from $20$ to $10$ dB increases the raw LLR of noisy reverberant
speech and deteriorates the $\Delta$LLR.

\subsection{Overall Performance Against the SNR}

This section investigates the overall performance of the KF algorithm
against the SNR. Figure 11 presents the algorithm's PESQ performance
compared to the baselines. Figures 11(a) and 11(b) depict the PESQ
improvement, $\Delta$PESQ, against the SNR for white noise when:
(a) $T_{60} = 0.40$ s and \text{DRR} = 0.17 dB, which is case L
in Table \ref{tab:abchhghahhaa}, and (b) $T_{60} = 0.73$ s and \text{DRR} = -2.1
dB, which is case R in Table \ref{tab:abchhghahhaa}. The indexes
L and R were chosen because they are not extreme cases and they have
different $T_{60}$ and DRR values. The ordering of the legends in
Fig. 11 matches that of the algorithms at high SNRs.

\begin{figure}[t]
\begin{minipage}[t]{0.48\columnwidth}%
\centering \includegraphics[bb=10bp 0bp 525bp 411bp,clip,width=1\columnwidth]{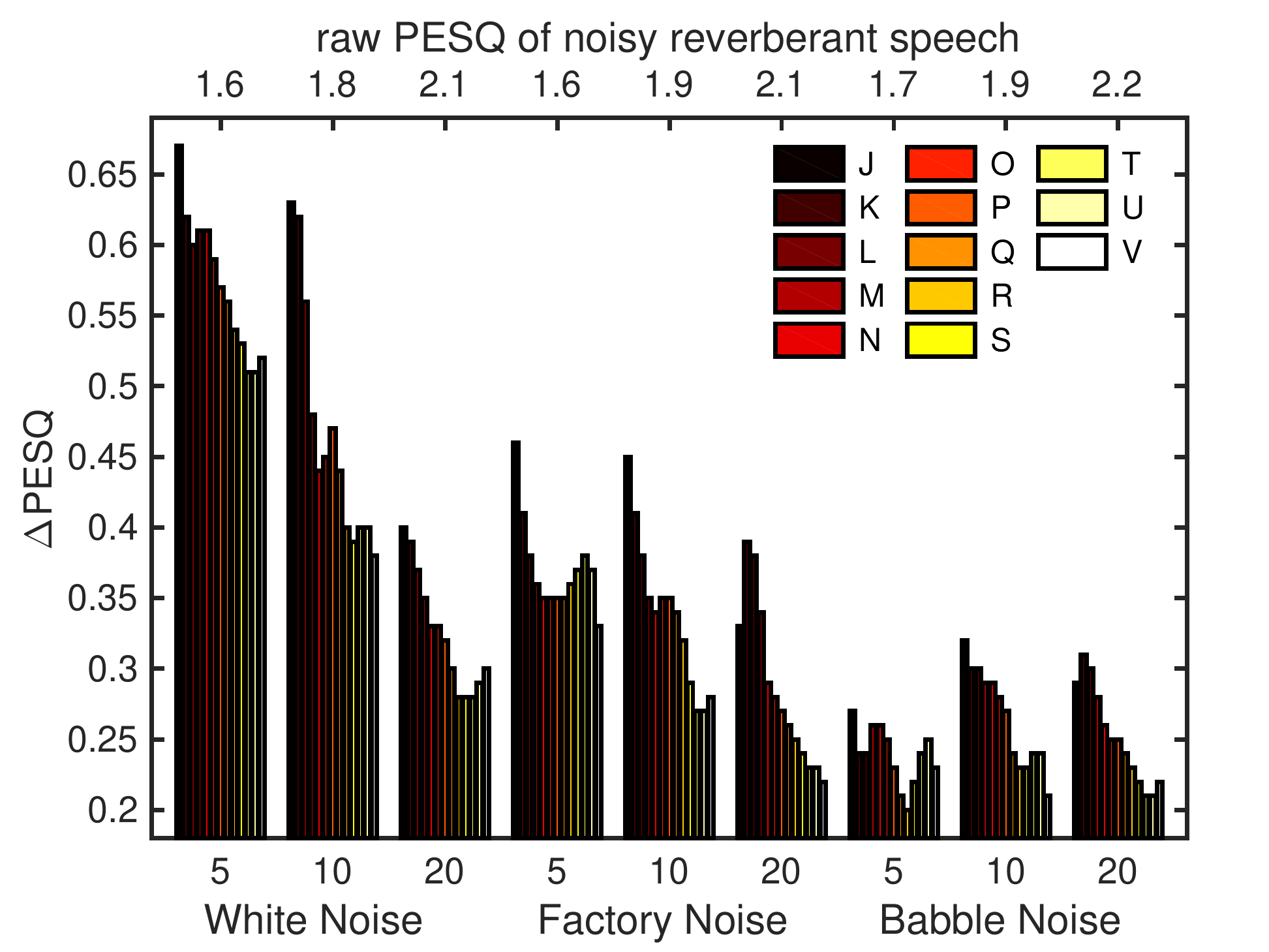}

{\footnotesize (a)}%
\end{minipage}\hfill{}%
\begin{minipage}[t]{0.48\columnwidth}%
\centering \includegraphics[bb=6bp 0bp 525bp 411bp,clip,width=1\columnwidth]{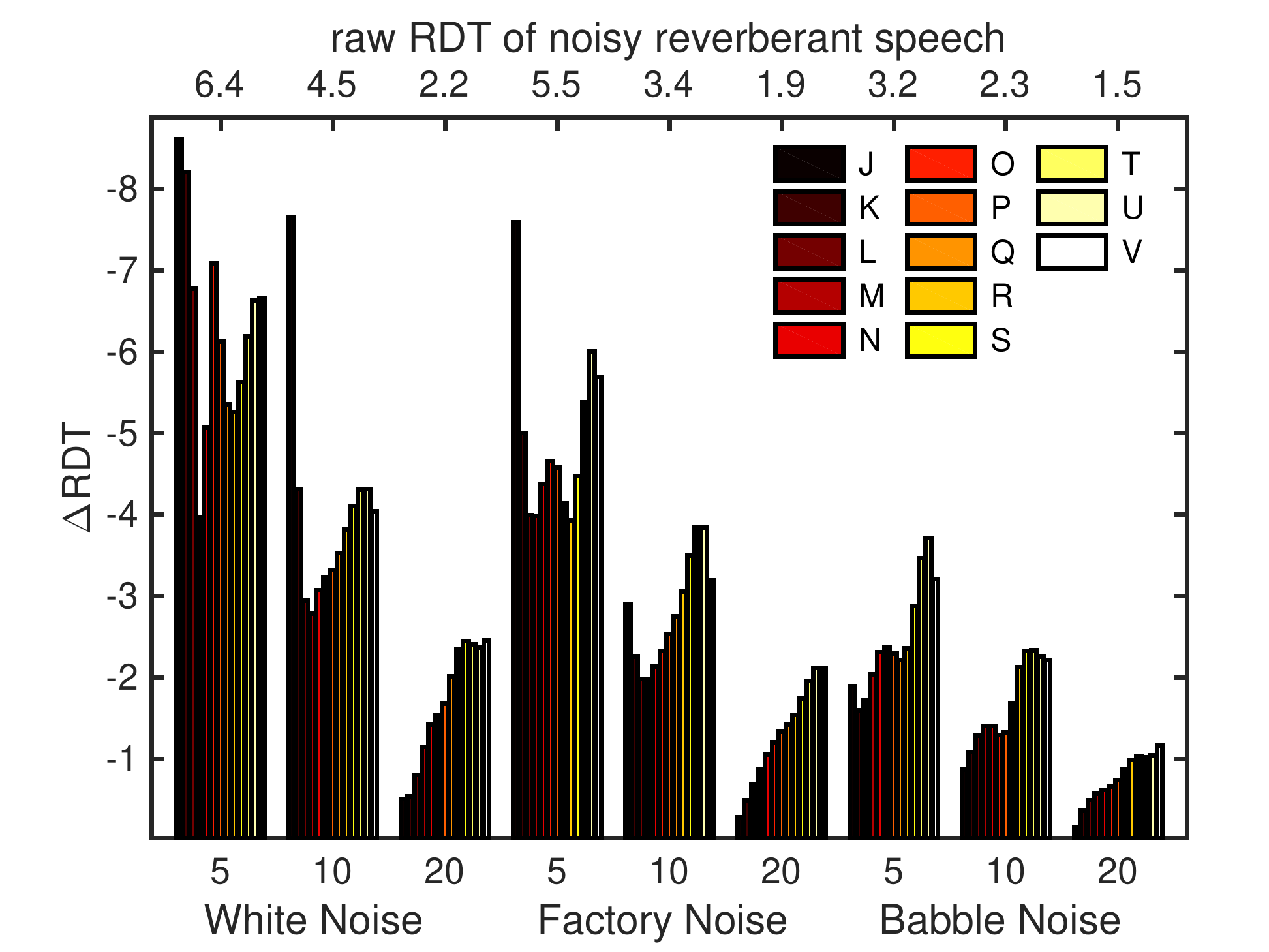}

{\footnotesize (b)}%
\end{minipage}\caption{Plot of (a) $\Delta$PESQ, and (b) $\Delta$RDT against SNR (dB) for
the KF algorithm for white, factory and babble noises. The indexes
J to V are shown.}
\end{figure}

Figures 11(c) and 11(d) show the $\Delta$PESQ against the SNR for
babble and factory noises, for cases L and R in Table \ref{tab:abchhghahhaa}
respectively, for the KF algorithm compared to the baselines. Babble
noise is used for the solid lines, the top axis and the legends while
factory noise is used for the dashed lines. The KF algorithm has a
$\Delta$PESQ that is dependent on the SNR and noise type and is,
most of the times, increasing with decreasing SNRs. Compared to the
unprocessed speech in Fig. 11(c), the algorithm has a $\Delta$PESQ
of about $0.35$ for factory noise for SNRs from $5$ to $20$ dB,
while decreasing for lower SNRs.

Figure 12 examines the RDT improvement, $\Delta$RDT, against the
SNR for white noise, for cases L and R in Table \ref{tab:abchhghahhaa},
in (a) and (b). Figure 12 also depicts the $\Delta$RDT against the
SNR for babble and factory noises, for cases L and R in Table \ref{tab:abchhghahhaa},
in (c) and (d). Babble noise is used for the solid lines, the top
axis and the legends. Factory noise is used for the dashed lines.
Figures 12(a)-(d) show that the $\Delta$RDT improves and the raw
RDT of noisy reverberant speech increases with decreasing SNR. The
$\Delta$RDT scores are better for stationary white noise than for
non-stationary factory and babble noises.

Figure 13 shows the $\Delta$PESQ and the $\Delta$RDT against the
SNR, from $5$ to $20$ dB, for the KF algorithm for white, factory
and babble noises. Figure 13 examines the acoustic conditions J to
V that correspond to the room $10 \times 7 \times 3$ m in Table \ref{tab:abchhghahhaa}.
Figure 13 differs from Figs. 11 and 12 in presenting the indexes J
to V and not only L and R. From Fig. 13(a), we observe that the $\Delta$PESQ
decreases as the $T_{60}$ increases. The $\Delta$PESQ improves with
decreasing SNR and the $\Delta$PESQ is higher for white noise than
for factory and babble noises. In Fig. 13(b), the $\Delta$RDT improves
as the $T_{60}$ increases and, moreover, the $\Delta$RDT improves
with decreasing SNR.

In summary, in this evaluation section, we have tested the KF algorithm
in different SNRs, noise types and acoustic conditions with different
$T_{60}$ and DRR values. The algorithm is effective in enhancing
distorted speech, decomposing noisy reverberant speech into speech,
reverberation and noise. Regarding noise robustness, Figs. 7-13 show
that the proposed KF algorithm achieves a significant performance
gain over different noise types and SNRs, compared to the unprocessed
noisy reverberant speech and to the examined baselines.

\section{Conclusion}

In this paper, we present a monaural speech enhancement algorithm
based on Kalman filtering in the log-magnitude spectral domain to
blindly suppress noise and reverberation while accounting for inter-frame
speech dynamics. The first two moments of the posterior distribution
of the speech log-magnitude spectrum are estimated in noisy reverberant
environments using a model that adaptively updates the $T_{60}$ and
DRR reverberation parameters. The non-linear KF algorithm updates
and tracks the two reverberation parameters of $\gamma_t$ and $\beta_t$
to further improve the estimation of the speech log-magnitude spectrum.
In this paper, we show by means of theoretical and experimental analyses
that Kalman filtering can be performed in challenging conditions by
performing the proposed signal decompositions into speech, reverberation
and noise, propagating backwards through the proposed signal model.
Experimental results using instrumental measures show improved performance
against both the $T_{60}$ and the SNR compared to the unprocessed
noisy reverberant speech and to alternative competing techniques that
perform blind denoising and dereverberation either in concatenation
or jointly.


\selectlanguage{british}%
\bibliographystyle{IEEEtran}
\addcontentsline{toc}{section}{\refname}\bibliography{sapref}

\selectlanguage{english}%
\end{document}